%% file: main.tex
\documentclass[preprint,12pt]{elsarticle}

\usepackage[
  a4paper,
  left=1.0in,
  right=1.0in,
  top=1.0in,
  bottom=1.0in
]{geometry}

\usepackage[utf8]{inputenc}
\usepackage[T1]{fontenc}
\usepackage{amsmath,amssymb}
\usepackage{graphicx}
\usepackage{cancel}
\usepackage{float}
\newif\ifreviewlines
\reviewlinesfalse
\ifreviewlines
  \usepackage{lineno}
\fi
\usepackage[hidelinks]{hyperref}

\journal{Journal of Computational Physics}

\newcommand{\uu}{\boldsymbol{u}}
\newcommand{\xx}{\boldsymbol{x}}

\usepackage{subcaption}

\begin{document}

\begin{frontmatter}

\title{A Conservative Hybrid Eulerian-Lagrangian Method with Persistent Structure Tracking for Multiscale Cavitation}

\author[wpi]{Mahdi Lavari}
\author[wpi]{Aswin Gnanaskandan\texorpdfstring{\corref{cor1}}{}}
\cortext[cor1]{Corresponding author.}
\ead{agnanaskandan@wpi.edu}

\affiliation[wpi]{organization={Department of Mechanical and Materials Engineering, Worcester Polytechnic Institute},
            addressline={100 Institute Road},
            city={Worcester},
            postcode={01609},
            state={Massachusetts},
            country={USA}}

\begin{abstract}
Hybrid Eulerian-Lagrangian methods for multiscale cavitation represent grid-resolved vapor structures in an Eulerian mixture formulation and unresolved bubbles in a Lagrangian description, requiring repeated transitions between the two representations as structures grow, collapse, and interact with resolved cavities. Existing hybrid approaches typically treat these transitions as local geometric operations, while the identity of a resolved vapor structure is reconstructed independently at each time step. This limits the ability to use structure history during representation changes and can introduce ambiguity in repeated switching, event detection, and parallel execution. This work presents a conservative hybrid Eulerian-Lagrangian framework with an embedded, on-the-fly persistent structure-tracking methodology. Resolved vapor structures are detected using connected-component labeling and reconciled across processor boundaries through a deterministic parallel labeling procedure. Successive observations are then associated using physics-informed prediction of structure position and size, allowing persistent identities, temporal histories, and birth, death, breakup, coalescence, and representation-transition events to be maintained during the simulation. The retained history is used to formulate history-aware Eulerian-to-Lagrangian transition criteria and to initialize newly created Lagrangian bubbles with the radius and radius-rate information required by the bubble-dynamics model. In addition, conservative Eulerian-to-Lagrangian and Lagrangian-to-Eulerian transfer operators are derived to preserve vapor mass and linear momentum while preventing simultaneous representation of the same vapor volume in both descriptions. The complete methodology is implemented within the parallel flow solver rather than as a post-processing procedure. Canonical verification cases demonstrate persistent identity preservation during fragmentation, coalescence, and close-crossing events, robust repeated representation switching, conservative transfer of mass and momentum, and deterministic behavior under parallel domain decomposition. The resulting framework provides a reproducible and conservative computational basis for tracking and transitioning vapor structures in multiscale cavitation simulations.
\end{abstract}


\end{frontmatter}

\ifreviewlines
  \linenumbers
\fi

\input{sections/01_introduction}
\input{sections/02_governing_equations}

\input{sections/04_e2l_transition}

\input{sections/03_l2e_transition}
\input{sections/05_conservation}
\input{sections/06_conclusions}

\section*{Acknowledgments}

This work is supported by the United States Office of Naval Research (ONR) under ONR Grant N00014-24-1-2291, monitored by Dr. Steve Martens, Dr. Raul Otero, and Dr. Craig Fernandes, and we acknowledge the support of WPI Academic \& Research Computing (ARC) in facilitating the use of the Turing Research Cluster in this work.

\section*{Disclosure}

During the preparation of this paper, we used ChatGPT to refine and improve the sentence structure of the text. We reviewed and edited the resulting content and take full responsibility for its contents.

\appendix
\input{sections/appendix_original_equations}

\input{sections/appendix_vortex_verification}

\input{sections/appendix_rp_derivation}
\input{sections/appendix_rp_verification}

\newpage

\input{jcp_references.tex}
\end{document}

%% file: sections/01_introduction.tex
\section{Introduction}
\label{sec:introduction}

Cavitation is a phase-change phenomenon in which vapor forms when the local liquid pressure falls below the saturation pressure. It occurs in hydrodynamic machinery, propulsion systems, injectors, biomedical applications, and erosion-prone flow devices, and it can produce broadband pressure fluctuations, vibration, performance loss, and material damage \cite{Karimi1986,kim2014,Feng2015,GNANASKANDAN20191743,KOKSAL2021}. Cavitating flows are inherently multiscale: vapor may appear as grid-resolved sheets and clouds that shed through re-entrant jet or condensation front mechanisms \cite{Kawanami1997,ganesh2016bubbly,LavariPOF2025,Dhruv2026}, or as a population of dispersed bubbles whose growth and collapse histories affect the later evolution of the flow.

Eulerian homogeneous mixture models \cite{GNANASKANDAN20191743,LavariPOF2025,Dhruv2026} represent resolved cavities efficiently but smear subgrid bubbles into continuum fields, whereas Lagrangian point-bubble models \cite{MAEDA2018994} resolve individual trajectories and radial dynamics but become prohibitively expensive for complex problems with many such bubbles. Hybrid Eulerian-Lagrangian strategies bridge this gap by representing large, grid-resolved cavities in the Eulerian frame and small, unresolved bubbles in the Lagrangian frame, converting between the two descriptions as structures grow or collapse. Coupled Eulerian-Lagrangian cavitation models were first demonstrated by AbdelMaksoud et al. \cite{ABDELMAKSOUD20101065} and Shams et al. \cite{Shams2011}, and later extended to marine and propulsion applications \cite{Yakubov2011,YAKUBOV2013365,Ma2015a,Ma2015b,MAEDA2018994}. Hybrid multiscale formulations that explicitly transition vapor structures between Eulerian and Lagrangian representations based on size were introduced by Apte et al. \cite{Apte2009} and formalized by Vallier \cite{Vallier2013}, and subsequently applied to sheet and tip-vortex cavitation on hydrofoils and propellers \cite{HSIAO2017102,Hsiao2015a,MA201768}, to propeller-noise prediction \cite{Lidtke2017}, and to compressible, four-way-coupled formulations that capture inter-bubble shielding, collisions, and breakup \cite{GHAHRAMANI2019339,Ghahramani2021,Madabhushi_Mahesh_2023,Zhao2024}. More recent studies have linked hybrid Eulerian-Lagrangian nuclei populations to cavitation noise, erosion risk, and water-quality effects \cite{Wang2024,WANG2025120398,Wang2025ijfe,Yang2024,QIN2025105142,Zhang2025}.

Despite this progress, representation transfer between the Eulerian and Lagrangian descriptions remains a fundamental numerical issue in hybrid methods. Removing a vapor structure from one representation and introducing it into the other requires transfer operators that preserve the quantities governed by the conservation laws. Without appropriately derived coupling terms, and without a mechanism that prevents a resolved cavity and an unresolved bubble from simultaneously representing the same vapor volume, a hybrid framework can introduce artificial changes in mass or momentum during a representation change even when the individual submodels are themselves accurate \cite{Ghahramani2021}.

A separate challenge arises from the temporal identity of the resolved Eulerian structures undergoing these transitions. Detecting vapor structures at a single instant is a well-established problem: connected-component labeling (CCL), coupled to Lagrangian point-particle tracking, has been used to identify resolved structures and transfer them to a dispersed-phase description, as in the coupling procedure of Herrmann \cite{Herrmann2010}. A CCL label, however, identifies an instantaneous connected component rather than a persistent physical structure. The numerical label may change between time steps because of mesh decomposition, processor traversal order, translation across the mesh, coalescence, breakup, or temporary loss of detection. Consequently, instantaneous labeling alone does not provide the temporal history required for history-dependent transition decisions or for reconstructing quantities such as the radius rate needed to initialize a newly created Lagrangian bubble. It can also make birth, death, breakup, coalescence, and repeated representation-transition records dependent on the numerical labeling procedure rather than on the evolution of the underlying physical structure.

A separate line of work has developed algorithms that link instantaneous observations into persistent bubble and droplet identities and lineages, predominantly as a post-processing step applied after a simulation has completed. Building on the optimal-network formulation of Gao et al. \cite{Gao2021ON}, which matches structures between adjacent snapshots using volume, position, and velocity constraints, subsequent tools have added lineage and breakup/coalescence statistics \cite{Chan2021Lineage}, kinematic and topological matching criteria \cite{Mangani2022,Bussmann2022Tracking,GonzalezDiaz2018Topological,Li2022Multiscale,HEINRICH2020,MENON2026134117}, optimal bipartite matching \cite{Liu2024Kuhn}, genealogy extraction \cite{Rubel2019Owkes}, and grid-level Eulerian label advection for conservative volume-based tracking \cite{Gaylo2022}. Most recently, Basak et al. \cite{Basak2026ON} introduced a parallel optimal-network algorithm that detects non-binary fragmentation and coalescence events using octree-refined shape descriptors, and showed that neglecting non-binary events can degrade the fraction of successfully traced droplet genealogies from over 85\% to below 2\% at practical output intervals. Several of these approaches, including that of Basak et al., are themselves parallelized to make offline tracking tractable for large structure counts.

The requirements of a hybrid Eulerian-Lagrangian solver, however, differ fundamentally from those of retrospective structure tracking. In an offline procedure, identities and lineages are reconstructed from stored snapshots after the corresponding flow states have already been computed. Every representation-transition decision has therefore already been made by the time this information becomes available. In a hybrid solver, structure identity must instead be determined online, using only information available at the current and preceding time steps, so that the accumulated history can directly inform representation-transition decisions. The present framework addresses this requirement by associating each new Eulerian observation with a prediction extrapolated from the accumulated history of an existing track. 

This paper presents a conservative hybrid Eulerian-Lagrangian framework in which persistent tracking of resolved vapor structures is embedded directly within the flow solver and coupled to the representation-transition procedure. Parallel connected-component labeling is used to identify resolved vapor structures, after which physics-informed prediction and deterministic association convert successive instantaneous observations into persistent tracks. The resulting histories support history-dependent transition eligibility, event logging, and reconstruction of the radius and radius rate required to initialize the bubble-dynamics model when a resolved Eulerian structure transitions to the Lagrangian representation. In parallel, conservative Eulerian-to-Lagrangian and Lagrangian-to-Eulerian transfer operators are formulated to preserve vapor mass and linear momentum while preventing simultaneous Eulerian and Lagrangian representation of the same vapor volume.


The principal contributions of this work are the following:
\begin{itemize}

    \item An on-the-fly persistent structure-tracking methodology is developed that converts instantaneous Eulerian connected components into persistent physical tracks using physics-informed prediction and deterministic association.

    \item The structure-tracking procedure is integrated directly into a parallel flow solver, including reconciliation of connected structures across processor boundaries, enabling structure detection, association, transition decisions, and event logging to be performed online and deterministically under domain decomposition.

    \item Persistent structure histories are used to formulate history-dependent representation-transition criteria and to reconstruct the dynamical information required to initialize newly created Lagrangian bubbles during Eulerian-to-Lagrangian transitions.

    \item Conservative Eulerian-to-Lagrangian and Lagrangian-to-Eulerian transfer operators are formulated to preserve vapor mass and linear momentum during representation changes while preventing simultaneous occupation of the same vapor volume by both representations.

    \item A dedicated verification suite assesses persistent-ID preservation, birth and death detection, fragmentation and coalescence handling, repeated representation switching, conservation of mass and momentum, and deterministic parallel execution.

\end{itemize}

%% file: sections/02_governing_equations.tex
\section{Governing equations}\label{sec:governing}\label{sec:gov_eqns}

The multiscale Eulerian-Lagrangian model considered in this study is formulated as a decomposition into resolved and unresolved components: a continuous liquid phase (resolved liquid) and a vapor phase represented in two forms, namely a resolved Eulerian (continuum) vapor field and an unresolved Lagrangian (dispersed) bubble. To ensure model robustness and internal consistency, the following assumptions are adopted:

\begin{enumerate}
    \item The resolved (Eulerian) and unresolved (Lagrangian) vapor phases are assigned identical thermophysical properties (e.g., density and viscosity).
    \item The material properties (density and viscosity) of the resolved phases are taken as constant, and each resolved phase is treated as individually incompressible. The resolved mixture is however, compressible in the sense that mixture density can change as a function of vapor volume fraction. 
    \item The resolved (Eulerian) vapor and unresolved (Lagrangian) vapor are enforced to be spatially non-overlapping; i.e., they are not permitted to simultaneously occupy the same computational control volume. In contrast, the resolved liquid may coexist with either resolved vapor or unresolved vapor within a control volume, consistent with an Eulerian mixture description.
    \item A no-slip constraint is imposed between the resolved liquid and resolved (Eulerian) vapor, whereas a slip (relative) velocity is permitted between the resolved liquid and the unresolved (Lagrangian) vapor structures.
\end{enumerate}

\noindent Consequently, transferring a vapor structure to the Lagrangian representation should not be interpreted as removing its contribution from the Eulerian framework. Rather, the unresolved vapor remains part of the overall Eulerian field through appropriate coupling terms, while its position and its mass and momentum evolution are advanced in the Lagrangian framework. Hence, this method can be seen as a unified multiscale framework.

\subsection{Unified Approach for the Multiscale Framework}

The homogeneous mixture model, used for modeling in the Eulerian framework, is based on the assumption that the liquid-vapor system can be treated as an effective single-fluid continuum. Accordingly, the governing material properties (e.g., density and viscosity) are defined as mixture quantities obtained from suitable phase averaging. The mixture composition is parameterized by the volume fraction of one of the phases; here, the liquid volume fraction $\alpha_{\mathrm{l}}$ is employed, defined as the ratio of the liquid volume to the total volume of a computational control volume. Under these assumptions, the mixture level conservation equations for mass and momentum can be written as

\begin{equation}
\frac{\partial \rho_{\mathrm{m}}}{\partial t} + \frac{\partial}{\partial x_i} \left(\rho_{\mathrm{m}} u_i\right) = 0,
\label{eqn:massE}
\end{equation}

\begin{equation}
\frac{\partial}{\partial t} (\rho_{\mathrm{m}} u_i) + \frac{\partial}{\partial x_j} (\rho_{\mathrm{m}} u_i u_j) = -\frac{\partial p}{\partial x_i} +\frac{\partial \tau_{ij}}{\partial x_j} + \rho_{\mathrm{m}} g_i.
\label{eqn:momentumE}
\end{equation}

\noindent Here, $\rho_m$ denotes the mixture density, $t$ is time, $x_i$ is the $i$-th spatial coordinate, and $u_i$ is the $i$-th component of the mixture velocity. Also, $p$ is the pressure, $\tau_{ij}$ is the (viscous) stress tensor, comprising the viscous stress and the capillary
(surface-tension) stress, and $g_i$ is the $i$-th component of the gravitational acceleration. Following Ghahramani et al. \cite{Ghahramani2021}, we introduce $\beta_{\mathrm{l}}$, the unresolved liquid volume fraction, to account for Lagrangian subgrid vapor content. In any computational cell that contains unresolved vapor represented in the Lagrangian framework, $\beta_{\mathrm{l}}$ satisfies $0 \le \beta_{\mathrm{l}} < 1$, whereas the resolved liquid volume fraction $\alpha_{\mathrm{l}}$ remains equal to unity (i.e., no resolved Eulerian vapor is present in that cell). Consequently, $\beta_{\mathrm{l}}$ can be written as

\begin{equation}
\quad \beta_{\mathrm{l}} \in [0, 1), \quad \text{where} \quad \alpha_{\mathrm{l}} = 1.
\label{eqn:betaaa}
\end{equation}

\noindent On the other hand, \(\alpha_{\mathrm{l}}\) is given by

\begin{equation}
\quad \alpha_{\mathrm{l}} \in [0, 1], \quad \text{where} \quad \beta_{\mathrm{l}} = 1.
\end{equation}

\noindent Using \(\alpha_{\mathrm{l}}\) and \(\beta_{\mathrm{l}}\) based on the assumptions stated before, the definitions of density, viscosity, and momentum are given by

\begin{equation}
\rho_{\mathrm{m}} = (\alpha_{\mathrm{l}} + \beta_{\mathrm{l}} - 1) \rho_{\mathrm{l}} + (1 - \alpha_{\mathrm{l}}) \rho_{\mathrm{v}} + (1 - \beta_{\mathrm{l}}) \rho_{\mathrm{v}},
\label{eqn:densityL}
\end{equation}

\begin{equation}
\mu_{\mathrm{m}} = (\alpha_{\mathrm{l}} + \beta_{\mathrm{l}} - 1) \mu_{\mathrm{l}} + (1 - \alpha_{\mathrm{l}}) \mu_{\mathrm{v}} + (1 - \beta_{\mathrm{l}}) \mu_{\mathrm{v}},
\label{eqn:viscosityL}
\end{equation}

\begin{equation}
\rho_{\mathrm{m}} u_i = (\alpha_{\mathrm{l}} + \beta_{\mathrm{l}} - 1) \rho_{\mathrm{l}} u_i + (1 - \alpha_{\mathrm{l}}) \rho_{\mathrm{v}} u_i + (1 - \beta_{\mathrm{l}}) \rho_{\mathrm{v}} u_{Li}.
\label{eqn:momentumL}
\end{equation}

\noindent Here, \(u_{Li}\) is the unresolved velocity in the Lagrangian framework, which differs from \(u_i\) in the Eulerian field based on the slip assumption. By substituting Eqn.~\eqref{eqn:densityL} and Eqn.~\eqref{eqn:momentumL} into Eqn.~\eqref{eqn:massE}, the mass conservation equation, expressed in terms of the velocity divergence, can be written as:

\begin{equation}
\begin{aligned}
\frac{\partial u_i}{\partial x_i} = &
\ \frac{\rho_{\mathrm{v}} - \rho_{\mathrm{l}}}{\rho_{\mathrm{v}}} 
\left[ 
\frac{\partial \alpha_{\mathrm{l}}}{\partial t} 
+ \frac{\partial (\alpha_{\mathrm{l}} u_i)}{\partial x_i} 
\right]  - \frac{1}{\beta_{\mathrm{l}} \rho_{\mathrm{l}}} 
\left[ 
\frac{\partial \left( (1-\beta_{\mathrm{l}}) \rho_{\mathrm{v}} \right)}{\partial t} 
+ \frac{\partial \left( (1-\beta_{\mathrm{l}}) \rho_{\mathrm{v}} u_{Li} \right)}{\partial x_i} 
\right] \\
& - \frac{1}{\beta_{\mathrm{l}}} 
\left[ 
\frac{\partial \beta_{\mathrm{l}}}{\partial t} 
+ u_i \frac{\partial \beta_{\mathrm{l}}}{\partial x_i} 
\right].
\end{aligned}
\label{eqn:hybrid-non-divergence-free}
\end{equation}


\noindent By substituting Eqs.~\eqref{eqn:densityL} and \eqref{eqn:momentumL} into Eqn.~\eqref{eqn:momentumE}, we obtain the momentum equation
given by

\begin{align}
& \left[ 
\frac{\partial \left( \alpha_{\mathrm{l}} \rho_{\mathrm{l}} + (1-\alpha_{\mathrm{l}}) \rho_{\mathrm{v}} \right) u_i}{\partial t} 
+ \frac{\partial \left( \alpha_{\mathrm{l}} \rho_{\mathrm{l}} + (1-\alpha_{\mathrm{l}}) \rho_{\mathrm{v}} \right) u_i u_j}{\partial x_j} 
\right] \notag \\
&\quad - \left[ 
\frac{\partial \left( (1-\beta_{\mathrm{l}}) \rho_{\mathrm{l}} u_i \right)}{\partial t} 
+ \frac{\partial \left( (1-\beta_{\mathrm{l}}) \rho_{\mathrm{l}} u_i u_j \right)}{\partial x_j} 
\right] \notag \\
&= - \left[ 
\frac{\partial \left( (1-\beta_{\mathrm{l}}) \rho_{\mathrm{v}} u_{Li} \right)}{\partial t} 
+ \frac{\partial \left( (1-\beta_{\mathrm{l}}) \rho_{\mathrm{v}} u_{Li} u_j \right)}{\partial x_j} 
\right] \notag \\
&\quad - \frac{\partial p}{\partial x_i} 
+ \frac{\partial \tau_{ij}}{\partial x_j} 
+ \rho_{\mathrm{m}} g_i.
\label{eqn:hybrid-mom}
\end{align}

The stress tensor $\tau_{ij}$ appearing in Eqns.~\eqref{eqn:momentumE} and~\eqref{eqn:hybrid-mom}
includes, in addition to the viscous stress, a capillary (surface-tension) contribution that is
retained here because several of the cases considered in this work are capillary-driven. This
contribution is modeled using the continuum surface force (CSF) approach of Brackbill et
al.~\cite{Brackbill1992}, in which the capillary part of the stress divergence
$\partial \tau_{ij}/\partial x_j$ is expressed as a volumetric body force localized to the
interface by the resolved liquid volume-fraction gradient,

\begin{equation}
f_{\sigma,i} = \sigma\,\kappa_{\sigma}\,\frac{\partial \alpha_{\mathrm{l}}}{\partial x_i},
\qquad
\kappa_{\sigma} = -\,\frac{\partial n_i}{\partial x_i},
\qquad
n_i = \frac{\partial \alpha_{\mathrm{l}}/\partial x_i}
{\left| \partial \alpha_{\mathrm{l}}/\partial x_k \right|},
\label{eqn:csf}
\end{equation}

\noindent Here, $f_{\sigma,i}$ is the capillary part of the stress divergence, $\sigma$ is the
surface-tension coefficient, $\kappa_{\sigma}$ is the local interface curvature, and $n_i$ is the
interface unit-normal vector obtained from the resolved liquid volume fraction
$\alpha_{\mathrm{l}}$. This contribution acts only on the resolved (Eulerian) vapor; capillary
effects on unresolved Lagrangian bubbles are represented separately through the bubble-dynamics
model. Eqn.~\eqref{eqn:hybrid-non-divergence-free} and Eqn.~\eqref{eqn:hybrid-mom} account for the presence of the unresolved bubbles, with terms containing \(\beta_{\mathrm{l}}\) representing their exact impact on the Eulerian framework. 
To close the above system of Eulerian governing equations, the temporal evolution of the phase volume fraction is obtained by solving a scalar transport equation, given by

\begin{equation}
\frac{\partial \alpha_{\mathrm{l}}}{\partial t} 
+ \frac{\partial (\alpha_{\mathrm{l}} u_i)}{\partial x_i} = \frac{\dot{m}}{\rho_{\mathrm{l}}}.
\label{eqn:alphal}
\end{equation}

\noindent Here, $\dot{m}$ denotes the volumetric mass-transfer source term associated with phase change, i.e., evaporation and condensation. To model these processes, finite-rate mass-transfer closures such as those proposed by Kunz et al. \cite{KUNZ2000}, Merkle et al. \cite{Merkle1998}, Singhal et al. \cite{Singhal2002}, and Saito et al. \cite{saito2007numerical} may be employed. In the present work, the Schnerr-Sauer model \cite{Schnerr2001} is adopted, in which cavitation is activated in regions where the local pressure drops below the saturation vapor pressure. The corresponding mass-transfer rate expressions are given by

\begin{equation}
    \dot{m}_c = C_c \alpha_{\mathrm{l}}(1 - \alpha_{\mathrm{l}}) \frac{3 \rho_{\mathrm{l}} \rho_{\mathrm{v}}}{\rho_{\mathrm{m}} R_{\text{B}}} \left( \frac{2}{3 \rho_{\mathrm{l}}} \left| p - p_{\text{sat}} \right| \right)^{1/2}, \quad p \geq p_{\text{sat}},
    \label{eqn:cccc}
\end{equation}

\begin{equation}
    \dot{m}_{\mathrm{v}} = C_{\mathrm{v}} \alpha_{\mathrm{l}} \left(1 + \alpha_{Nc} - \alpha_{\mathrm{l}} \right) \frac{3 \rho_{\mathrm{l}} \rho_{\mathrm{v}}}{\rho_{\mathrm{m}} R_{\text{B}}} \left( \frac{2}{3 \rho_{\mathrm{l}}} \left| p - p_{\text{sat}} \right| \right)^{1/2}, \quad p < p_{\text{sat}}.
    \label{eqn:vvvvv}
\end{equation}

\noindent Here, \( \dot{m}_c \) and \( \dot{m}_{\mathrm{v}} \) represent the mass transfer rates due to condensation and vaporization, respectively. Unless specified otherwise, the coefficients \( C_c =1 \) and \( C_{\mathrm{v}} = 1\), following Lavari and Gnanaskandan \cite{LavariPOF2025}, are empirical parameters that control the rates of condensation and vaporization. The variables \( p \) and \( p_{\text{sat}} \) refer to the local static pressure and the saturation vapor pressure, respectively. \( R_{\text{B}} \) and \( \alpha_{Nc} \) denote the radius and void fraction of bubble nuclei in the liquid. These parameters are determined as

\begin{equation}
    \alpha_{Nc} = \frac{\pi n_0 d_{Nc}^3}{6} \left( 1 + \frac{\pi n_0 d_{Nc}^3}{6} \right)^{1/2},
\end{equation}

\begin{equation}
    R_B = \left( \frac{3}{4 \pi n_0} \frac{1 + \alpha_{Nc} - \alpha}{\alpha} \right)^{1/3}.
\end{equation}

\noindent Here, \( n_0 = 1.6 \times 10^{13} \) and \( d_{N_c} = 2 \times 10^{-6} \) are user-defined parameters, following Schnerr and Sauer \cite{Schnerr2001}, signifying the number of nuclei per liquid cubic meter and the diameter of nucleation sites, respectively. 





\subsection{Lagrangian Source Terms within the Unified Framework}

The definition of total derivative can be utilized to connect the Lagranigan terms with the Eulerian framework. The volumetric rate of change of the unresolved (Lagrangian) vapor mass within a computational cell is given as

\begin{equation}
\begin{aligned}
\dot{m}_{\text{b}} = \frac{1}{V} \frac{\text{D} m_{\text{b}}}{\text{D}t} &= \left[ \frac{\partial (1 - \beta_{\mathrm{l}}) \rho_{\mathrm{v}}}{\partial t} + \frac{\partial (1 - \beta_{\mathrm{l}}) \rho_{\mathrm{v}} u_{L_i}}{\partial x_i} \right].
\end{aligned}
\label{eqn:mbubble2}
\end{equation}

\noindent Following the same approach, the rate of change of momentum per unit volume, i.e., the total force per unit volume acting on the unresolved vapor, is given by


\begin{equation}
\frac{F}{V} = \frac{1}{V}\frac{\text{D}(m_{\text{b}} u_{L_i})}{\text{D}t} =  \frac{1}{V} \frac{\text{D} V (1 - \beta_{\mathrm{l}}) \rho_{\mathrm{v}} u_{L_i}}{\text{D}t} \\
= \frac{\partial \!\big((1-\beta_{\mathrm{l}})\rho_{\mathrm{v}} u_{L_i}\big)}{\partial t}
   + \frac{\partial \!\big((1-\beta_{\mathrm{l}})\rho_{\mathrm{v}} u_{L_i} u_j\big)}{\partial x_j}\,.
\label{eqn:force2}
\end{equation}

\noindent The final hybrid Eulerian-Lagrangian governing equations are given by

\begin{equation}
\frac{\partial u_i}{\partial x_i} = \left( \frac{1}{\rho_{\mathrm{l}}} - \frac{1}{\rho_{\mathrm{v}}} \right) \dot{m}  -\frac{1}{\beta_{\mathrm{l}} \rho_{\mathrm{l}}} \left[ \dot{m}_{\text{b}} \right] 
- \frac{1}{\beta_{\mathrm{l}}} \left[ \frac{\text{D} \beta_{\mathrm{l}}}{\text{D} t} \right],
\label{eqn:hybrid-non-divergence-free2}
\end{equation}

\begin{align}
& \left[ 
\frac{\partial \left( \alpha_{\mathrm{l}} \rho_{\mathrm{l}} + (1-\alpha_{\mathrm{l}}) \rho_{\mathrm{v}} \right) u_i}{\partial t} 
+ \frac{\partial \left( \alpha_{\mathrm{l}} \rho_{\mathrm{l}} + (1-\alpha_{\mathrm{l}}) \rho_{\mathrm{v}} \right) u_i u_j}{\partial x_j} 
\right] \notag \\
&\quad - \left[ 
\frac{\partial \left( (1-\beta_{\mathrm{l}}) \rho_{\mathrm{l}} u_i \right)}{\partial t} 
+ \frac{\partial \left( (1-\beta_{\mathrm{l}}) \rho_{\mathrm{l}} u_i u_j \right)}{\partial x_j} 
\right] \notag \\
&= - \left[ \frac{F}{V} \right] 
- \frac{\partial p}{\partial x_i} 
+ \frac{\partial \tau_{ij}}{\partial x_j} 
+ \rho_{\mathrm{m}} g_i.
\label{eqn:hybrid-mom2}
\end{align}

\noindent The above equations correspond to the conservation of mass and momentum, respectively, and ensure a consistent coupling between the Eulerian and Lagrangian representations. Within the Lagrangian framework, bubble dynamics are modeled to obtain the interfacial mass-transfer rate $\dot{m}_{\mathrm{b}}$ and the material rate of change $\mathrm{D}\beta_{\mathrm{l}}/\mathrm{D}t$. To further assess consistency with established formulations in the literature, a consistency analysis of the governing equations is shown to result in the conventional homogeneous mixture model and conventional resolved liquid-subgrid gas Eulerian-Lagrangian models in \ref{appA}.


\subsubsection{Lagrangian Bubble Kinematics and Dynamics}

Within the Lagrangian framework, each bubble is characterized by its centroid position $\boldsymbol{x}_b(t)$, translational velocity $\mathbf{u}_b(t)$, and radius $R(t)$. The centroid position is advanced in time according to the kinematic relation

\begin{equation}
    \frac{\text{D} \boldsymbol{x}_b}{\text{D}t} = \mathbf{u}_b,
    \label{eq:dbm_kinematics}
\end{equation}

\noindent where $\mathbf{u}_b$ is determined by solving Newton's second law of motion for the bubble. A collective set of hydrodynamic forces affects the bubble's velocity, imposed by the surrounding liquid. The formulation for spherical bubbles may be written as

\begin{equation}
     \frac{\text{D} (m_b \mathbf{u}_b) }{\text{D}t} =
    \mathbf{F}_d + \mathbf{F}_l + \mathbf{F}_a + \mathbf{F}_p + \mathbf{F}_b + \mathbf{F}_g.
    \label{eq:dbm_momentum}
\end{equation}

\noindent Here, $m_b$ denotes the bubble mass, $\mathbf{F}_d$ the drag force, $\mathbf{F}_l$ the lift force, $\mathbf{F}_a$ the added-mass force, $\mathbf{F}_p$ the pressure-gradient force, $\mathbf{F}_b$ the buoyancy force, and $\mathbf{F}_g$ the gravitational force. This vector force balance is distinct from the scalar per-volume exchange term $F$ introduced in Eqn.~\eqref{eqn:force2}, which represents the bulk momentum source transferred to the Eulerian field rather than the force on an individual bubble. In this formulation, each bubble is subjected to a set of hydrodynamic and body-force contributions that govern its translational motion and, consequently, its trajectory. The reduced order models for each of the hydrodynamics forces follow Lavari et al. \cite{Lavari2026SNH, Lavari2026USNCTAM}.
Altogether, these drag, lift, added mass, pressure gradient, buoyancy, and gravity terms describe the full spectrum of hydrodynamic and body forces on a bubble. Their explicit resolution in the Lagrangian framework enables capturing the detailed dynamics of cavitation bubbles that would otherwise be smeared out in Eulerian mixture approaches. A verification of the resulting bubble-kinematics implementation against a classical vortex-entrainment benchmark is given in \ref{appVortex}.

\begin{figure}[t]
  \centering
  \includegraphics[width=0.3\linewidth]{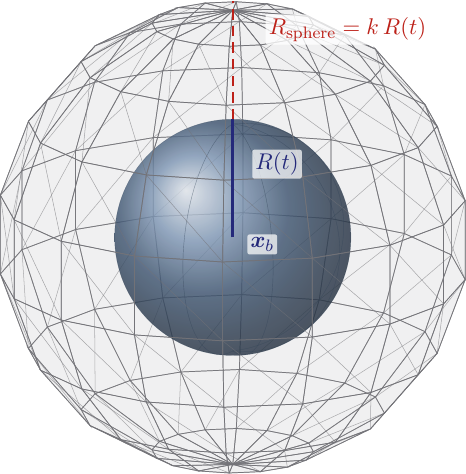}
  \caption{Equal-area spherical-shell sampling used to evaluate the localized forcing $p(kR)$. The bubble (inner, opaque sphere) of radius $R(t)$ is centered at $\boldsymbol{x}_b$. The transparent outer sampling shell of radius $R_{\text{sphere}} = k\,R(t)$ is partitioned into approximately equal-area panels with sampling points placed at each panel center to reduce mesh dependence and numerical noise.}
  \label{fig:sphere_sampling}
\end{figure}

For bubble dynamics, we consider a localized Rayleigh-Plesset model in which the bubble forcing is given by the resolved pressure on a finite spherical shell of radius $r = kR(t)$ with $k>1$. This choice incorporates local hydrodynamic effects while preserving the analytical tractability of the original Rayleigh-Plesset equation \cite{Plesset1977} structure driven by pressure at infinity. The derivation and a consistency analysis of this equation with the existing literature are presented in \ref{appB}. This localized Rayleigh-Pleseet equation is given by

\begin{equation}
  \Big(1 - \frac{1}{k}\Big) R \ddot{R}
  \;+\;
  \Big(\frac{3}{2} - \frac{2}{k} + \frac{1}{2k^4}\Big) \dot{R}^{\,2}
  \;=\; \frac{1}{\rho}\left(
      p_b - \frac{2\sigma}{R} - \frac{4\mu\,\dot R}{R}
      - p(kR)
  \right),
  \label{eq:finite-RP}
\end{equation}

\noindent where \( k \) denotes the non-dimensional distance of the measuring sphere (or the location where the pressure is evaluated) relative to the bubble radius. 
In order to evaluate the localized pressure forcing term $p(kR)$, we compute a spherical-shell average of the surrounding Eulerian fields (pressure, density, and viscosity) on a surface of radius $r = kR(t)$ centered at the bubble centroid, as shown in Figure~\ref{fig:sphere_sampling}. This procedure ensures that the forcing reflects the resolved near-field flow environment rather than the value at a single grid point, thereby reducing sensitivity to numerical noise and mesh dependence.

\noindent The spherical surface is discretized into $N_\theta$ azimuthal and $N_\phi$ polar divisions and corresponding sampling angles are chosen to achieve approximately uniform coverage of the surface area:

\begin{align}
  \theta_i &= \frac{2\pi\,(i+0.5)}{N_\theta}, \qquad i = 0,\dots,N_\theta-1,\\
  \phi_j &= \arccos\!\left(1 - \frac{2(j+0.5)}{N_\phi}\right), \qquad j = 0,\dots,N_\phi-1.
\end{align}

\noindent This angular discretization corresponds to an equal-area partitioning of the sphere, which minimizes sampling bias toward the poles. Each sampling location is then obtained as

\begin{equation}
  \boldsymbol{x}_{ij} \;=\; \boldsymbol{x}_b + R_{\text{sphere}}
  \begin{bmatrix}
    \sin\phi_j \cos\theta_i \\[4pt]
    \sin\phi_j \sin\theta_i \\[4pt]
    \cos\phi_j
  \end{bmatrix},
\end{equation}

\noindent where $\boldsymbol{x}_b$ is the instantaneous bubble centroid position. At each valid sampling location $\boldsymbol{x}_{ij}$, the Eulerian fields are interpolated from the underlying computational mesh

\begin{equation}
  \rho_{ij} = \rho(\boldsymbol{x}_{ij}),\quad
  \mu_{ij} = \mu(\boldsymbol{x}_{ij}),\quad
  p_{ij} = p(\boldsymbol{x}_{ij}).
\end{equation}

\noindent Finally, the spherical averages of the sampled fields are evaluated as

\begin{align}
  \rho(kR) &= \frac{1}{N_s}\sum_{i,j}\rho_{ij},\\
  \mu(kR) &= \frac{1}{N_s}\sum_{i,j}\mu_{ij},\\
  p(kR) &= \frac{1}{N_s}\sum_{i,j}p_{ij},
\end{align}
where $N_s = N_\theta N_\phi$ is the total number of sampling points. This averaging provides a smooth and physically consistent estimate of the pressure forcing term acting on the bubble surface. Since Ghahramani et al. \cite{Ghahramani2021} similarly performed simulations with $k=2$, we adopt $k=2$ for the remainder of the present study. A verification of the integration scheme used to solve this equation against classical benchmark solutions in the limit where $k$ goes to infinity is given in \ref{subsec:rp_kinf_verification}.

\subsection{Implementation of the Hybrid Model}

The governing equations described above are solved using OpenFOAM-v2406, following the tensorial finite-volume approach of Weller et al. \cite{weller1998tensorial}. To treat multiphase flows with phase change, the simulations employ the PIMPLE procedure, a hybrid pressure-velocity coupling strategy that combines the Pressure-Implicit with Splitting of Operators (PISO) and the Semi-Implicit Method for Pressure-Linked Equations (SIMPLE). This framework provides a consistent iterative update of the liquid volume fraction, velocity field, and pressure within each time step. In the adopted solution sequence, the liquid volume fraction $\alpha_{\mathrm{l}}$ is advanced first by integrating the corresponding transport equation \cite{Lavari2024}, written as

\begin{equation}
\frac{\partial \alpha_{\mathrm{l}}}{\partial t} + \nabla \cdot (\alpha_{\mathrm{l}} \mathbf{U}) +\alpha_{\mathrm{l}}\nabla \cdot \mathbf{U} = \frac{\dot{m}}{\rho_{\mathrm{l}}} - \alpha_{\mathrm{l}}\left( \frac{1}{\rho_{\mathrm{l}}} - \frac{1}{\rho_{\mathrm{v}}} \right)  \dot{m}.
\label{eqn:alphal2}
\end{equation}

\noindent The temporal derivative in Eqn.~\eqref{eqn:alphal2} is discretized using a first-order implicit Euler scheme. For the spatial discretization of the liquid volume-fraction transport, the convective term is approximated using a TVD van Leer scheme, while the additional compression-related term is discretized using Gauss linear scheme. To enforce boundedness of the liquid volume fraction, a multidimensional universal limiter with explicit solution (MULES) \cite{greenshieldsweller2022} is applied with two corrector iterations and two sub-cycles per time step, and 10 limiter iterations. After updating $\alpha_{\mathrm{l}}$, the mixture properties are recomputed using Eqn.~\eqref{eqn:densityL} and Eqn.~\eqref{eqn:viscosityL} for the mixture density and mixture viscosity, respectively. With these updated properties, the discretized and linearized momentum equation can be written as

\begin{equation}
a_P^u u_P + \sum_N a_N^u u_N = r - \nabla p.
\label{eqn:momentum}
\end{equation}

\noindent Here, \(a_P^u\) and \(a_N^u\) are the diagonal and non-diagonal parts of the coefficient matrix, respectively, while \(r\) represents the source terms (including the source term from the Lagrangian framework) in Eqn.~\eqref{eqn:hybrid-mom2}. We introduce \(\mathbf{H(u)}\) which includes the source terms in addition to the non-diagonal terms of the coefficient matrix as

\begin{equation}
 \mathbf{H(u)}  = r - \sum_N a_N^u u_N.
\label{eqn:Hoperator}
\end{equation}

\noindent Substituting Eqn.~\eqref{eqn:Hoperator} into Eqn.~\eqref{eqn:momentum} yields

\begin{equation}
a_P^u u_P = \mathbf{H(u)} - \nabla p,
\end{equation}
or
\begin{equation}
u_P = (a_P^u)^{-1} \left( \mathbf{H(u)} - \nabla p\right).
\label{eqn:Up}
\end{equation}

\noindent The convective term in the momentum equation is discretized using a second-order upwind-biased linear-upwind scheme that leverages upstream gradients. The resulting linear system for the momentum predictor is solved using the symmetric Gauss-Seidel approach with a tolerance of \(10^{-10}\). After this momentum step, the predicted velocity field of the flow must satisfy the continuity condition, Eqn.~\eqref{eqn:hybrid-non-divergence-free2}. Substituting Eqn.~\eqref{eqn:Up} into Eqn.~\eqref{eqn:hybrid-non-divergence-free2} yields the pressure correction equation, given as

\begin{equation}
\nabla \cdot \left[ (a_P^u)^{-1} \nabla p \right] = \nabla \cdot \left[ (a_P^u)^{-1} \mathbf{H(u)} \right] + \left( \frac{1}{\rho_{\mathrm{l}}} - \frac{1}{\rho_{\mathrm{v}}} \right) \dot{m} -\frac{1}{\beta_{\mathrm{l}} \rho_{\mathrm{l}}} \left[ \dot{m}_{\text{b}} \right] 
- \frac{1}{\beta_{\mathrm{l}}} \left[ \frac{\text{D} \beta_{\mathrm{l}}}{\text{D} t} \right].
\label{eqn:nonfree}
\end{equation}

\begin{figure}[]
\centering
\includegraphics[width=0.96\linewidth,height=0.95\textheight,keepaspectratio]{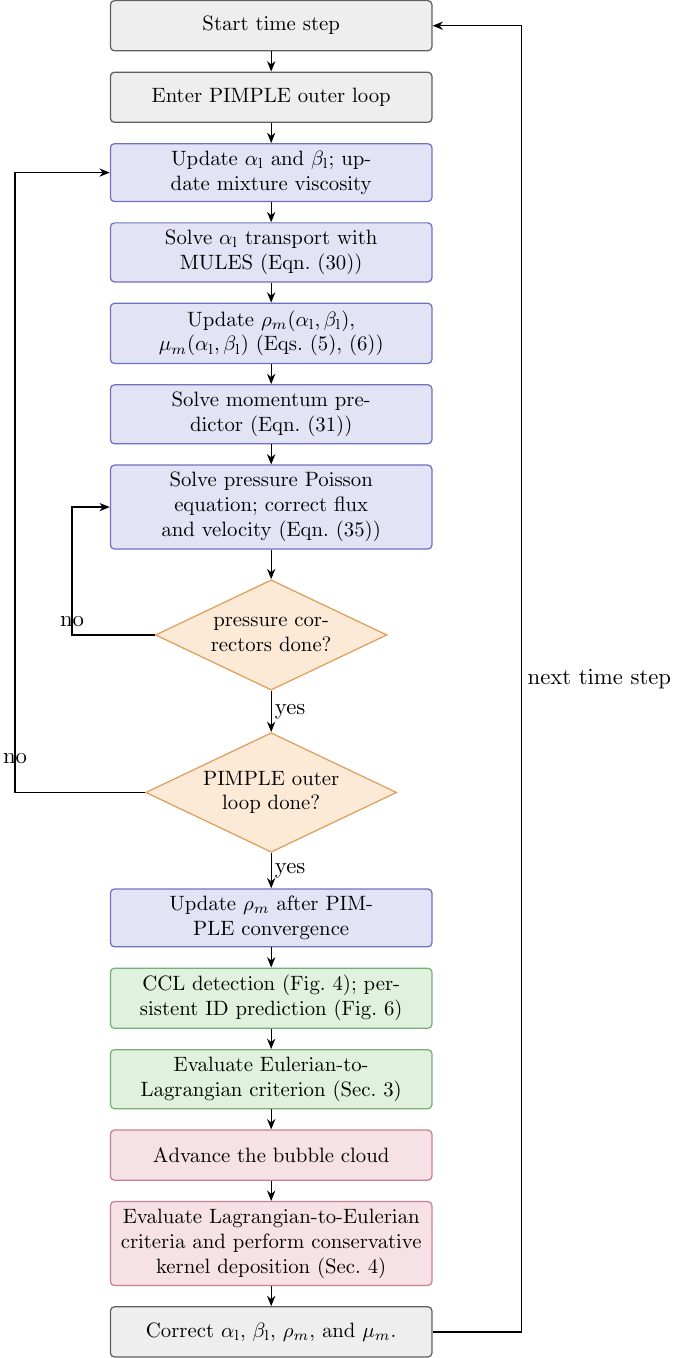}
\caption{Flowchart of the numerical solution within a single time step.}
\label{fig:solution-flowchart}
\end{figure}

\noindent The pressure Poisson equation, Eqn.~\eqref{eqn:nonfree}, is solved using a preconditioned conjugate-gradient method (PCG) with a diagonal incomplete-Cholesky (DIC) preconditioner \cite{Nhan2021}, with a tolerance of \(10^{-10}\). Within each time step, the pressure-velocity coupling is performed using the PIMPLE algorithm: three outer PIMPLE iterations are executed, and within each outer iteration the pressure correction is applied three times. In addition, one non-orthogonal correction pass is used to improve the treatment of mesh non-orthogonality parts. These momentum and pressure updates are repeated until the specified convergence criteria are met. The combination of numerical schemes used provides a balance of numerical stability (upwind bias, limiter corrections, bounded \(\alpha_{\mathrm{l}}\)) and accuracy (second-order spatial schemes where appropriate, tight convergence tolerances), making them well-suited for studying highly transient cavitating flows. 
Figure~\ref{fig:solution-flowchart} shows the complete sequence of solution steps within a single time step, including the nested PIMPLE outer and pressure-corrector loops and the loop back to the next time step. 


%% file: sections/04_e2l_transition.tex
\section{Eulerian-to-Lagrangian transition}
\label{sec:e2l}

\begin{figure}[b]
  \centering
  \includegraphics[width=0.5\textwidth]{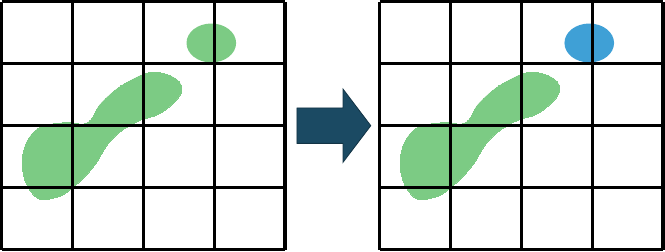}
  \caption{Transition from the Eulerian to the Lagrangian framework. The resolved vapor structure contracts below a prescribed size threshold relative to the local grid resolution and is subsequently converted into an unresolved Lagrangian bubble. Green color is resolved Eulerian structure while blue is unresolved Lagrangian structure}
  \label{fig:euler-to-lagrangian}
\end{figure}

Resolved Eulerian vapor structures may contract to length scales below the local grid resolution, at which point they can no longer be represented efficiently or accurately within the Eulerian description. As illustrated in Figure~\ref{fig:euler-to-lagrangian}, such under-resolved vapor regions are then transferred to the Lagrangian framework. The conversion is performed on connected vapor structures rather than on isolated cells. The following subsections describe how these connected structures are detected, assigned persistent identities, and converted while preserving their physical history and parallel consistency.


\subsection{Resolved vapor tracking strategy}
\label{subsec:ccl}

At each time step, the solver must identify resolved vapor structures in the Eulerian field and reduce each to the compact observation record used by the persistent ID tracker described in Section~\ref{subsec:pid}. 
Two requirements make this harder in parallel than in serial: every cell belonging to the same physical structure must end up with the same label, and that label must not depend on how many processors the structure happens to span. The implementation addresses this in four stages: a local flood fill, an iterative parallel correction that provably converges to the correct global grouping, a lightweight compaction step, and a sparse reduction of the observation set. A resolved cavity is defined as a connected set of cells for which

\begin{equation}
    \alpha_{\mathrm{l}} < \alpha_{EL},
    \label{eq:ccl_threshold}
\end{equation}

\noindent where $\alpha_{EL}$ is the Eulerian connected-component threshold; in this work, $\alpha_{EL}=0.99$. This threshold should be chosen close to unity so that a cell is classified as part of a resolved vapor structure as soon as it carries any non-negligible vapor content, capturing the full extent of a structure including its diffuse interface rather than only its most vapor-rich core. Each cell satisfying this criterion that has not yet been labeled is seeded with

\begin{equation}
    s_i^{(0)} = g(i) + 1,
    \label{eq:global_seed_label}
\end{equation}

\noindent where $g(i)$ is a global cell index, unique across all processors. Seeding from a global rather than a local cell index is what allows every processor to grow structures independently without two processors ever choosing the same temporary label for two physically different structures. Starting from the seed cell, a flood fill over the cell adjacency graph propagates this label to every vapor cell reachable from it within the local subdomain. This local phase requires no communication, but it can leave a single physical structure carrying several different labels wherever it crosses onto neighboring processors.

Reconciling these locally consistent but globally fragmented labels is the central parallel step. The solver repeats two coupled operations until no rank reports further change. First, every coupled processor-patch face compares the labels on both sides and, where both sides are vapor, adopts the smaller label. Second, because a label arriving at a shared face can still need to propagate further into the interior of a subdomain, the local flood fill is repeated so that this smaller value spreads through the rest of its connected regions. Formally, both operations are instances of the same relaxation,

\begin{equation}
    s_i^{(m+1)} = \min\!\left(s_i^{(m)},\ \min_{j\in\mathcal{N}(i)} s_j^{(m)}\right),
    \qquad m=0,1,2,\ldots,
    \label{eq:parallel_min_label}
\end{equation}

\noindent where $\mathcal{N}(i)$ denotes the local cell-adjacency neighbors and, for boundary cells, the coupled neighbors across processor patches. After each such sweep, every rank reports whether it changed any label, and a single global reduction with a maximum operator tells all ranks whether to repeat the cycle. Because a label can only ever decrease and is bounded below, this process is monotone and is guaranteed to terminate at the unique minimum label reachable within each physical structure.

\begin{figure}[ht]
  \centering

  \begin{subfigure}{0.68\linewidth}
    \centering
    \caption{Single-subdomain connected-component labeling steps.}
    \includegraphics[width=\linewidth]{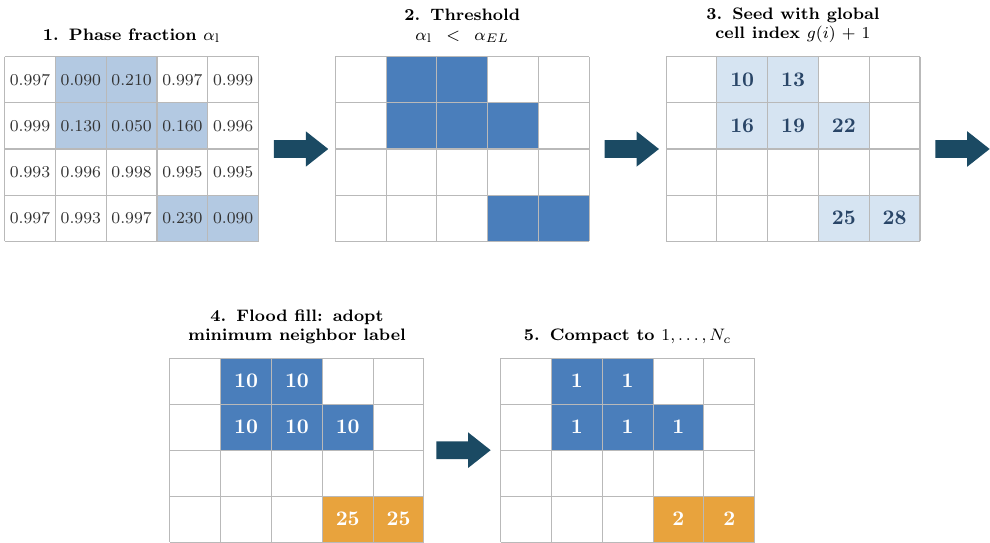}
    \label{fig:ccl_combined1}
  \end{subfigure}

  \begin{subfigure}{0.68\linewidth}
    \centering
    \caption{Parallel correction across a processor boundary.}
    \includegraphics[width=\linewidth]{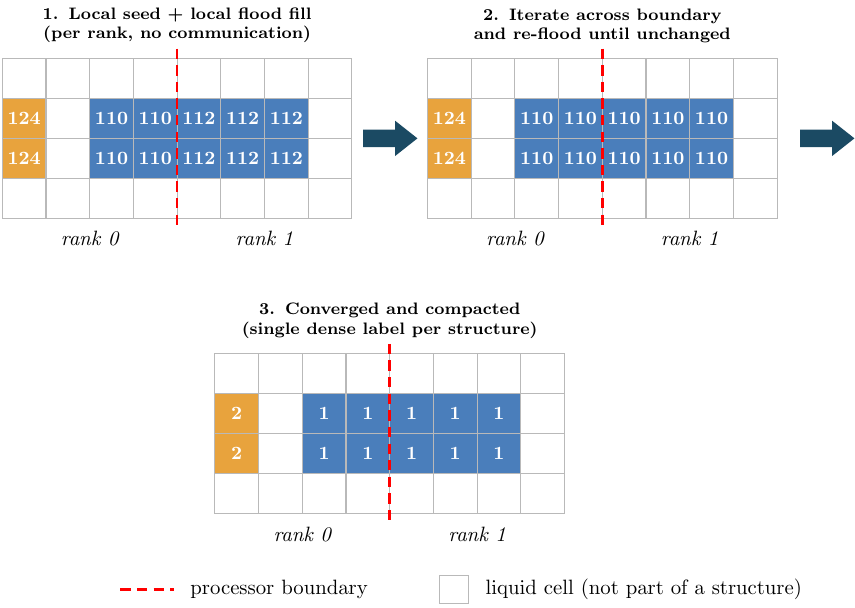}
    \label{fig:ccl_combined2}
  \end{subfigure}

  \caption{Connected-component labeling procedure. 
  (a) Single-subdomain steps: threshold, seed with a global cell index, flood-fill to the minimum label, and compact to $1,\ldots,N_c$. 
  (b) Parallel correction across a processor boundary (Eqn.~\eqref{eq:parallel_min_label}): two sides of a shared structure start with different labels and converge to a common minimum through iterative exchange and re-flooding before compaction.}
  \label{fig:ccl}
\end{figure}

Once the labels have converged, they are still an arbitrary, sparse set of global cell numbers and must be compacted into a dense sequence before they can index the observation arrays. Instead of scanning the whole mesh for a global maximum, each rank collects only the distinct label values it still holds, typically a small number equal to its local component count, and these short lists are gathered to the master rank. The master assigns each distinct label a dense index $1,\ldots,N_c$ and broadcasts the resulting remap table, which every rank applies locally to the Eulerian field. Because the communicated quantity is a list of component labels rather than a per-cell or per-processor-pair message, the communication volume associated with this compaction step scales with the number of resolved structures rather than with the mesh size or the processor count. For each compacted component $k$, the reduced observation


\begin{equation}
    \mathcal{O}_k = \left( \xx_k,\uu_k,V_k,R_k,N_k,s_k \right),
    \label{eq:obs}
\end{equation}

\noindent is assembled from

\begin{equation}
    N_k = \sum_{i\in \Omega_k} 1,
    \qquad
    V_k = \sum_{i\in \Omega_k} (1-\alpha_{\mathrm{l},i})\Delta V_i,
    \label{eq:parallel_volume_reduce}
\end{equation}

\noindent where $\Omega_k$ is the set of cells assigned to component $k$ and $\Delta V_i$ is the volume of cell $i$,

\begin{equation}
    \xx_k = \frac{1}{V_k}\sum_{i\in\Omega_k}(1-\alpha_{\mathrm{l},i})\Delta V_i\,\xx_i,
    \qquad
    \uu_k = \frac{1}{V_k}\sum_{i\in\Omega_k}(1-\alpha_{\mathrm{l},i})\Delta V_i\,\uu_i ,
    \label{eq:parallel_centroid_reduce}
\end{equation}

\noindent where $\xx_k$ is the vapor-volume-weighted centroid, $\uu_k$ is the vapor-volume-weighted mean velocity, $V_k$ is the resolved vapor volume, $R_k=(3V_k/4\pi)^{1/3}$ is the equivalent radius, $N_k$ is the number of cells in the component, and $s_k$ is the compacted component label. Figure~\ref{fig:ccl} shows in part (a) the CCL steps in this algorithm and in part (b) how the parallel part is working: these instantaneous labels are not yet persistent physical identities, and the same physical structure can receive a different label from one time step to the next, which is precisely the gap that persistent tracking addresses in Section \ref{subsec:pid}.

\subsection{Persistent identity tracking}
\label{subsec:pid}

\begin{figure}[ht]
  \centering
  \includegraphics[width=0.85\linewidth]{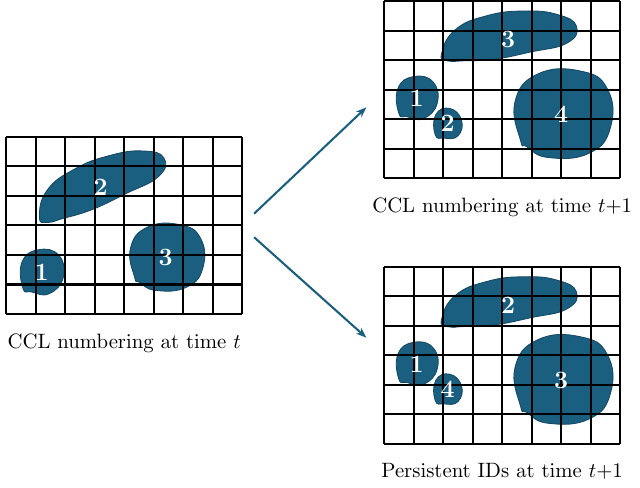}
  \caption{Motivation for persistent IDs. Raw CCL numbering at $t+1$ relabels persisting structures and cannot distinguish them from a genuinely new one. Persistent IDs at $t+1$ correctly preserve the identity of each persisting structure and assign a new ID only to the structure that is actually new.}
  \label{fig:pid_motivation}
\end{figure}

\begin{figure}[ht]
 \centering
 \includegraphics[width=0.64\linewidth]{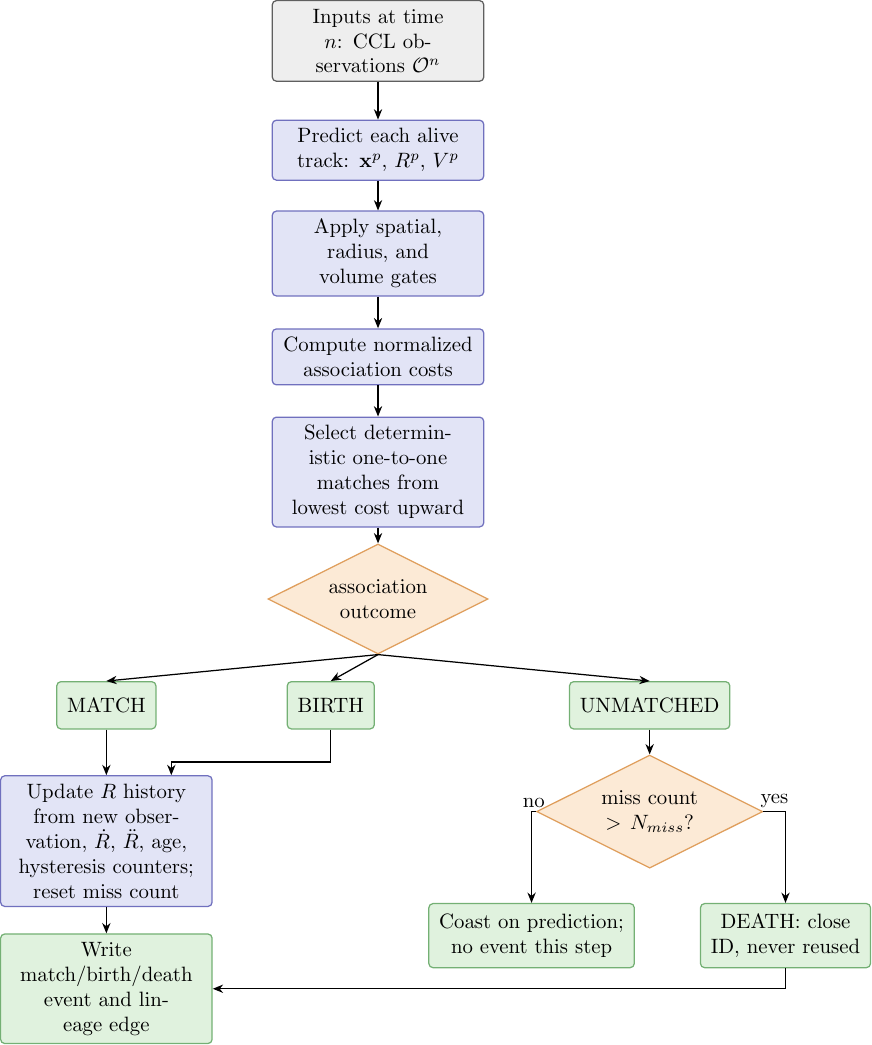}
 \caption{Persistent-ID tracking workflow implemented. A matched or newborn track updates its history and is logged immediately. An unmatched track increments its miss count and coasts silently on its predicted state; only once the miss count exceeds $N_{miss}$ is it declared dead and logged.}
 \label{fig:persistent_id_flowchart}
 \end{figure}

The connected-component labels produced in Section~\ref{subsec:ccl} are reconstructed from scratch at every time step and carry no memory: traversal order, mesh decomposition, or a small topological change can all relabel the same physical structure differently from one step to the next. Figure~\ref{fig:pid_motivation} explains this concept. At time $t$, three resolved vapor structures are labeled $1$, $2$, and $3$. By time $t+1$, the elongated structure has drifted and is arbitrarily relabeled $3$ by a fresh CCL traversal, the round structure has grown and is relabeled $4$, and a genuinely new small structure has appeared close to the original structure $1$; raw CCL assigns the labels $1$ and $2$ to this pair of nearby small blobs with no regard for which one is the original and which is new. A bookkeeping scheme built directly on these raw labels could not tell the persisting structures from the new one and would misread this step as a mix of false deaths and false births. The persistent-identity layer removes this ambiguity by tracking each structure through its predicted physical state rather than through the label a given flood-fill happened to assign it: the elongated and round structures correctly keep IDs $2$ and $3$, the original small structure keeps ID $1$, and only the genuinely new structure receives the new ID $4$. This layer turns a sequence of unrelated instantaneous observations into continuous structure histories that the transition operators depend on. Figure~\ref{fig:persistent_id_flowchart} summarizes the per-time-step workflow described below.

\begin{figure}[t]
  \centering

  \begin{subfigure}[t]{0.4\linewidth}
    \centering
    \includegraphics[height=4cm]{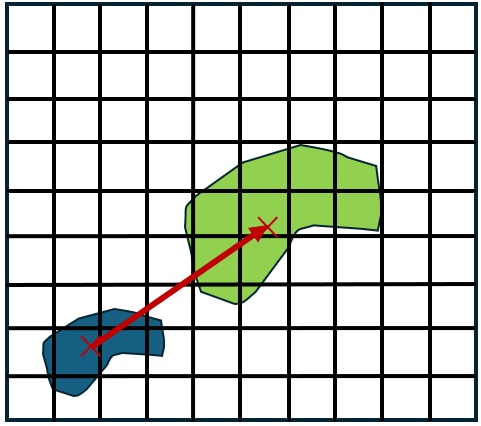}
    \caption{Prediction of the next location and size of a structure based on its state at time $t^n$.}
    \label{fig:pid_prediction}
  \end{subfigure}
  \hfill
  \begin{subfigure}[t]{0.4\linewidth}
    \centering
    \includegraphics[height=4cm]{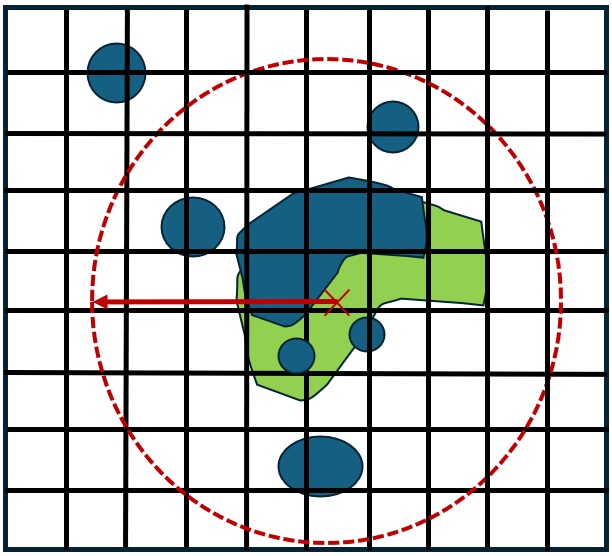}
    \caption{Gating of admissible matches: the red circle centered at the predicted centroid defines the search region given by Eqn.~\eqref{eq:gate}.}
    \label{fig:pid_gate_region}
  \end{subfigure}

  \caption{Geometric ingredients used by the tracker.}
  \label{fig:pid_gate}
\end{figure}

\noindent For each alive track $m$, the next centroid and radius are predicted, shown in Figure~\ref{fig:pid_prediction}, from the current state before any observation is examined,

\begin{equation}
    \xx_m^{p} = \xx_m^n + \uu_m^n \Delta t,
    \label{eq:x_predict}
\end{equation}
\begin{equation}
    R_m^{p} = \max\left(R_m^n + \dot{R}_m^n \Delta t + \frac{1}{2}\ddot{R}_m^n\Delta t^2,0\right).
    \label{eq:r_predict}
\end{equation}

\noindent The radius derivatives are obtained from the stored radius history rather than from the current step alone. Once two previous values are available, a second-order backward difference is used,

\begin{equation}
    \dot{R}^n = \frac{3R^n-4R^{n-1}+R^{n-2}}{2\Delta t},
    \qquad
    \ddot{R}^n = \frac{R^n-2R^{n-1}+R^{n-2}}{\Delta t^2},
    \label{eq:bdf2_radius}
\end{equation}

\noindent and a first-order two-point fallback is used until that history has accumulated. An observation $k$ is a candidate match for track $m$ only if it survives a set of geometric and kinematic gates, illustrated in Figure~\ref{fig:pid_gate_region}. The centroid must lie within a search radius built from the predicted structure size and its predicted motion,

\begin{equation}
    \|\xx_k-\xx_m^p\| \le c_R R_{ref}+c_U \|\uu_m\|\Delta t,
    \qquad c_R=1, \quad c_U=1,
    \label{eq:gate}
\end{equation}

\noindent where $R_{ref}=\max(R_m^p,0.25\bar{R})$ and $\bar{R}$ is the mean equivalent radius over all current observations; the floor at $0.25\bar{R}$ keeps the search radius from collapsing to zero for a very small predicted structure. Candidates are further rejected if the relative jump in radius exceeds $100\%$ between the prediction and the observation. For each pair that survives gating, a normalized cost function is computed,

\begin{equation}
    C_{mk} = w_x \left(\frac{\|\xx_k-\xx_m^p\|}{R_{ref}}\right)^2
    + w_R \left(\frac{|R_k-R_m^p|}{R_m^p}\right)^2
    + w_U \left(\frac{\|\uu_k-\uu_m\|}{\|\uu_m\|}\right)^2.
    \label{eq:association_cost}
\end{equation}

\noindent The weights are set to $w_x=1.0$, $w_R=0.5$, and $w_U=0.25$ to reflect the intended relative importance of the association criteria. The sensitivity of the association results to these weights is
examined using the adjacent-structure benchmark in Section~\ref{subsec:crossing_case}. The use of positional and size consistency in the association cost is supported by Basak et al~\cite{Basak2026ON}. The positional term receives the largest weight because the tracking procedure uses the predicted centroid and a search region determined by the predicted structure size and motion to identify admissible matches. Within this region, proximity to the predicted centroid is prioritized. Agreement with the predicted radius provides a complementary size-based constraint and receives an intermediate weight. The velocity term receives the smallest weight to reduce the influence of local velocity-field fluctuations on the association decision. These weights represent a physically motivated implementation choice. Candidate pairs are then committed from lowest cost to highest cost: the cheapest remaining pair is accepted, both its track and its observation are removed from further consideration, and the process repeats until no feasible pair remains. This greedy rule is deterministic and gives each track and each observation at most one match per time step.

Association resolves every track into exactly one of three outcomes, based on the lowest-cost greedy matching defined by the association cost in Eqn.~\eqref{eq:association_cost}. A track that is matched to an observation inherits that observation's position, velocity, and size, and its radius history, derivatives, and hysteresis counters are advanced accordingly; its miss count is reset to zero. An observation that matches no alive track creates a new track with a fresh, monotonically increasing ID that is never reused, and its radius history is initialized entirely from that first observation. A track that matches no observation is not declared dead immediately: it coasts on its predicted state and its miss count is incremented, and only once that count exceeds a threshold $N_{miss}$ is the track finally closed and its ID retired. This coasting behavior matters in practice because CCL detection can miss a genuine structure for a step or two, for example when a structure's occupancy fraction briefly crosses the resolvability threshold in both directions during interface advection; without it, a single dropped detection would needlessly terminate a track and fragment its history into two IDs. Independent of these three outcomes, a representation transition terminates a Eulerian track when its structure is converted to an unresolved bubble and links the new bubble to that track, or creates or updates an Eulerian structure linked to the source bubble on the reverse conversion.

\begin{figure}[t]
  \centering
  \includegraphics[width=0.9\linewidth]{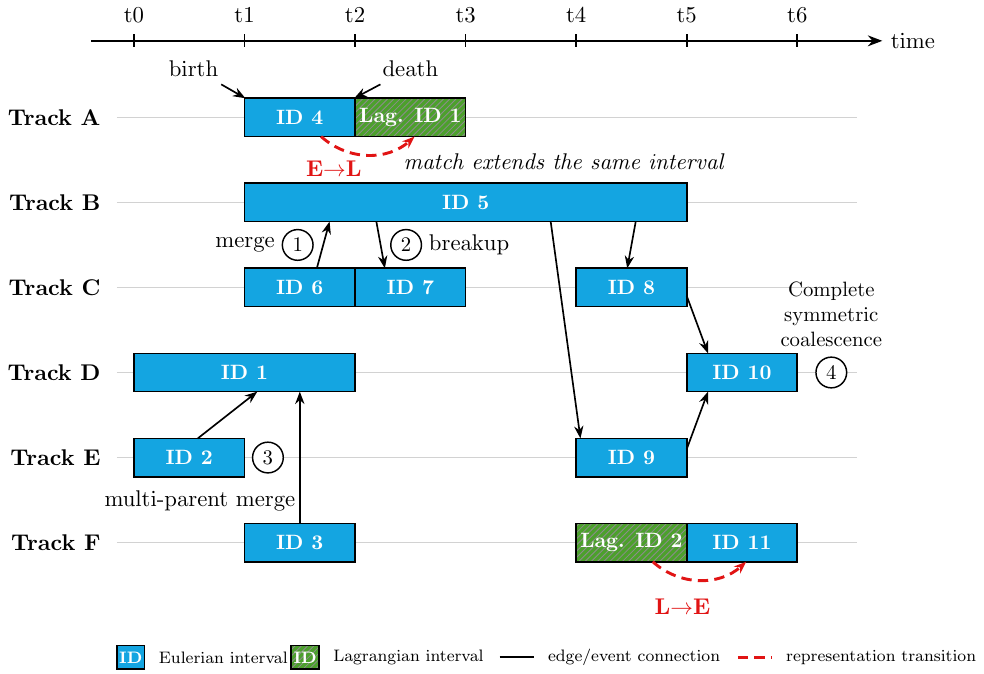}
  \caption{Timeline summary of the persistent-ID history. Each colored bar is a structure interval. A match extends the same interval, a birth starts a new interval, a death closes an interval, a merge or breakup is recovered from the event pattern between intervals, and a representation transition links an Eulerian interval to a Lagrangian interval corresponding to the same physical structure.}
  \label{fig:lineage_graph}
\end{figure}

This one-to-one association is used only to establish persistence when the topology of
a structure remains unchanged. Topology-changing events are treated separately in the
lineage graph because a merger or breakup does not admit an unambiguous continuation
of a single physical identity. Each vertex of the lineage graph represents a structure interval: a persistent ID together
with the time span over which that structure exists without a topological change. A
successful one-to-one match extends the same interval forward in time, whereas an
unmatched observation with no identified parent initiates a new birth and an unmatched
track is eventually closed as a death after the prescribed missed-detection tolerance. When a single parent structure fragments into multiple connected observations, the
parent interval is terminated and each resulting child is assigned a new persistent ID.
Lineage edges are then created from the parent interval to all child intervals. Similarly,
when multiple parent structures coalesce into a single connected observation, all parent
intervals are terminated and the merged structure is assigned a new persistent ID, with
lineage edges linking each parent to the new child. This convention avoids assigning
physical identity through a topological event based solely on an arbitrary association
cost or geometric proximity. Representation transitions are handled differently because the physical structure itself
is intended to persist while only its mathematical representation changes. An
Eulerian-to-Lagrangian transition therefore links the terminating Eulerian interval to the
corresponding Lagrangian interval of the same physical structure, and the reverse
Lagrangian-to-Eulerian transition creates the corresponding Eulerian continuation. In this
way, ordinary temporal persistence, topological lineage, and representation changes are
distinguished explicitly within the event history.
Figure~\ref{fig:lineage_graph} summarizes the same information in a cleaner timeline form, showing how persistent-ID intervals are extended, terminated, merged, split, and transferred between Eulerian and Lagrangian representations.

\subsection{Transition eligibility}
\label{subsec:e2l_hysteresis}

The Eulerian-to-Lagrangian operator is not triggered by a single instantaneous threshold crossing. This is precisely where the persistent track history, rather than the instantaneous observation, becomes essential. In a cavitating flow, a connected component detected by CCL can transiently grow or shrink by a cell or two because of interface advection, compression, or parallel relabeling, so converting on a raw threshold crossing would let a structure switch between representations every time it wobbles across the threshold boundary. The implementation avoids this by requiring a structure to first become unambiguously resolved/unresolved before it is allowed to count toward conversion.

\begin{figure}[t]
  \centering
  \includegraphics[width=1\linewidth]{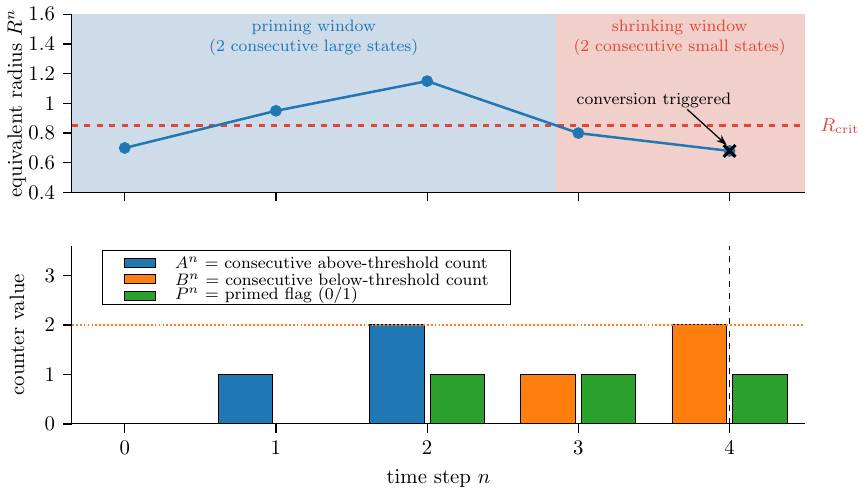}
  \caption{History-based Eulerian-to-Lagrangian eligibility logic. The radius-based criterion is shown; the cell-based criterion is identical with $R$ replaced by the cell count $N$.}
  \label{fig:e2l_history_schematic}
\end{figure}

For each persistent track $m$, two running counts are maintained alongside the history needed ($R_m$ and $\dot{R}_m$ for Rayleigh-Plesset equation initial conditions): $A_m^n$, the number of consecutive time steps for which the track's size has been observed above the resolution threshold, and $B_m^n$, the number of consecutive time steps for which it has been observed below the threshold since the track was first primed. A third flag, $P_m^n\in\{0,1\}$, records whether the track has ever satisfied the above-threshold requirement. In this work, both counts are set to two: a track becomes primed ($P_m^n = 1$) once it is observed above the threshold on two consecutive time steps, and it is converted once it is subsequently observed below the threshold on two consecutive time steps, as shown in Figure~\ref{fig:e2l_history_schematic}.

The monitored size quantity is selected at runtime: in radius-based mode, $S_m^n=R_m^n$ is compared against $S_{\mathrm{crit}}=R_{\min}$, a user-defined minimum radius, and in cell-based mode, $S_m^n=N_m^n$ is compared against $S_{\mathrm{crit}}=N_{\min}$, a user-defined minimum cell count. A structure is classified as large when $S_m^n\ge S_{\mathrm{crit}}$ and small otherwise, and the history logic that follows is identical for either choice. The large-state counter and the primed flag evolve as

\begin{equation}
A_m^n=
\begin{cases}
A_m^{n-1}+1, & S_m^n\ge S_{\mathrm{crit}},\\[1mm]
0, & S_m^n< S_{\mathrm{crit}},
\end{cases}
\qquad
P_m^n = P_m^{n-1} \;\lor\; \left(A_m^n\ge N_{\mathrm{above}}\right),
\label{eq:e2l_aboveCount}
\end{equation}

\noindent where $N_{\mathrm{above}}=2$. Only once a track is primed is it allowed to accumulate a small-state streak,

\begin{equation}
B_m^n=
\begin{cases}
0, & P_m^n=0,\\[1mm]
B_m^{n-1}+1, & P_m^n=1\ \text{and}\ S_m^n<S_{\mathrm{crit}},\\[1mm]
0, & P_m^n=1\ \text{and}\ S_m^n\ge S_{\mathrm{crit}},
\end{cases}
\label{eq:e2l_belowCount}
\end{equation}

\noindent and the Eulerian structure is converted to an unresolved bubble only once both conditions hold simultaneously,

\begin{equation}
\mathcal{I}_m^n = \left(P_m^n=1\right) \wedge \left(B_m^n\ge N_{\mathrm{below}}\right),
\qquad N_{\mathrm{below}}=2,
\label{eq:e2l_inject_condition}
\end{equation}

This priming-and-shrinking logic has three practical consequences. A newly born small structure is not injected immediately, because $P_m^n=0$ forces $B_m^n$ to remain zero. A structure that merely fluctuates around the threshold without first being genuinely resolved cannot chatter between representations. And once a structure has been resolved and is later observed to shrink for long enough, the conversion is triggered robustly and reproducibly. The question the algorithm asks is therefore never whether a component is small right now, but whether its recent history is consistent with a genuine resolved-to-unresolved transition. Figure~\ref{fig:e2l_history_schematic} illustrates this priming-and-shrinking logic.

\subsection{Verification}
\label{subsec:e2l_testcase}

The verification cases in this section are arranged as a progressive test of the persistent-ID algorithm. We begin with breakup, where one connected Eulerian structure produces many fragments and the tracker must close the parent interval while assigning consistent child identities. We then consider coalescence, the complementary many-to-one event. Finally, we examine a close-crossing event in which no topological change occurs; this provides the most direct test that the prediction and gating model prevents ID exchange when candidate search regions overlap.

The verification cases presented below are designed to assess the numerical correctness of the persistent tracking and Eulerian-Lagrangian transition algorithms independently of the physical cavitation model. Accordingly, the verification problems retain the same Eulerian vapor volume-fraction formulation used throughout the proposed framework, but the macroscopic liquid-vapor phase-change source term is disabled. The resolved vapor structures therefore evolve solely through the resolved transport equations, without vapor generation or condensation. This isolates the computational aspects of connected-component identification, persistent structure tracking, event detection, and representation transitions from the additional complexity introduced by cavitation physics.


\subsubsection{Persistent-ID tracking through capillary-driven fragmentation of a resolved vapor filament}\label{subsec:ring_fragmentation}

This verification case examines the capillary-driven fragmentation of a resolved vapor filament ring \cite{Basak2026ON}. An initially toroidal vapor filament is seeded in an otherwise quiescent liquid domain. Surface tension acts on the entangled ring and causes Plateau-Rayleigh instability: azimuthal perturbations grow along the filament until the ring pinches apart into a necklace of discrete structures. This case is a severe test of the tracker because a single connected component fragments into more than a dozen simultaneous children in one step, well beyond the pairwise breakups.

The initial filament is constructed by assigning the liquid volume fraction, applied as the initial condition. Writing $r=\sqrt{x^2+y^2}$ for the radial distance from the ring axis (taken here as the $z$-axis) and $\theta=\mathrm{atan2}(y,x)$ for the azimuthal angle, the toroidal filament of major radius $R_{maj}$ and minor (tube) radius $R_{min}$, seeded with an azimuthal corrugation of wavenumber $m$ and relative amplitude $a_0$, is the region enclosed by

\begin{equation}
    \phi(\xx) = \left(r-R_{maj}\right)^2 + z^2 - R_{min}^2\left[1+a_0\cos(m\theta)\right],
    \label{eq:torus_sdf}
\end{equation}

\noindent with the initial liquid volume fraction assigned as

\begin{equation}
    \alpha_{\mathrm{l}}(\xx,0) =
    \begin{cases}
        0, & \phi(\xx) \le 0 \quad \text{(inside the filament)},\\
        1, & \phi(\xx) > 0 \quad \text{(outside the filament)}.
    \end{cases}
    \label{eq:torus_ic}
\end{equation}

\begin{figure}[t!]
  \centering
  \includegraphics[width=0.65\linewidth]{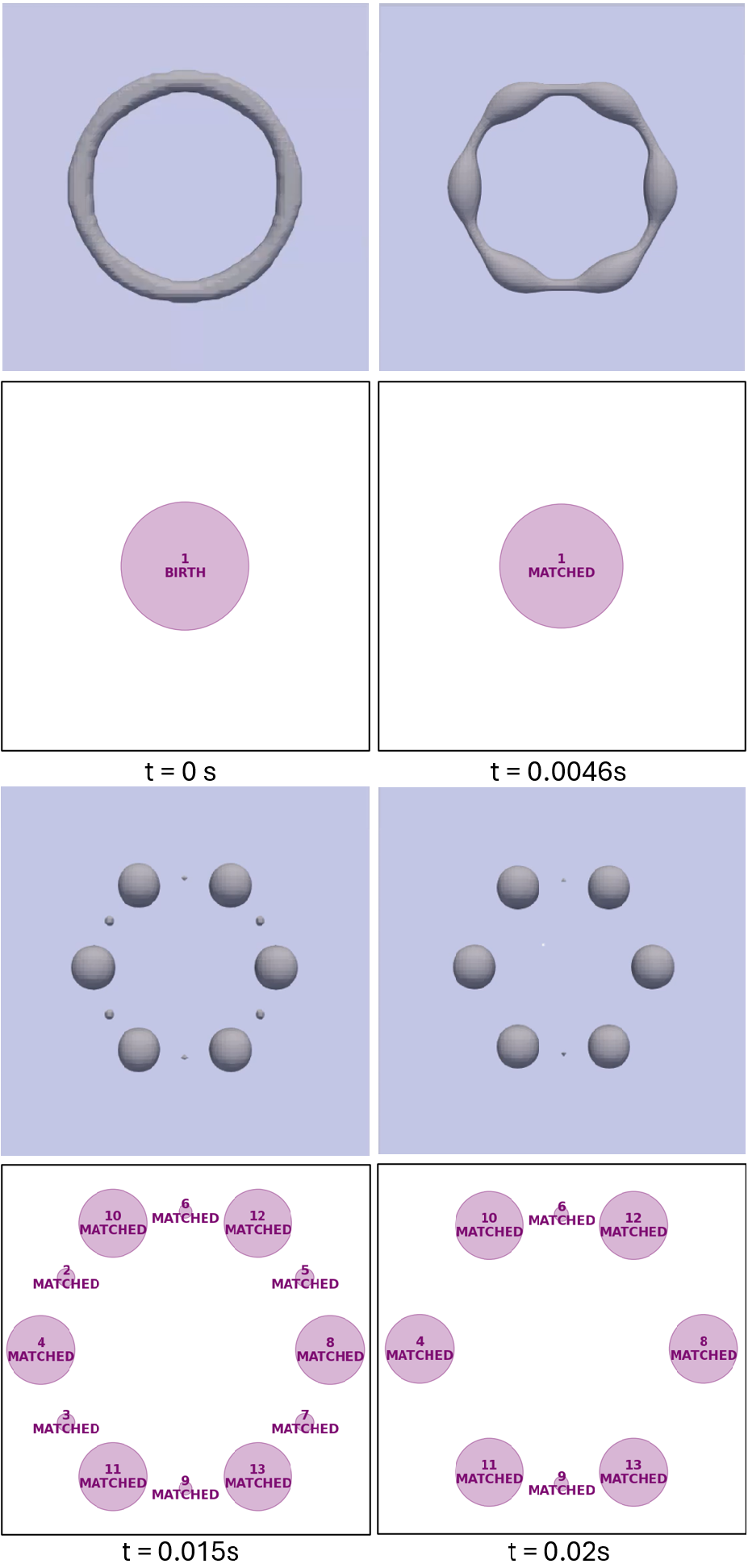}
  \caption{Capillary-driven fragmentation of a vapor filament ring. The ring is born as ID~1 and matched while it remains a single component; at pinch-off, ID~1 closes and IDs 2-13 are born together in one breakup event; by $t=0.02~\mathrm{s}$, the transient satellites (IDs 4, 6, 8, and 9) have closed with death events while the remaining IDs continue to be matched.}
  \label{fig:ring_fragmentation}
\end{figure}

\noindent For this case, $R_{maj}=0.90~\mathrm{mm}$, $R_{min}=0.10~\mathrm{mm}$, $m=6$, and $a_0=0.2$, so the ring is seeded with a hexagonal corrugation from $t=0$. The computational domain is a cubic box of size $4~\mathrm{mm}\times4~\mathrm{mm}\times4~\mathrm{mm}$, discretized with $128\times128\times128$ uniform cells. The properties are $\rho_{\mathrm{l}}=1000~\mathrm{kg/m^3}$, $\nu_{\mathrm{l}}=1\times10^{-6}~\mathrm{m^2/s}$, $\rho_{\mathrm{v}}=1~\mathrm{kg/m^3}$, and $\nu_{\mathrm{v}}=1\times10^{-5}~\mathrm{m^2/s}$, with surface tension $\sigma=0.00126~\mathrm{N/m}$; gravity is disabled and the flow is treated as laminar so that the fragmentation is driven by capillarity rather than by buoyancy or turbulence. The simulation is run to $t=0.03~\mathrm{s}$. The mesh is partitioned into $40$ processor subdomains, so the ring, and every fragment it produces, repeatedly crosses processor boundaries as it evolves. This makes the case a demanding test of the parallel implementation: the labels must stay consistent and the gating must keep the correct identity attached to each fragment even as the ring and its fragments repeatedly cross processor boundaries, as described in Section~\ref{subsec:pid}.

Figure~\ref{fig:ring_fragmentation} shows the evolution together with the corresponding persistent-ID record. At $t=0~\mathrm{s}$ the ring is a single connected structure and is recorded as a birth event, receiving persistent ID~1. At $t=0.0046~\mathrm{s}$ the filament has developed a visible azimuthal, hexagonal-like corrugation but is still one connected component, so the tracker emits a match event that simply extends ID~1 lifetime. Between $t=0.0046~\mathrm{s}$ and $t=0.015~\mathrm{s}$ the corrugation grows to the point of pinch-off, and the ring separates into six primary structures together with several smaller satellite structures in the necks between them. The tracker records this as a single one-to-many breakup: ID~1 is closed, and twelve new IDs, 2 through 13, are born simultaneously, one for each fragment resolved by CCL at that step. By $t=0.02~\mathrm{s}$, four of the transient satellite droplets, IDs 4, 6, 8, and 9, have shrunk below the resolvable footprint and are closed with death records, while the remaining IDs are carried forward as match events. This sequence is exactly the kind of one-to-many lineage event described in Section~\ref{subsec:pid}, here produced by the test case's physics.

To assess sensitivity to parallel domain decomposition, the same fragmentation case was repeated using 20, 40, and 80 ranks. The mesh, initial conditions, physical properties, and solver settings were unchanged. The comparison covers $t=0.0001$-$0.03~\mathrm{s}$ with $\Delta t=10^{-4}~\mathrm{s}$. Table~\ref{tab:ring_parallel} summarizes the recorded events and total wall-clock times; each match denotes one successful association at one time step. All three runs produce 13 births and five deaths, leaving eight selected tracks at the final time. The identical event counts and association counts across the three decompositions demonstrate decomposition-invariant tracking for this test case. 

\begin{table}[H]
  \centering
  \caption{Rank comparison for ring fragmentation. Death denotes a \texttt{KILL} record. Wall-clock times
  refer to the complete simulation.}
  \label{tab:ring_parallel}
  \begin{tabular}{r r r r r}
    \hline
    MPI ranks & Births & Matches & Deaths & Wall time [s] \\
    \hline
    20 & 13 & 2261 & 5 & 5566 \\
    40 & 13 & 2261 & 5 & 2054 \\
    80 & 13 & 2261 & 5 &  684 \\
    \hline
  \end{tabular}
\end{table}

\subsubsection{Persistent-ID tracking through resolved vapor structure coalescence}
\label{subsec:coalescence_case}

The second case study in the persistent-identity algorithm concerns the coalescence of two approaching vapor structures, applying the merge pattern in the opposite direction of Section~\ref{subsec:ring_fragmentation}. Instead of one structure fragmenting into many, two independent structures combine into one. The case is designed to exhibit the complete birth-match-death-birth cycle in a single, easily inspected event: two structures are born, matched while they approach (completely symmetrical), closed by death at the moment of coalescence, and replaced by a single newborn structure that is subsequently matched as it relaxes toward a sphere.

\begin{figure}[t!]
  \centering
  \includegraphics[width=0.65\linewidth]{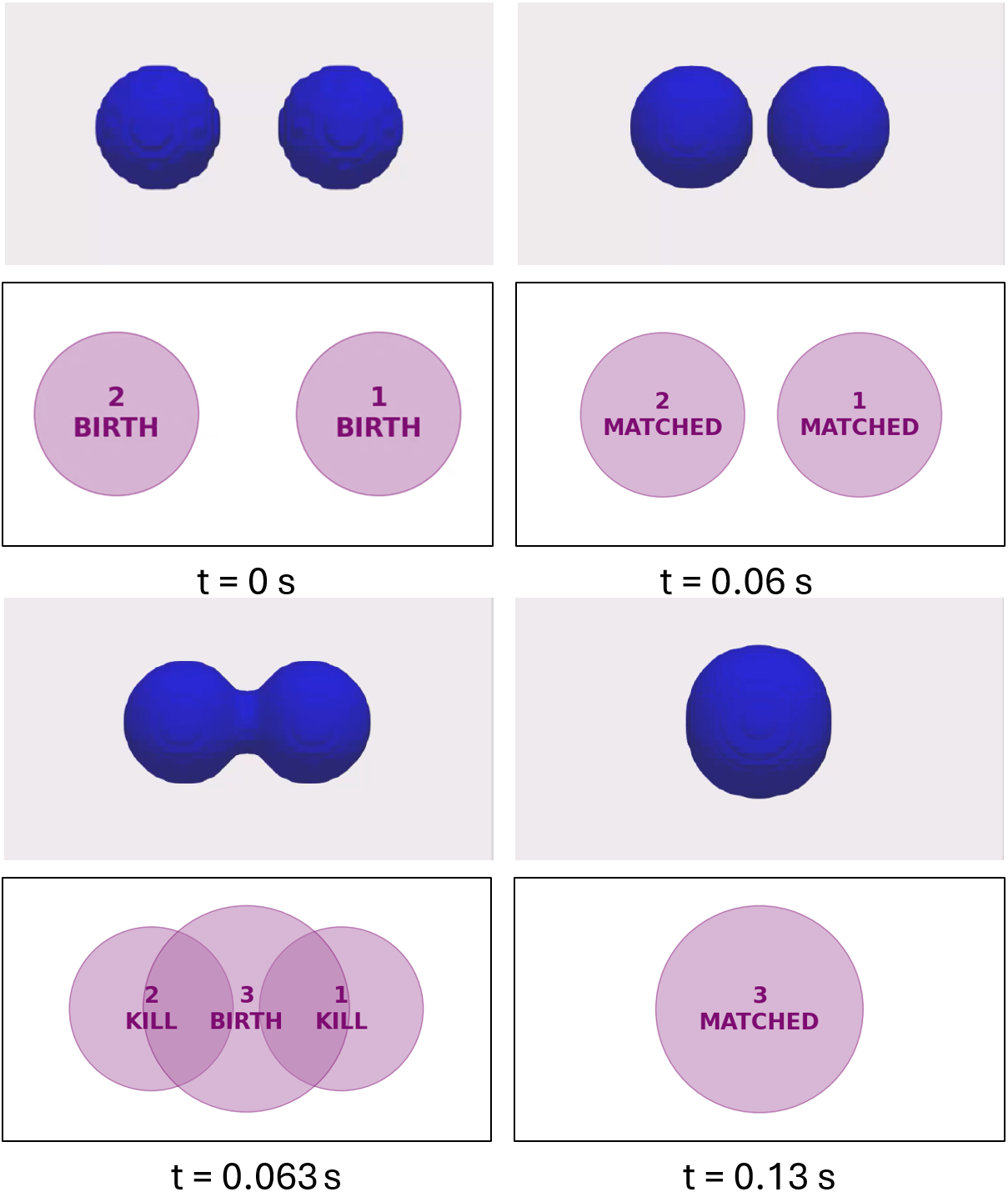}
  \caption{Persistent-ID tracking through bubble coalescence, decomposed across 128 processor subdomains. Top row: resolved vapor iso-surface at $t=0$, $0.06$, $0.07$, and $0.13~\mathrm{s}$. Bottom row: the corresponding persistent-ID record. The two approaching bubbles are born as IDs~1 and~2 and matched while they remain separate; at coalescence, both parent IDs close with death events and a new ID~3 is born, storing both parents; ID~3 is then matched as the merged bubble relaxes toward a sphere.}
  \label{fig:bubble_coalescence}
\end{figure}

The computational domain is a three-dimensional cube of side length $30~\mathrm{mm}$, discretized with $120\times120\times120$ uniform cells. The properties are $\rho_{\mathrm{l}}=1000~\mathrm{kg/m^3}$, $\nu_{\mathrm{l}}=1\times10^{-6}~\mathrm{m^2/s}$, $\rho_{\mathrm{v}}=1~\mathrm{kg/m^3}$, and $\nu_{\mathrm{v}}=1\times10^{-5}~\mathrm{m^2/s}$, with surface tension $\sigma=0.072~\mathrm{N/m}$. Two spherical vapor bubbles are initialized with a small separation between their surfaces so that they drift together and coalesce under surface tension. The mesh is partitioned into $128$ processor subdomains, a substantially finer parallel split than the ring-fragmentation case, so that the coalescence event, like the breakup event of Section~\ref{subsec:ring_fragmentation}, is resolved consistently even though the merging pair and the resulting structure repeatedly cross processor boundaries.

Figure~\ref{fig:bubble_coalescence} shows the sequence together with the corresponding persistent-ID record. At $t=0~\mathrm{s}$ the two bubbles are detected as separate components and are recorded as two \textsc{birth} events, receiving persistent IDs~1 and~2. At $t=0.06~\mathrm{s}$ the bubbles have drifted closer together but remain two distinct components, so both IDs are carried forward as \textsc{match} events. Between $t=0.06~\mathrm{s}$ and $t=0.07~\mathrm{s}$ the two interfaces touch and coalesce into a single connected structure: IDs~1 and~2 are closed with \textsc{death} records, and a new ID~3 is born, consistent with the merge rule of Section~\ref{subsec:pid}, under which the structure resulting from a merge is treated as a new vertex in the lineage graph rather than a continuation of either parent. By $t=0.07~\mathrm{s}$ the merged structure has already been associated once and appears as \textsc{match}; at $t=0.13~\mathrm{s}$, after the combined bubble has relaxed from its post-coalescence dumbbell shape toward a sphere, ID~3 is still carried forward as \textsc{match}, confirming that identity is preserved through the shape relaxation that follows coalescence and is not disturbed by the transient neck seen at $t=0.063~\mathrm{s}$.

\subsubsection{Persistent-ID tracking through adjacent resolved vapor structures}
\label{subsec:crossing_case}

Another important component of the persistent-ID algorithm is the gating step, Eqs.~\eqref{eq:gate} and ~\eqref{eq:association_cost}, which must keep the correct IDs attached to physical structures even when several structures come within each other's search distance. In particular, the tracker must not switch IDs simply because neighboring candidates are momentarily close together. To verify this behavior, we consider a case study in which three resolved vapor structures of different sizes approach one another, pass through a close crossing event with overlapping search neighborhoods, and then separate again without merging. The expected outcome is that each physical structure retains its original ID throughout the event, independent of the parallel mesh decomposition.

The computational domain is the same three-dimensional cube of side length $30~\mathrm{mm}$ used in Section~\ref{subsec:coalescence_case}, discretized with $120\times120\times120$ uniform cells with the same liquid and gas properties. 
The mesh is partitioned into $40$ processor subdomains, so each structure, and the region in which the three come closest together, is not guaranteed to remain on a single rank throughout the event.

\begin{figure}[t!]
  \centering
  \includegraphics[width=0.8\linewidth]{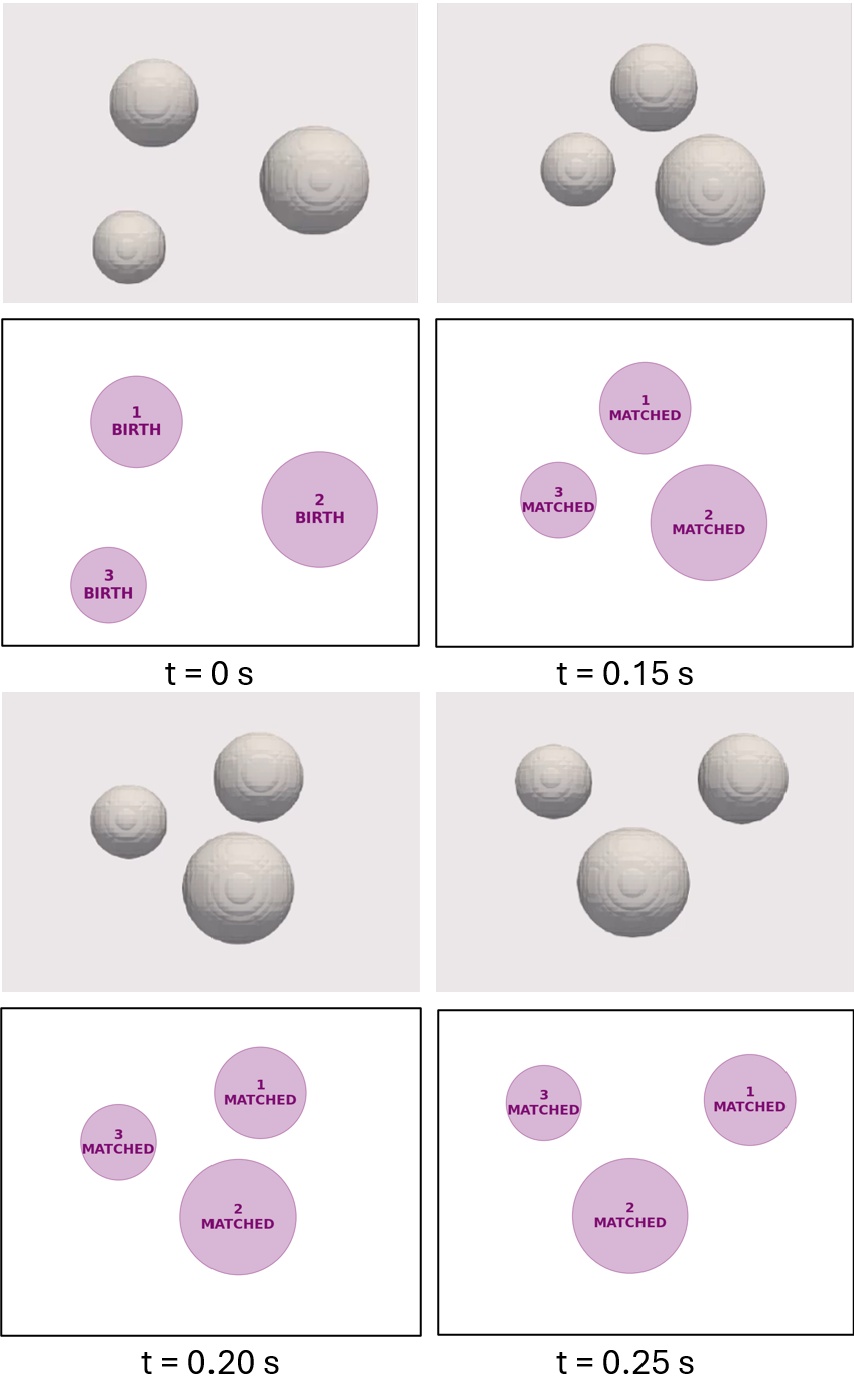}
  \caption{Persistent-ID tracking through parallel bubble crossing, decomposed across 40 processor subdomains. Top row: rendered bubbles at $t=0$, $0.15$, $0.20$, and $0.25~\mathrm{s}$. Bottom rows: the corresponding persistent-ID record. All three bubbles are born as IDs~1, 2, and~3 and are matched at every subsequent step; despite the close approach and crossing of their paths between $t=0.15~\mathrm{s}$ and $t=0.20~\mathrm{s}$, no identity is exchanged between structures.}
  \label{fig:three_bubble_crossing}
\end{figure}

Figure~\ref{fig:three_bubble_crossing} shows the sequence together with the corresponding persistent-ID record. At $t=0~\mathrm{s}$ the three structures are detected as separate components and are recorded as three \textsc{birth} events, receiving persistent IDs~1, 2, and~3. As they approach one another, each remains a distinct connected structure at every step; the tracker never records a death or a birth over the course of the event, and IDs~1, 2, and~3 are carried forward as \textsc{match} events at $t=0.15$, $0.20$, and $0.25~\mathrm{s}$. Between $t=0.15~\mathrm{s}$ and $t=0.20~\mathrm{s}$ the three structures are at their closest mutual separation, and their relative screen-space ordering changes as their paths cross, yet each structure is matched to the track whose predicted state, rather than its previous position, is closest to it, so every ID is retained by the same physical bubble throughout and none are exchanged between structures. This is the outcome the geometric and kinematic gates of Section~\ref{subsec:pid} are designed to guarantee because association is resolved from the predicted state $\mathbf{x}_m^p$ rather than the last observed position, a close pass does not by itself create an ambiguous match, as described in Section~\ref{subsec:e2l_hysteresis}, the result is unaffected by which of the $40$ subdomains each structure momentarily occupies during the crossing.

A sensitivity study was performed for the adjacent resolved vapor structures using the five weight combinations listed in Table~\ref{tab:association_weight_sensitivity}. Each configuration was evaluated over $t=0$-$0.5~\mathrm{s}$, covering 4,999 association time steps and 16,412 accepted track-observation pairs. The table reports the mean accepted cost, obtained by averaging the pairwise association costs over all accepted matches throughout this interval. All tested weight combinations yielded identical track-observation pairings and no ID exchanges. The results therefore indicate that the association outcome is insensitive to the examined weight variations. For a given weight combination, a smaller cost indicates closer agreement between the predicted track state and the selected observation according to the weighted criteria. Case B, $(w_x,w_R,w_U)=(1,0.5,0.25)$, yielded the lowest mean accepted cost, $3.89\times10^{-4}$, while producing the same final pairings as all other tested configurations. These results support retaining Case B because its lower cost indicates greater confidence in pair selection.

\begin{table}[htbp]
    \centering
    \caption{Sensitivity of the adjacent-structure association results
    to the cost-function weights.}
    \label{tab:association_weight_sensitivity}
    \begin{tabular}{cccc}
        \hline
        Case & $(w_x,w_R,w_U)$ & Accepted pairs
              & Mean accepted cost \\
        \hline
        A & $(1,1,1)$       & 16,412 & $1.13\times10^{-3}$ \\
        B & $(1,0.5,0.25)$  & 16,412 & $3.89\times10^{-4}$ \\
        C & $(1,0.5,1)$     & 16,412 & $1.10\times10^{-3}$ \\
        D & $(1,1,0.25)$    & 16,412 & $4.20\times10^{-4}$ \\
        E & $(1,1,0.5)$     & 16,412 & $6.56\times10^{-4}$ \\
        \hline
    \end{tabular}
\end{table}

\subsection{Verification of Eulerian-to-Lagrangian transition}
\label{subsec:e2l_transition_verification}

\label{subsec:collapse}

A further three-dimensional test case is performed to examine the collapse of a single spherical bubble in an initially quiescent liquid and to verify the correct prescription of initial conditions for the Rayleigh-Plesset equation once the bubble is transferred to the unresolved (Lagrangian) representation. In this study, a radius-based switching criterion is employed to trigger the transition based on the predicted collapse dynamics. The computations are carried out in a cubic domain of side length $8~\mathrm{mm}$ centered at the origin, providing a far-field distance of approximately ten times the initial bubble radius and thereby reducing boundary-induced effects during collapse. A mesh with $40$ cells per direction is used. A refined core region is embedded at the center, yielding a characteristic cell size of approximately $53~\mu\mathrm{m}$, while the outer region is substantially coarser. All six boundaries are treated as far-field pressure boundaries with fixed pressure $p=1\times 10^{5}~\mathrm{Pa}$, and the velocity field is assigned a zero-gradient condition.

\begin{figure}[t]
\centering

\begin{minipage}[t]{0.45\textwidth}\vspace{0pt}
\centering
\includegraphics[width=\linewidth]{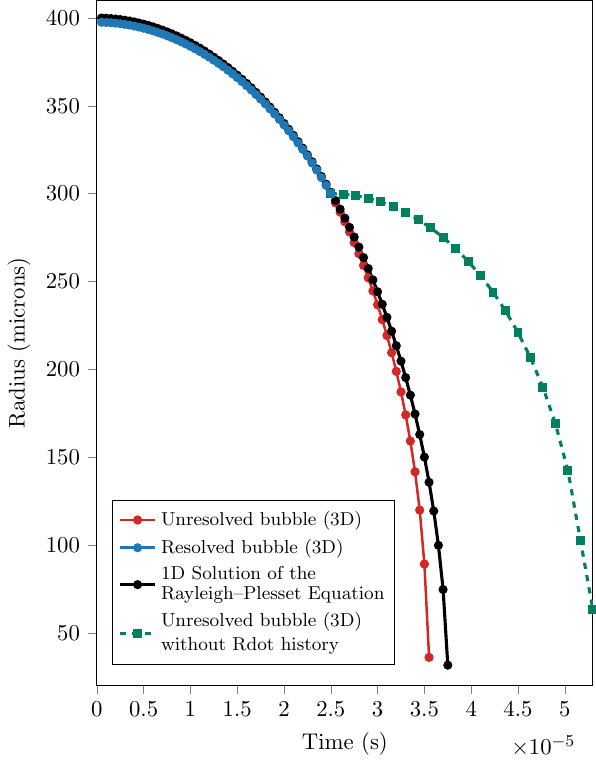}
\end{minipage}%
\hfill
\begin{minipage}[t]{0.54\textwidth}\vspace{0pt}
\centering

\begin{minipage}[t]{0.32\textwidth}\vspace{20pt}\centering
\includegraphics[width=\linewidth]{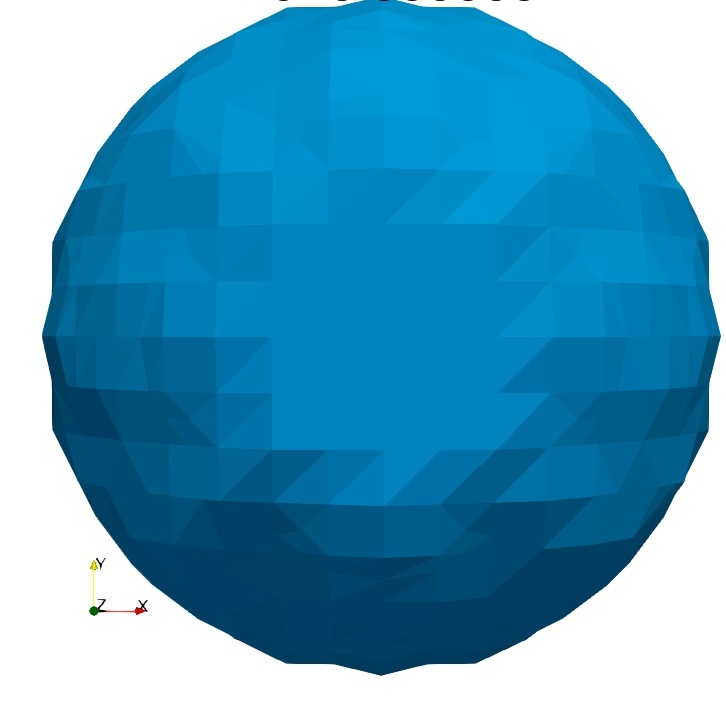}\\[-2pt]{\small $t=0.0~\mathrm{s}$}
\end{minipage}\hfill
\begin{minipage}[t]{0.32\textwidth}\vspace{20pt}\centering
\includegraphics[width=\linewidth]{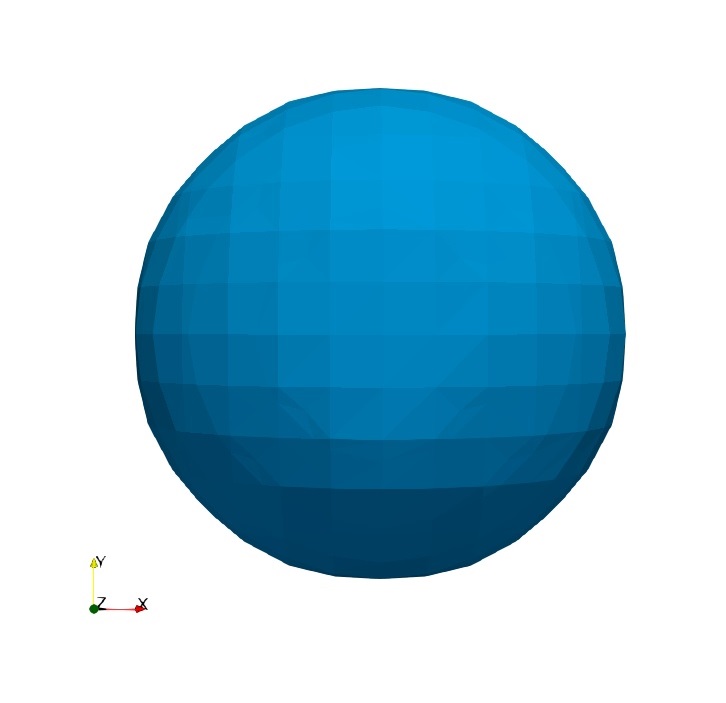}\\[-2pt]{\small $t=2.5\times10^{-5}~\mathrm{s}$}
\end{minipage}\hfill
\begin{minipage}[t]{0.32\textwidth}\vspace{20pt}\centering
\includegraphics[width=\linewidth]{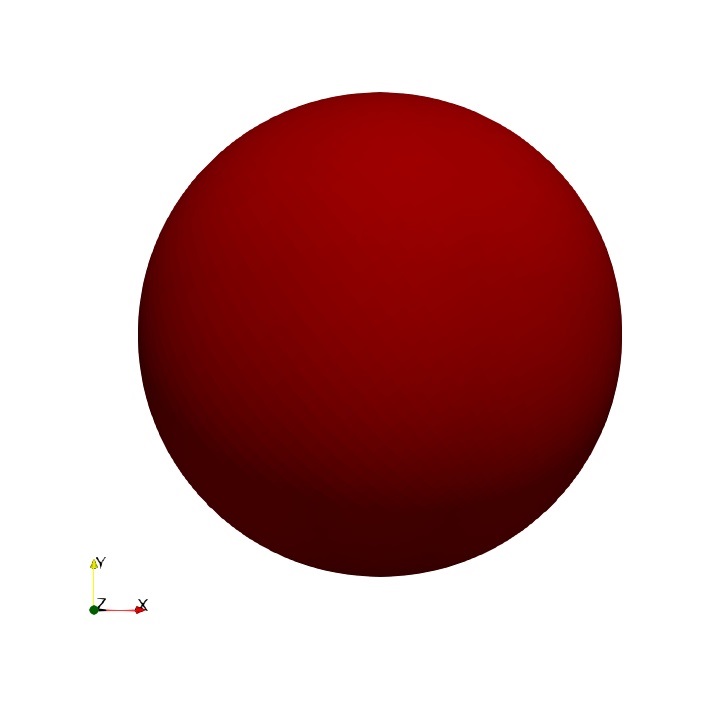}\\[-2pt]{\small $t=2.6\times10^{-5}~\mathrm{s}$}
\end{minipage}

\vspace{0.6em}

\begin{minipage}[t]{0.32\textwidth}\vspace{20pt}\centering
\includegraphics[width=\linewidth]{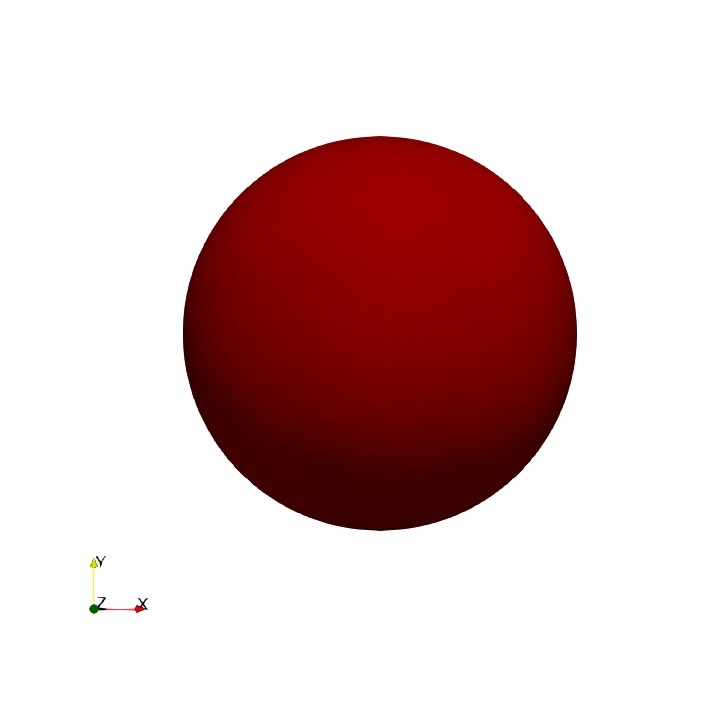}\\[-2pt]{\small $t=3.0\times10^{-5}~\mathrm{s}$}
\end{minipage}\hfill
\begin{minipage}[t]{0.32\textwidth}\vspace{20pt}\centering
\includegraphics[width=\linewidth]{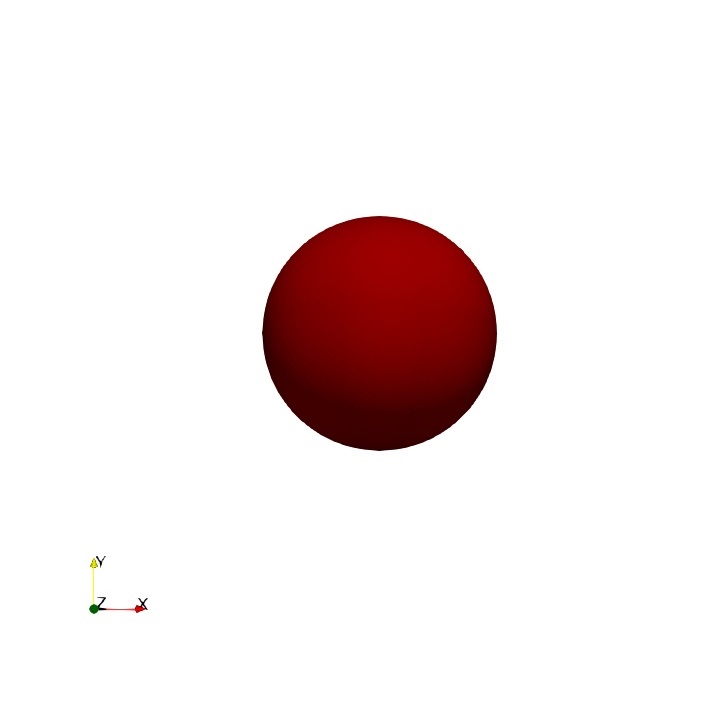}\\[-2pt]{\small $t=3.4\times10^{-5}~\mathrm{s}$}
\end{minipage}\hfill
\begin{minipage}[t]{0.32\textwidth}\vspace{20pt}\centering
\includegraphics[width=\linewidth]{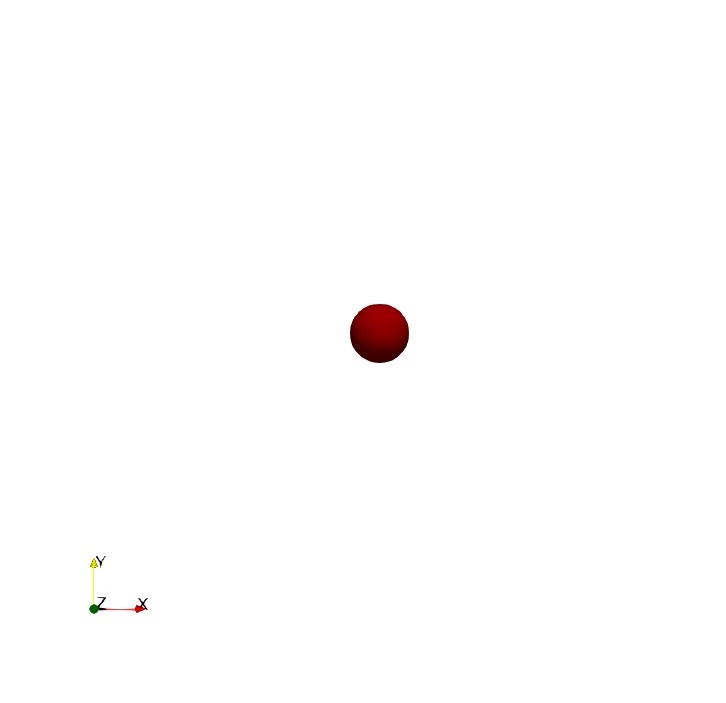}\\[-2pt]{\small $t=3.6\times10^{-5}~\mathrm{s}$}
\end{minipage}

\end{minipage}

\caption{Threshold-based Eulerian-to-Lagrangian transition during the collapse of a single spherical bubble. Left: radius history comparing the hybrid simulation with the Rayleigh-Plesset reference. Right: three-dimensional snapshots showing the resolved bubble collapse and the subsequent continuation as an unresolved Lagrangian bubble after the switch at $t=2.6\times10^{-5}~\mathrm{s}$. Blue denotes the resolved Eulerian representation and red denotes the unresolved Lagrangian representation. The green curve initialized with \(\dot R=0\) demonstrates the error produced when the Eulerian radius history is not retained.}
\label{fig:bubble_side_by_side}
\end{figure}

The two-phase system consists of water and vapor. Water is modeled with density $\rho_{\mathrm{l}}=1000~\mathrm{kg/m^3}$ and kinematic viscosity $\nu_{\mathrm{l}}=9\times 10^{-7}~\mathrm{m^2/s}$, whereas the vapor has density $\rho_{\mathrm{v}}=2.308\times 10^{-2}~\mathrm{kg/m^3}$ and kinematic viscosity $\nu_{\mathrm{v}}=4.273\times 10^{-4}~\mathrm{m^2/s}$. Surface tension is set to $\sigma=0.072~\mathrm{N/m}$, and the saturation pressure used in the cavitation model is $p_v=2300~\mathrm{Pa}$. The governing equations are solved using the Schnerr-Sauer closure (Eqn.~\eqref{eqn:cccc}). For this canonical verification problem, \(C_c=3\) is selected to obtain a resolved Eulerian collapse trajectory comparable to the one-dimensional Rayleigh–Plesset reference before the representation transition. The purpose of this case is to assess the transition and initialization procedure rather than to validate the Schnerr–Sauer closure.  The flow is initialized at rest with zero velocity and a spatially uniform far-field pressure of $1\times10^{5}~\mathrm{Pa}$. The initial bubble is prescribed at the domain center as a spherical region of radius $R_0=400~\mu\mathrm{m}$ in which the liquid volume fraction is set to $\alpha_{\mathrm{l}}=0$, surrounded by water. Time integration is performed with a constant time step of $\Delta t=5\times10^{-8}~\mathrm{s}$ up to $t=3.6\times10^{-5}~\mathrm{s}$. The conversion between Eulerian and Lagrangian representations is governed by a radius-based criterion with a single threshold of $300~\mu\mathrm{m}$. When the Eulerian-resolved cavity shrinks below this radius, it is converted into a Lagrangian bubble, as shown in Figure~\ref{fig:bubble_side_by_side} at $t=2.6\times10^{-5}~\mathrm{s}$.

Figure~\ref{fig:bubble_side_by_side} shows that the three-dimensional simulation exhibits a smooth, nearly spherical collapse during the Eulerian phase, with the bubble radius decreasing monotonically from the initial $400~\mu\mathrm{m}$. Over the resolved interval, the predicted radius evolution agrees closely, with only minor deviations, with the one-dimensional Rayleigh-Plesset solution. After the transition at $R=300~\mu\mathrm{m}$, the temporal evolution deviates slightly further from the one-dimensional reference. This residual discrepancy is attributed to small differences between the resolved Eulerian radius evolution and the idealized one-dimensional solution prior to switching, together with the difference between the radius rate reconstructed from the preceding Eulerian history and that of the idealized one-dimensional Rayleigh-Plesset reference at the same radius. The importance of transferring the radial-velocity history is further illustrated by the green curve, for which the unresolved-bubble calculation is initialized with $\dot{R}=0$ at the transition. This initialization produces a substantially smaller decrease in radius and delays the minimum recorded radius to approximately $53~\mu\mathrm{s}$. In contrast, initialization using the preceding Eulerian radius history maintains much closer agreement with the one-dimensional reference. This comparison highlights that preserving the bubble radius alone at the transition is insufficient; the inward radial velocity must also be transferred to retain the ongoing collapse dynamics. Overall, this test demonstrates that the hybrid Eulerian-Lagrangian methodology can reproduce Rayleigh-Plesset-consistent collapse dynamics while the bubble remains sufficiently resolved and can initialize the unresolved-bubble model consistently at the switching instant.

%% file: sections/03_l2e_transition.tex
\section{Lagrangian-to-Eulerian transition}
\label{sec:l2e}
\label{subsec:l2e_criteria}

\begin{figure}[t]
  \centering
  
  \begin{subfigure}[b]{\linewidth}
    \centering
    \includegraphics[width=0.6\linewidth]{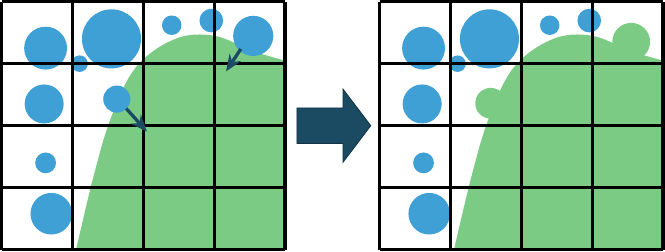}
    \caption{Coalescence-triggered Lagrangian to Eulerian transition.}
    \label{fig:l2e_growth}
  \end{subfigure}
  
  \vspace{0.5cm} 
  
  \begin{subfigure}[b]{\linewidth}
    \centering
    \includegraphics[width=0.6\linewidth]{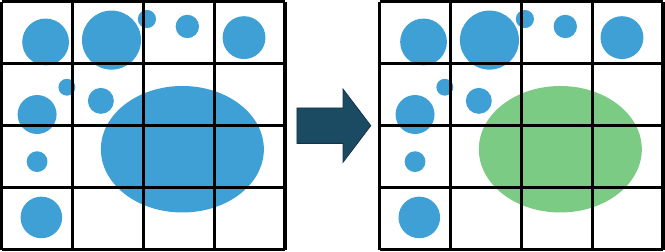}
    \caption{Growth-triggered Lagrangian to Eulerian transition.}
    \label{fig:l2e_coalescence}
  \end{subfigure}
  
  \caption{Lagrangian-to-Eulerian transition mechanisms.}
  \label{fig:l2e_schematic}
\end{figure}

When an unresolved Lagrangian entity grows to a grid-resolvable scale, or when it interacts with a nearby resolved Eulerian vapor region, it is converted back to the Eulerian representation. Two mechanisms drive this geometric criterion. The first is a threshold-based transition: conversion is triggered once the unresolved vapor entity exceeds a prescribed resolvability criterion, expressed as either a minimum number of occupied cells or a prescribed radius. The second is an interface-based transition: coalescence or direct interaction between an unresolved bubble and a resolved Eulerian vapor structure initiates the conversion once the bubble approaches an existing Eulerian interface. 
Figure~\ref{fig:l2e_schematic} illustrates these two physical mechanisms: size growth, which triggers the threshold-based transition, and coalescence, which triggers the interface-based transition.

Once the criterion is satisfied, the unresolved bubble's vapor volume is transferred to the Eulerian field through the kernel-based vapor deposition procedure described in Section~\ref{subsec:kernel}, which distributes the volume onto the surrounding cells as $\beta_{\mathrm{l}}$. The bubble is then removed from the Lagrangian framework, and $\beta_{\mathrm{l}}$ is transferred into $\alpha_{\mathrm{l}}$ so that the total vapor mass is conserved across the transition; the same conservative treatment is applied to the associated momentum source term.

\subsection{Kernel-based deposition of vapor volume and momentum}
\label{subsec:kernel}

The most important numerical operation in the Lagrangian-to-Eulerian conversion is the deposition of a finite unresolved bubble volume onto a finite-volume Eulerian mesh. When a bubble is smaller than a computational cell, its influence can be deposited into the host cell as an effective point source. When the bubble size becomes comparable to or larger than the local grid spacing, the coupling must instead account for a finite interaction footprint by spreading the bubble volume over a compact neighborhood of cells, followed by a normalization step that enforces exact volume conservation.

\begin{figure}[t]
  \centering
  \includegraphics[width=0.75\linewidth]{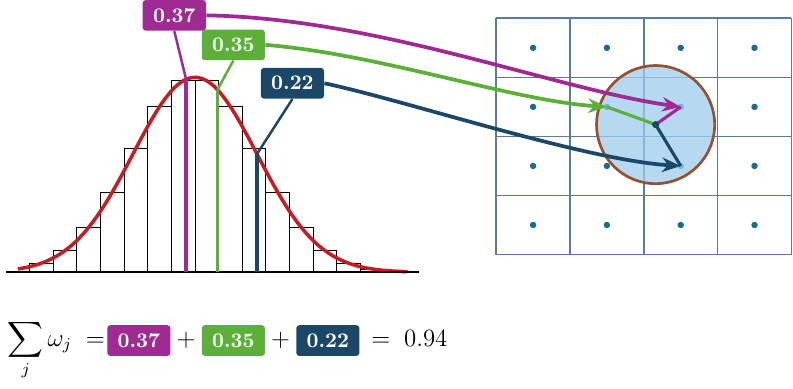}
  \caption{Problems caused by non-normalized kernel weights. The raw Gaussian weights sampled at the cell centers within the local neighborhood sum to $0.94$, not unity, so depositing the bubble volume without normalization would not conserve mass.}
  \label{fig:kernel1}
\end{figure}

A naive discretization of a Gaussian kernel over a finite set of cells can produce raw weights that do not sum to unity, resulting in non-conservative deposition unless an explicit correction is applied. Figure~\ref{fig:kernel1} illustrates this problem for a representative stencil, where the unnormalized weights sum to only 0.94 rather than 1. Even with normalized weights, a sufficiently large bubble relative to local cell volumes may overfill one or more cells, driving $\beta_{\mathrm{l}}$ outside its admissible bound $\beta_{\mathrm{l}}\in[0,1)$ established in Eqn.~\eqref{eqn:betaaa}. The implementation therefore uses a normalized and capacity-bounded redistribution procedure that ensures conservative spreading within the local neighborhood while enforcing a minimum admissible liquid fraction and redistributing any excess vapor volume to nearby cells with remaining capacity.

\begin{figure}[t]
  \centering
  \includegraphics[width=0.3\linewidth]{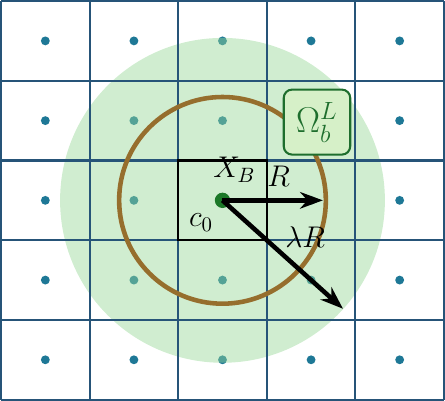}
  \caption{Bubble area of influence, constructed by graph traversal from the host cell $c_0$, collecting cells whose centers lie within the coverage radius $\lambda R_b$ of the bubble centroid $\boldsymbol{x}_b$. The green shaded region denotes the Eulerian coupling neighborhood $\Omega_b^{L}$.}
  \label{fig:kernel2}
\end{figure}

Starting from the bubble center's cell, $c_0$, the code traverses cell-to-cell connectivity and collects neighboring cells whose centers lie inside the coverage radius

\begin{equation}
    r_i=\|\mathbf{x}_i-\mathbf{x}_b\| \le \lambda R_b,
    \label{eq:l2e_support}
\end{equation}

\noindent as shown in Figure~\ref{fig:kernel2}. Here $\lambda$ is a user-defined coverage factor; in this work, $\lambda=1.2$, consistent with the approximate spreading radius of $1.16R$ estimated by Ghahramani et al.~\cite{Ghahramani2021} from the random close-packing limit. The set $\Omega_b^{L}$ denotes the Eulerian neighborhood formed by the cells whose centers satisfy Eqn.~\eqref{eq:l2e_support}. This graph traversal is preferable to an index-based stencil because it works on unstructured meshes and naturally follows local connectivity. The raw distance weights are Gaussian inside the local neighborhood,

\begin{equation}
    \widetilde{w}_i =
    \begin{cases}
    1, & |\Omega_b^{L}|=1,\\[2mm]
    \exp\!\left[-\dfrac{1}{2}\left(\dfrac{r_i}{\sigma_b}\right)^2\right], & |\Omega_b^{L}|>1,
    \end{cases}
    \qquad
    \sigma_b=\frac{R_b}{\gamma},
    \label{eq:kernel_raw}
\end{equation}

\noindent where $\gamma$ is a user-defined standard-deviation factor; in this work, $\gamma=0.3$, following Ghahramani et al.~\cite{Ghahramani2021}. The weights are normalized on the selected discrete support,

\begin{equation}
    w_i=\frac{\widetilde{w}_i}{\sum_{j\in\Omega_b^{L}}\widetilde{w}_j},
    \qquad
    \sum_{i\in\Omega_b^{L}}w_i=1.
    \label{eq:kernel}
\end{equation}

\noindent This normalization is essential: without it, truncating the Gaussian at a finite radius would create mesh- and position-dependent volume loss or gain, exactly the problem illustrated in Figure~\ref{fig:kernel1}. The provisional deposited vapor volume and the corresponding liquid-occupancy decrement are

\begin{equation}
    V_i^\star=w_i V_b,
    \qquad
    v_i^\star = \frac{V_i^\star}{V_i},
    \qquad
    V_b=\frac{4\pi R_b^3}{3},
    \label{eq:target_vapor_fraction}
\end{equation}

\noindent where $R_b$ is the bubble radius and $V_i$ is the cell volume. The implementation then enforces boundedness. The available vapor capacity of a receiving cell is

\begin{equation}
    A_i = \left(\beta_{{\mathrm{l}},i}^{-}-\beta_{\min}\right)V_i,
    \label{eq:l2e_capacity}
\end{equation}

\noindent where the superscript $-$ denotes the state before deposition and $\beta_{\min}$ is the minimum admissible liquid fraction retained in a cell; in this work, $\beta_{\min}=0.3$, allowing a maximum bubble volume fraction of $0.7$. This capacity constraint follows the packing-based rationale of Ghahramani et al.~\cite{Ghahramani2021}. The accepted first-pass deposition is

\begin{equation}
    \Delta V_i^{(0)}=\min\!\left(V_i^\star,A_i\right),
    \qquad
    \beta_{{\mathrm{l}},i}^{(0)}=\beta_{{\mathrm{l}},i}^{-}-\frac{\Delta V_i^{(0)}}{V_i}.
    \label{eq:l2e_volume}
\end{equation}

\noindent The excess volume produced by capacity clipping is

\begin{equation}
    V_{ex}=\sum_{i\in\Omega_b^{L}} \left(V_i^\star-\Delta V_i^{(0)}\right)_+,
    \label{eq:excess_volume}
\end{equation}

\noindent and it is redistributed to cells in the local neighborhood that still have capacity. If the initial neighborhood contains only the owner cell, its immediate neighbors are added as secondary redistribution candidates. The final deposited volume is therefore

\begin{equation}
    \Delta V_i = \Delta V_i^{(0)} + \Delta V_i^{(\mathrm{red})},
    \qquad
    \beta_{{\mathrm{l}},i}^{+}=\beta_{{\mathrm{l}},i}^{-}-\frac{\Delta V_i}{V_i}.
    \label{eq:l2e_beta_update}
\end{equation}

\begin{figure}[t!]
  \centering
  \includegraphics[width=\linewidth]{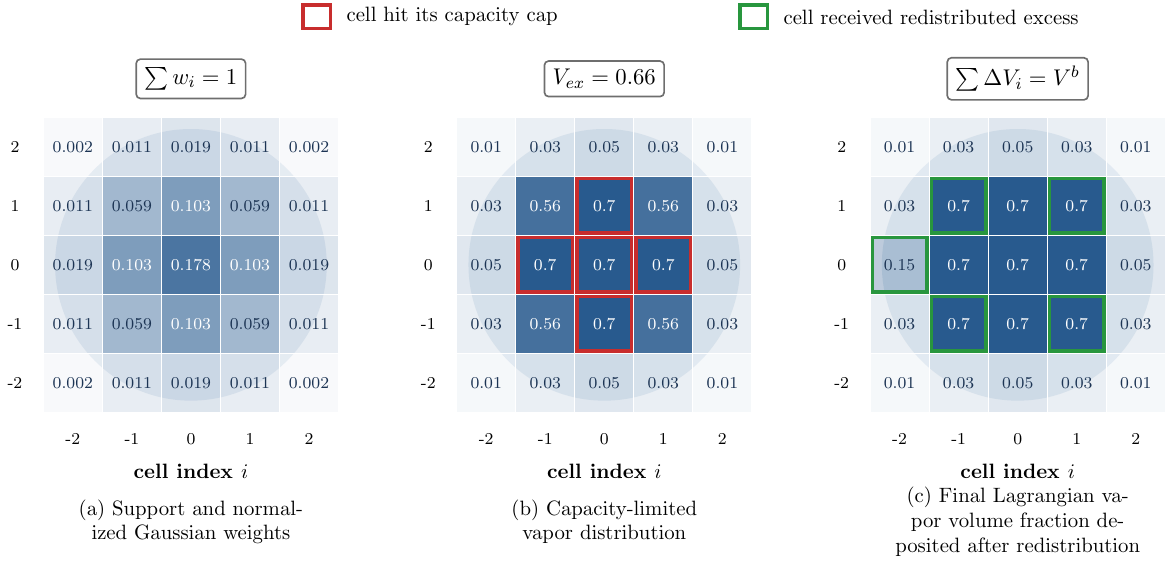}
  \caption{Visualization of the conservative volume-deposition kernel used during Lagrangian-to-Eulerian conversion. }
  \label{fig:kernel_volume_deposition_3d}
\end{figure}

\noindent Here, $\Delta V_i^{(\mathrm{red})}$ denotes the additional vapor volume assigned to cell $i$ during the nearest-first redistribution of the excess volume remaining after the capacity-limited first pass. The conversion is exactly conservative when the neighborhood and the secondary redistribution candidates have sufficient remaining capacity. This process has been shown in Figure~\ref{fig:kernel_volume_deposition_3d}.  Figure~\ref{fig:kernel_volume_deposition_3d}a shows the local neighborhood and the normalized Gaussian weights on the receiving cells. Figure~\ref{fig:kernel_volume_deposition_3d}b shows the first-pass deposition before boundedness enforcement, where some cells are clipped by their vapor-capacity constraint. Figure~\ref{fig:kernel_volume_deposition_3d}c shows the final conservative deposition after redistribution of the excess volume to cells with remaining capacity.

Away from an existing resolved Eulerian interface, receiving cells are forced to $\alpha_{\mathrm{l}}=1$ after the unresolved vapor has been deposited into $\beta_{\mathrm{l}}$. This enforces non-overlap between resolved Eulerian vapor and unresolved Lagrangian vapor. In cells where the resolved-vapor volume fraction exceeds the prescribed threshold, set to $0.9$ in this work, the update preserves the interface-aware behavior required for coalescence of an unresolved bubble with a resolved Eulerian cavity. 

In addition to the conservative volume-deposition kernel used for the representation change, the solver employs a second kernel for the recurrent exchange of bubble-induced source terms with the Eulerian field. These source terms include the momentum-exchange contribution $\boldsymbol{F}/V$ in Eqn.~\eqref{eqn:hybrid-mom2} and the bubble mass source $\dot{m}_{\mathrm{B}}$ in Eqn.~\eqref{eqn:hybrid-non-divergence-free2}. If these terms are deposited with a purely Gaussian kernel, most of the forcing is concentrated in the host cell and its immediate neighbors. Although this choice is local, it can increase mesh sensitivity and produce unnecessarily stiff, cell-local forcing.
A blended source kernel is therefore used for the recurrent source exchange. For the actual Lagrangian-to-Eulerian conversion, the deposited vapor volume should remain strongly concentrated near the bubble center while still respecting boundedness and capacity limits. For the repeated exchange of momentum and mass source terms during bubble evolution, however, a less concentrated distribution over the same neighborhood helps reduce localized source peaks and the associated numerical stiffness. Compared with a nearest-cell delta kernel, the present construction avoids discontinuous jumps when the bubble crosses cell faces. Compared with an unbounded or truncated Gaussian, it preserves strict normalization on a compact graph-based stencil. Compared with a purely uniform top-hat, it retains a physically meaningful preference for cells near the bubble center. The result is a conservative, compact, and numerically robust source-deposition operator. The uniform weights on the same neighborhood are

\begin{equation}
    w_i^{U}=\frac{1}{|\Omega_b^{L}|},
    \qquad
    \sum_{i\in\Omega_b^{L}}w_i^{U}=1.
    \label{eq:blend_uniform}
\end{equation}

\noindent The blended kernel is then defined as the convex combination

\begin{equation}
    \widehat{w}_i^{\mathrm{blend}}(\eta_s)
    =(1-\eta_s)w_i+\eta_s w_i^{U},
    \qquad 0\leq\eta_s\leq1.
    \label{eq:source_blend_kernel}
\end{equation}

\noindent Because both constituent kernels are non-negative and normalized, the blended kernel also remains non-negative and normalized,

\begin{equation}
    \sum_{i\in\Omega_b^{L}}\widehat{w}_i^{\mathrm{blend}}(\eta_s)
    =(1-\eta_s)\sum_{i\in\Omega_b^{L}}w_i
    +\eta_s\sum_{i\in\Omega_b^{L}}w_i^{U}=1.
    \label{eq:blend_conservation}
\end{equation}

\noindent Thus, $\eta_s=0$ recovers purely Gaussian deposition, whereas $\eta_s=1$ gives a uniform distribution over the selected stencil. Intermediate values broaden the support while retaining the spatial preference of the Gaussian kernel. In the present work, $\eta_s=0.5$.

During the Lagrangian update over a time step $\Delta t$, each bubble accumulates carrier-phase source contributions, primarily (i) the momentum exchange $\boldsymbol{F}$ and (ii) the net bubble mass-transfer rate $\dot{m}_{\mathrm{B}}$. These quantities are distributed to every Eulerian control volume $i\in\Omega_b^L$ according to

\begin{align}
    \boldsymbol{F}_{i} &= \widehat{w}_i^{\mathrm{blend}}(\eta_s)\,\boldsymbol{F},
    \label{eq:blend_ucoeff}\\
    \dot{m}_{\mathrm{B},i} &= \widehat{w}_i^{\mathrm{blend}}(\eta_s)\,\dot{m}_{\mathrm{B}}.
    \label{eq:blend_mdot}
\end{align}

\begin{figure}[H]
  \centering
  \begin{minipage}[t]{0.4\linewidth}
    \centering
    \includegraphics[width=\linewidth]{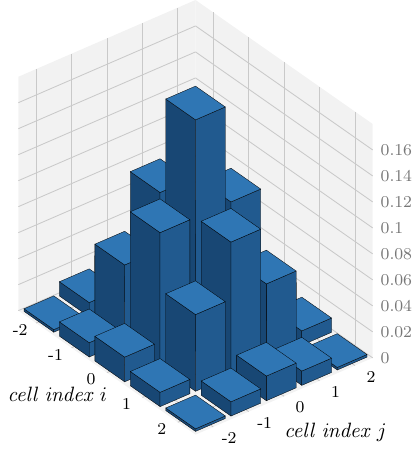}
  \end{minipage}\hfill
  \begin{minipage}[t]{0.4\linewidth}
    \centering
    \includegraphics[width=\linewidth]{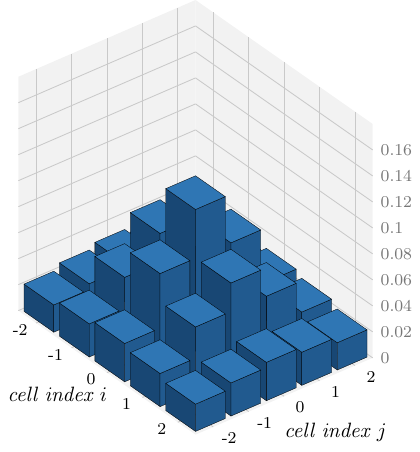}
  \end{minipage}
  \caption{Comparison of the peaked normalized Gaussian used for volume deposition (left) with the flatter blended kernel used to distribute momentum source terms during ordinary bubble tracking (right).}
  \label{fig:kernel_blended_3d}
\end{figure}

\noindent Here, $\boldsymbol{F}_{i}$ and $\dot{m}_{\mathrm{B},i}$ are the contributions assigned to cell $i$. Because $\sum_i \widehat{w}_i^{\mathrm{blend}}=1$, the sum of the cellwise contributions exactly reproduces the bubble-level source totals. In this way, the overall coupling remains consistent: the representation change itself is handled by the capacity-limited conservative kernel of Figure~\ref{fig:kernel_volume_deposition_3d}, whereas the ongoing source exchange during bubble evolution is handled by the smoother blended kernel. Figure~\ref{fig:kernel_blended_3d} compares the blended kernel at $\eta_s=0.5$ with the sharper Gaussian used for conservative volume transfer.

\subsection{Verification}
\label{subsec:l2e_verification}

The Lagrangian-to-Eulerian transition is verified using two complementary configurations. The first case tests threshold-based conversion of an advected and growing bubble using a cell-count criterion.  The second case examines the coalescence of a single unresolved Lagrangian bubble with a resolved Eulerian vapor volume. Together, these cases verify the transition trigger, the removal of the Lagrangian bubble, and the transfer of its vapor contribution to the Eulerian volume-fraction field.

\subsubsection{Threshold-Based Lagrangian-to-Eulerian Transition}
\label{subsec:l2e_threshold_verification}

The unresolved Lagrangian bubble is converted to the resolved Eulerian representation once its spatial extent exceeds a user-defined resolvability threshold, as illustrated in Figure~\ref{fig:l2e_growth}. 
The computational domain is a three-dimensional rectangular box of size $0.3~\mathrm{m}\times0.1~\mathrm{m}\times0.1~\mathrm{m}$. It is discretized using $60\times20\times20$ uniform cells, and a cell edge length of $0.005~\mathrm{m}$. An inlet is prescribed on the left boundary, an outlet on the right boundary, and no-slip walls on the remaining faces. A streamwise pressure gradient is imposed by decreasing the pressure from $2000~\mathrm{Pa}$ at the inlet to $1000~\mathrm{Pa}$ at the outlet.  The Lagrangian bubble grows while being advected downstream. The transition threshold is set to 44 occupied cells; once the unresolved vapor footprint exceeds this value, the bubble is converted to the Eulerian representation.

\begin{figure}[t]
    \centering
    \includegraphics[width=0.7\linewidth]{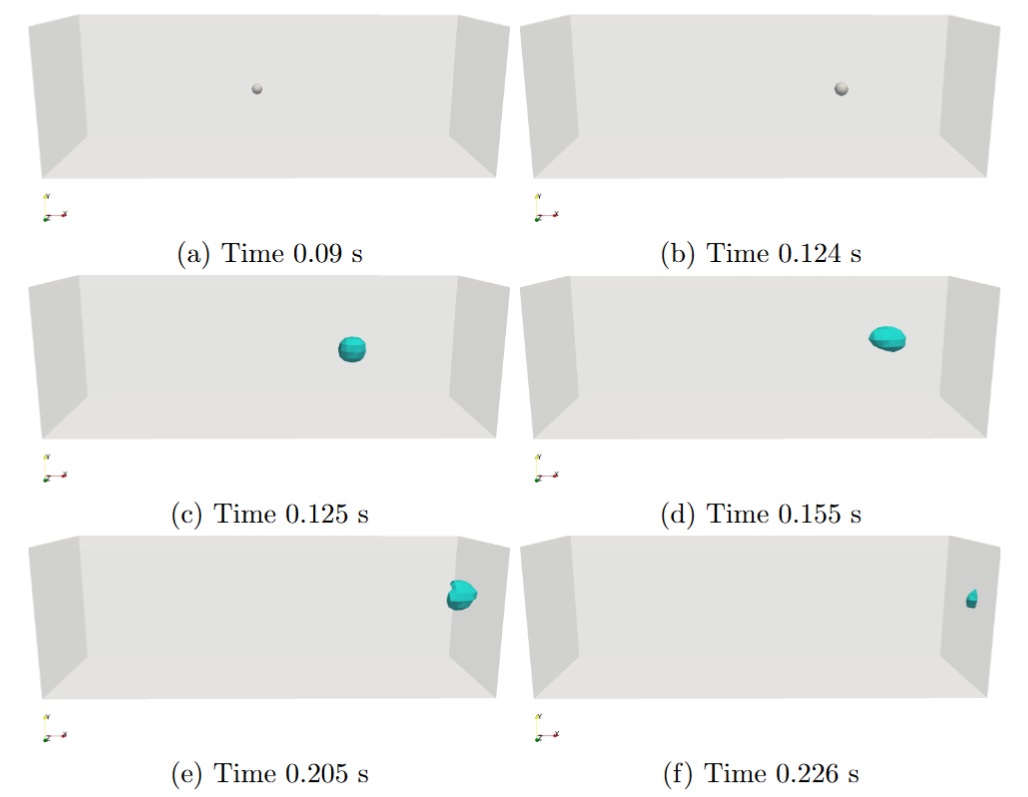}
    \caption{Threshold-based Lagrangian-to-Eulerian transition of an advected and growing vapor nucleus. The bubble expands under pressure condition, exceeds the resolvability threshold, and is transferred to the Eulerian volume-fraction field. Adapted from Lavari and Gnanaskandan~\cite{LavariFEDSM2025}.}
    \label{fig:l2e_threshold_sequence}
\end{figure}

The simulation is advanced to $t=0.25~\mathrm{s}$ using a time step of $\Delta t=5\times10^{-4}~\mathrm{s}$. At $t=0.02~\mathrm{s}$, a bubble with an initial diameter of $0.001~\mathrm{m}$ and an initial velocity of $2~\mathrm{m/s}$ is injected at the inlet. Figure~\ref{fig:l2e_threshold_sequence} shows that the bubble travels from left to right while expanding. At $t=0.124~\mathrm{s}$, its unresolved vapor footprint exceeds the prescribed threshold, and at $t=0.125~\mathrm{s}$ the bubble is converted to the Eulerian framework. The Lagrangian bubble is then removed, while its vapor contribution is retained in the Eulerian liquid volume-fraction field $\alpha_{\mathrm{l}}$ and is subsequently transported by the carrier flow under the imposed pressure gradient.

\subsubsection{Coalescence of a Single Lagrangian Bubble with an Eulerian Vapor Volume}
\label{subsec:l2e_coalescence_verification}

To demonstrate the ability of the hybrid approach to transform a vapor volume from the Lagrangian framework to the Eulerian framework during interaction with an already resolved structure, we simulated the coalescence of a small Lagrangian bubble with a larger, nearly spherical Eulerian vapor volume. To visualize the transformation between the two representations as clearly as possible, the solution of the equations of bubble motion and bubble dynamics was deactivated for the Lagrangian bubble, and its velocity was prescribed so that it moved with a constant speed toward the Eulerian vapor structure.

The computational domain is a rectangular box of size $1.2\times10^{-3}~\mathrm{m}\times6.0\times10^{-4}~\mathrm{m}\times6.0\times10^{-4}~\mathrm{m}$ discretized with a uniform $160\times80\times80$ mesh. All boundaries are walls. At the initial time, a resolved Eulerian vapor sphere of radius $1.2\times10^{-4}~\mathrm{m}$ is placed at the domain center by setting $\alpha_{\mathrm{l}}=0$ inside the sphere, while the surrounding liquid is initialized with $\alpha_{\mathrm{l}}=1$. A single unresolved Lagrangian bubble is placed at $(-2.20\times10^{-4},0,0)$ with an initial diameter of $2.0\times10^{-5}~\mathrm{m}$ and a prescribed translational velocity of $(1,0,0)~\mathrm{m/s}$. The simulation uses a constant time step of $\Delta t=10^{-7}~\mathrm{s}$ and is advanced to $t=2.0\times10^{-4}~\mathrm{s}$. In this configuration, the interface-based transition criterion is the relevant mechanism: once the unresolved bubble reaches the surface of the resolved Eulerian vapor structure, it is transferred to the Eulerian representation and then merges with the existing vapor volume.

\begin{figure}[t]
    \centering
    \begin{minipage}[t]{0.4\linewidth}
        \centering
        \includegraphics[width=\linewidth]{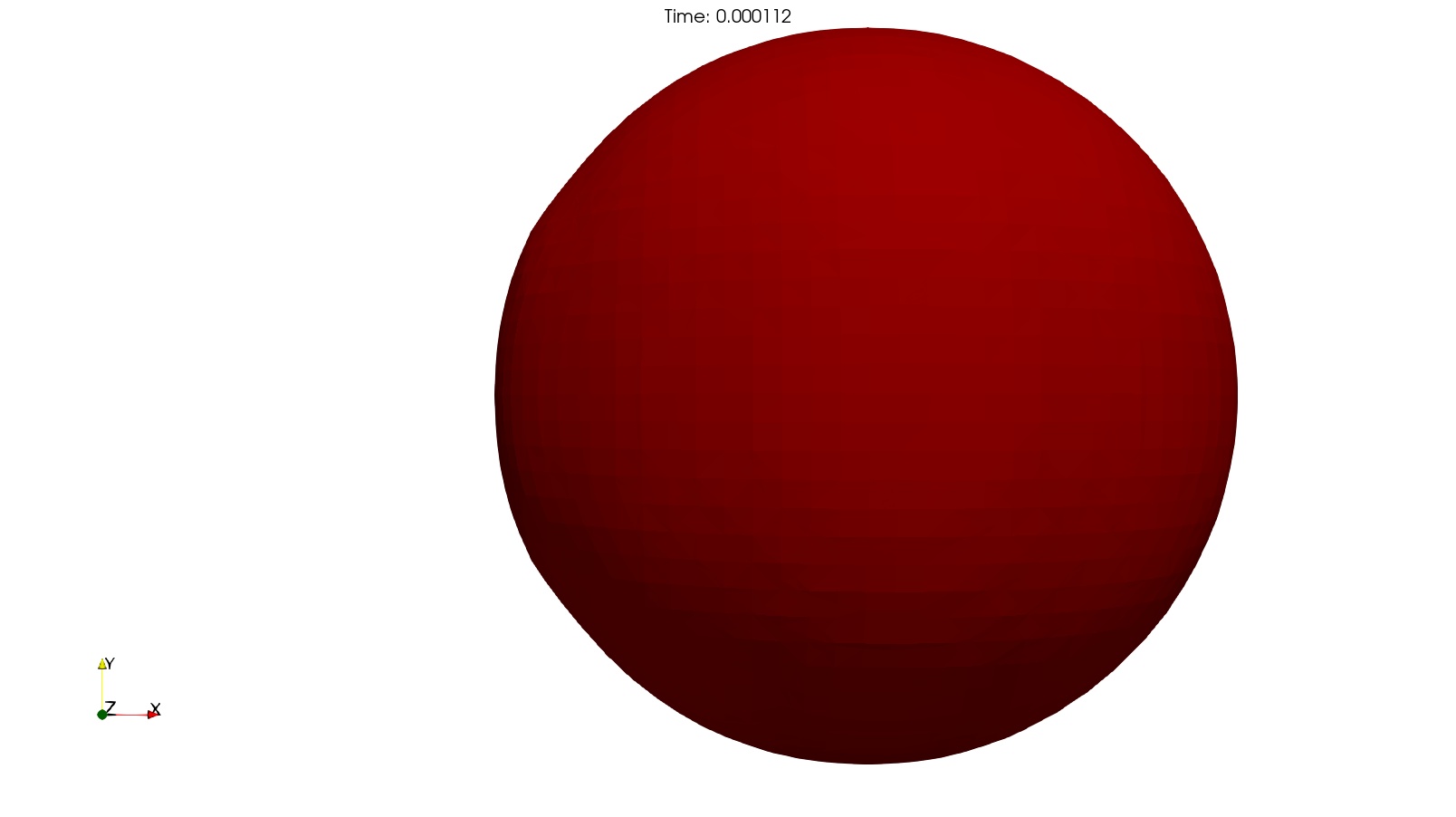}\\[-2pt]
        {\scriptsize (a) $t=0~\mathrm{s}$}
    \end{minipage}\hfill
    \begin{minipage}[t]{0.4\linewidth}
        \centering
        \includegraphics[width=\linewidth]{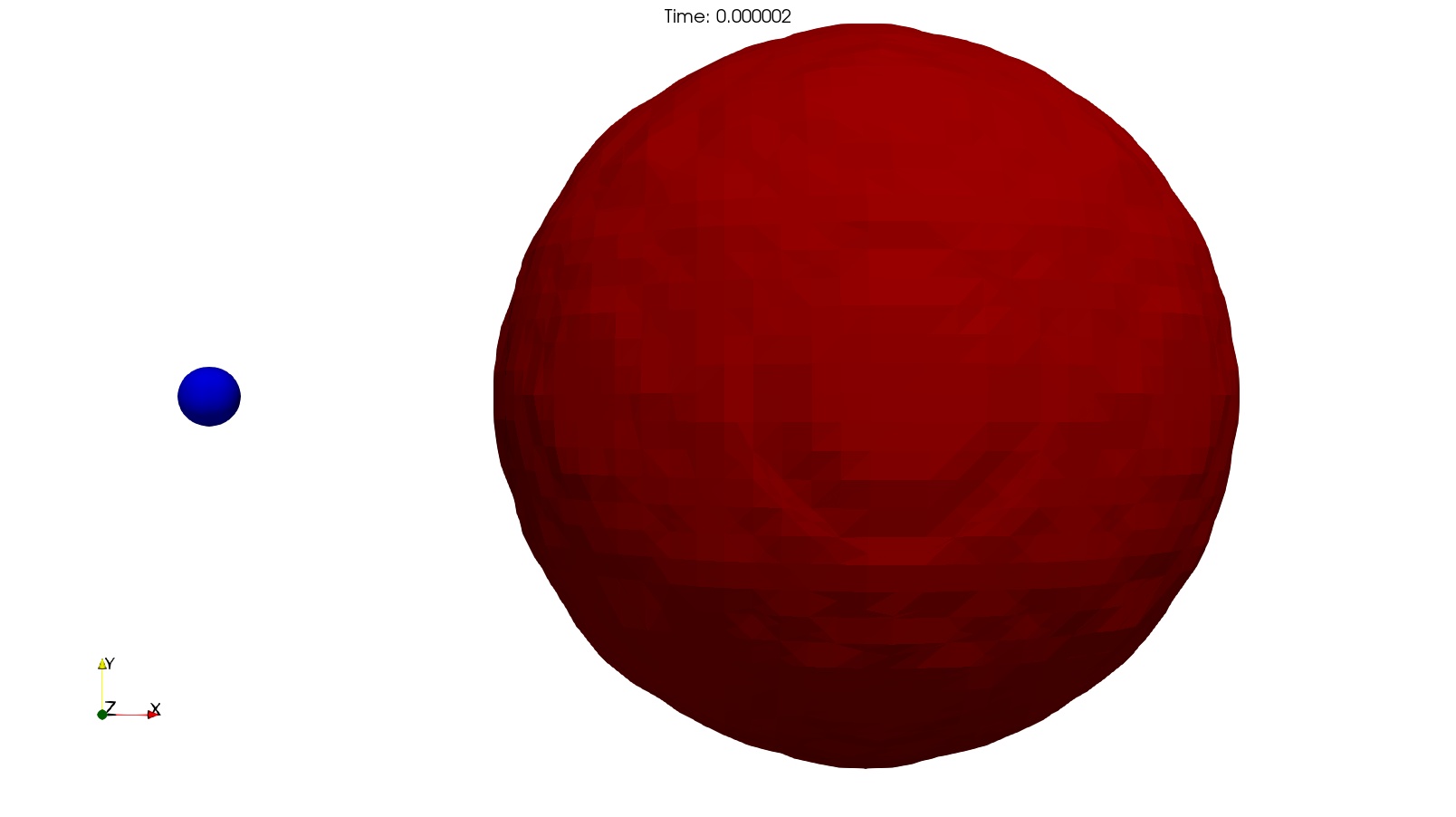}\\[-2pt]
        {\scriptsize (b) $t=0.000052~\mathrm{s}$}
    \end{minipage}

    \vspace{0.5em}

    \begin{minipage}[t]{0.4\linewidth}
        \centering
        \includegraphics[width=\linewidth]{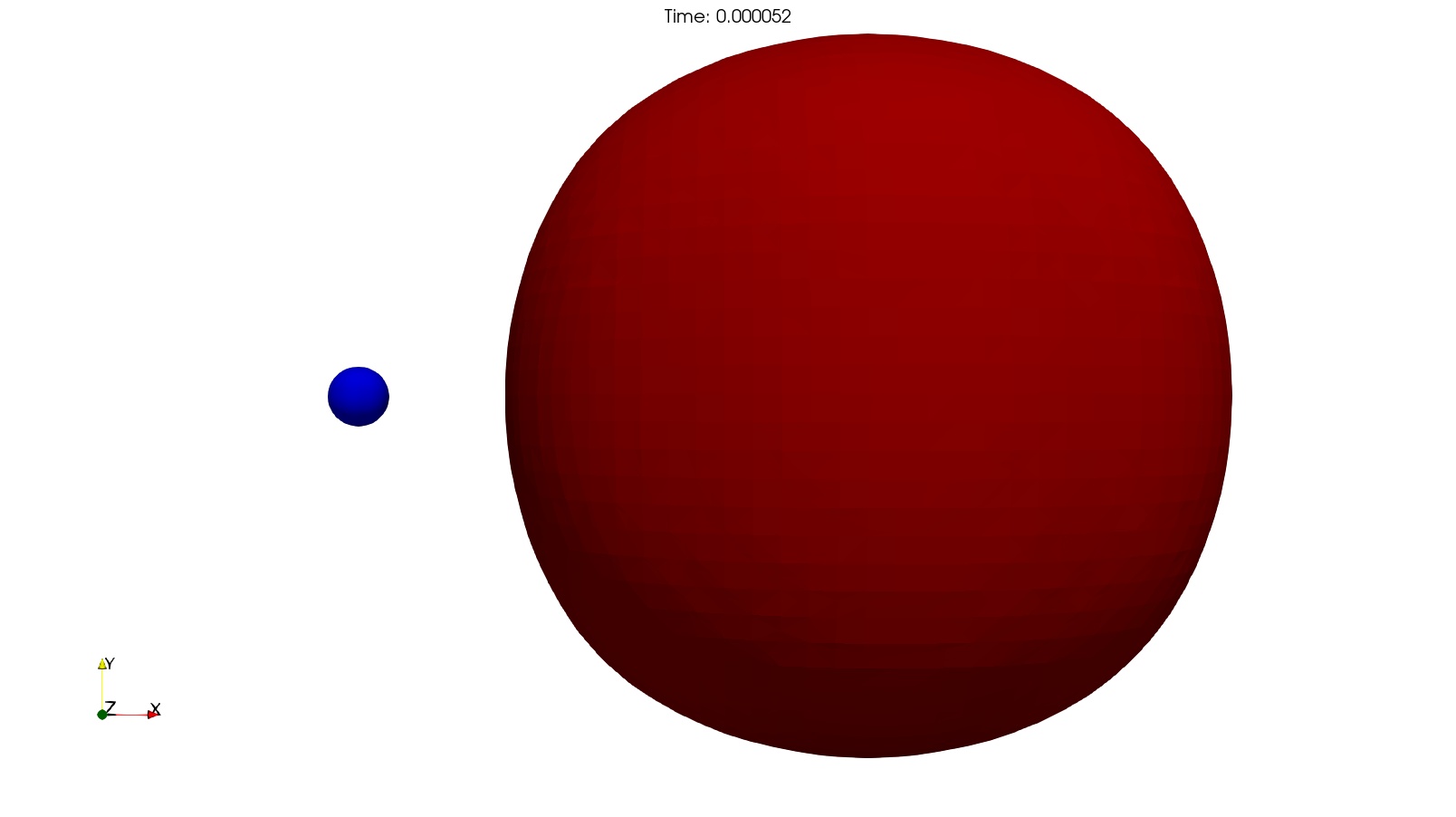}\\[-2pt]
        {\scriptsize (c) $t=0.000085~\mathrm{s}$}
    \end{minipage}\hfill
    \begin{minipage}[t]{0.4\linewidth}
        \centering
        \includegraphics[width=\linewidth]{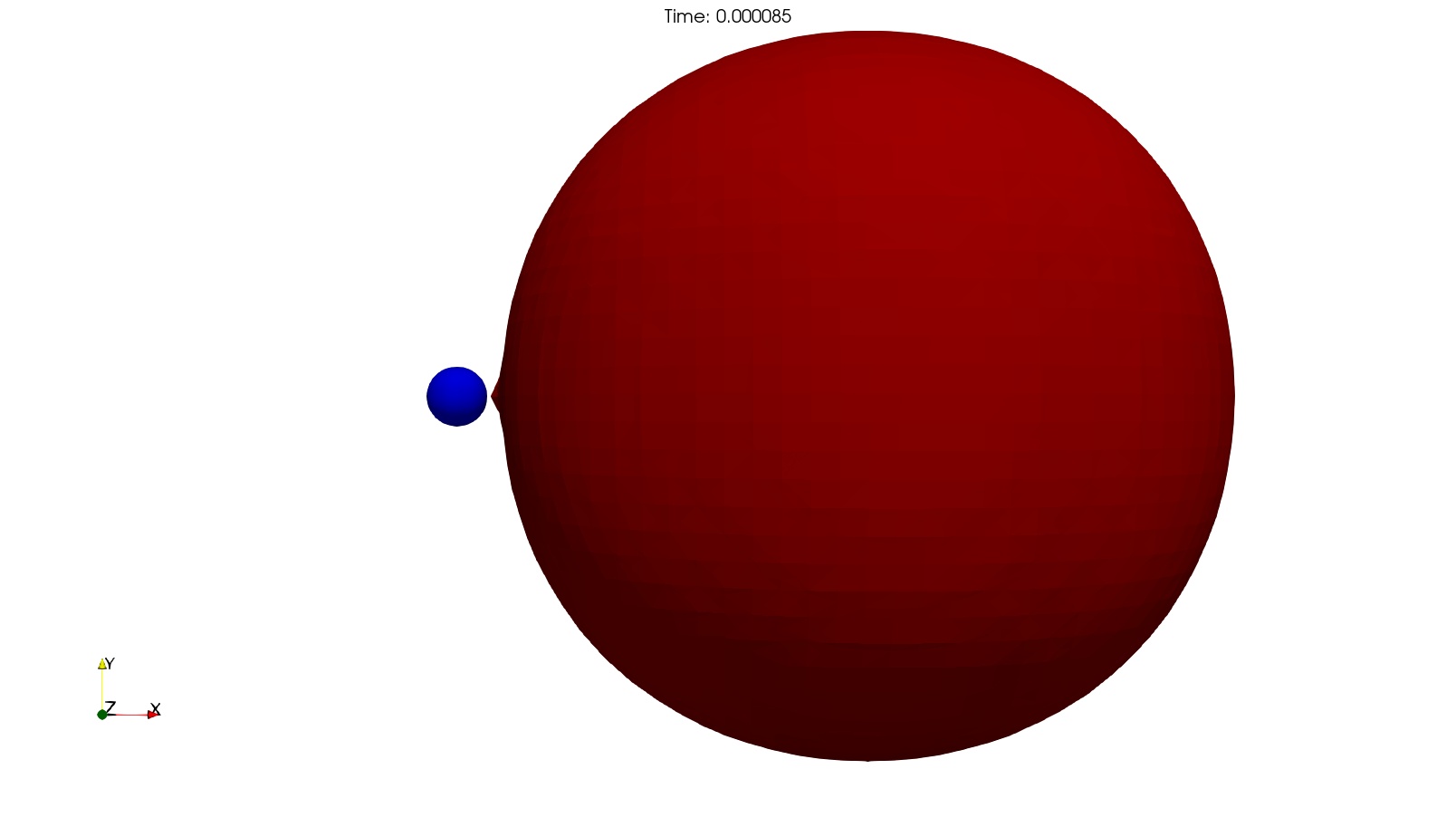}\\[-2pt]
        {\scriptsize (d) $t=0.000086~\mathrm{s}$}
    \end{minipage}

    \vspace{0.5em}

    \begin{minipage}[t]{0.4\linewidth}
        \centering
        \includegraphics[width=\linewidth]{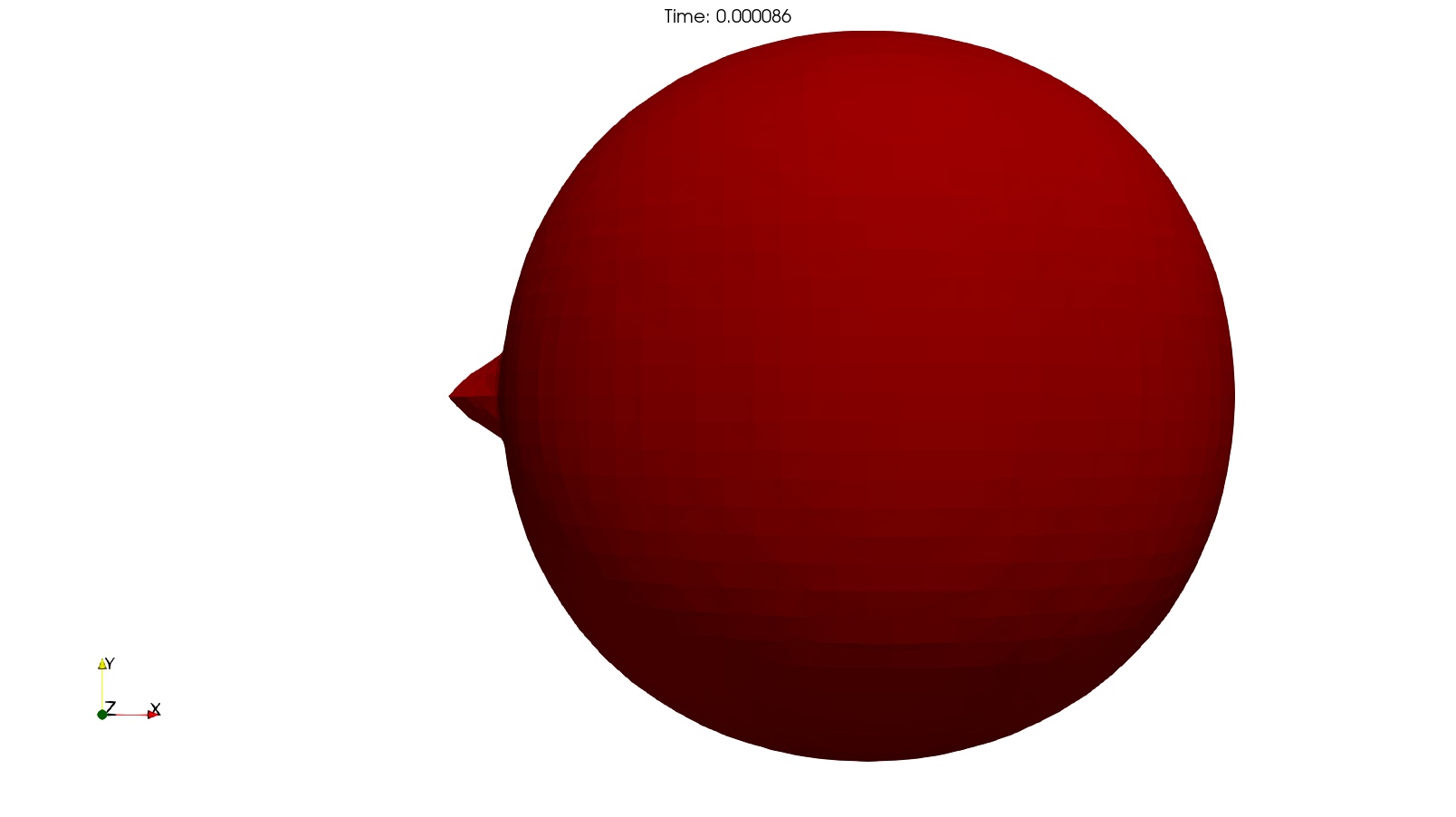}\\[-2pt]
        {\scriptsize (e) $t=0.000089~\mathrm{s}$}
    \end{minipage}\hfill
    \begin{minipage}[t]{0.4\linewidth}
        \centering
        \includegraphics[width=\linewidth]{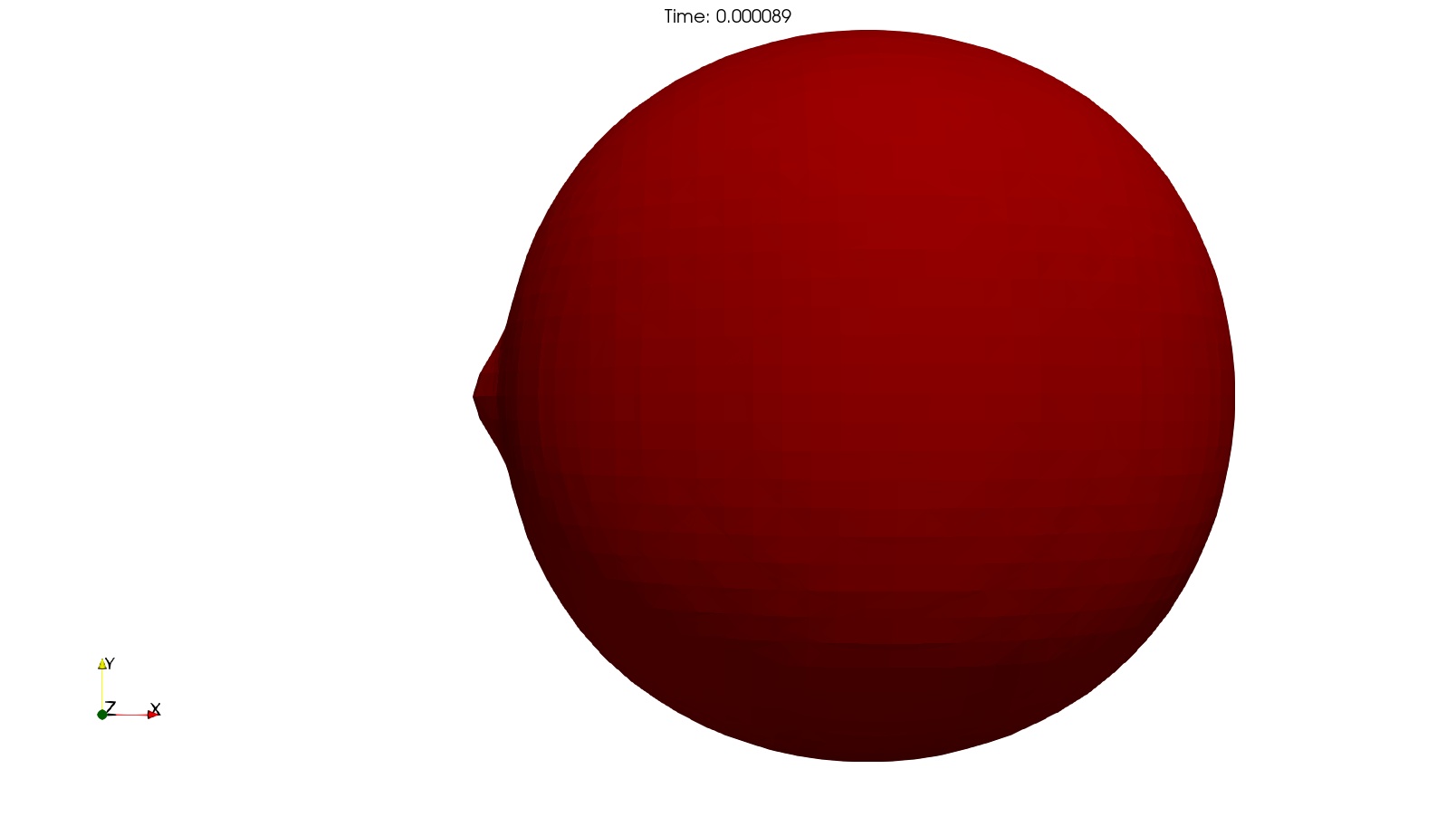}\\[-2pt]
        {\scriptsize (f) $t=0.000112~\mathrm{s}$}
    \end{minipage}
    \caption{Coalescence of a single Lagrangian bubble with a resolved Eulerian vapor volume.}
    \label{fig:l2e_coalescence_sequence}
\end{figure}

Figure~\ref{fig:l2e_coalescence_sequence} shows the complete sequence. The bubble approaches the Eulerian vapor volume while remaining unresolved at $t=0$, $5.2\times10^{-5}~\mathrm{s}$, and $8.5\times10^{-5}~\mathrm{s}$. Between $t=8.5\times10^{-5}~\mathrm{s}$ and $t=8.6\times10^{-5}~\mathrm{s}$, the unresolved bubble touches the surface of the Eulerian vapor volume, satisfies the interface-based transition criterion, and is converted to the Eulerian description. The resulting vapor protrusion then coalesces with the resolved vapor structure, producing a single merged Eulerian cavity by $t=1.12\times10^{-4}~\mathrm{s}$. This case therefore verifies that the model can detect contact with a resolved interface, remove the Lagrangian bubble at the correct instant, and transfer its vapor content consistently into the Eulerian field during coalescence.

%% file: sections/05_conservation.tex
\section{Conservation of mass and momentum}
\label{sec:conservation}

This section presents two conservative verification cases for the hybrid formulation. Section~\ref{subsec:mass_conservation_transition} verifies mass conservation under repeated Eulerian-Lagrangian switching. Section~\ref{subsec:mass_momentum_translation_transition} then verifies the simultaneous preservation of mass and momentum for a moving bubble, including the parallel execution setup used for the transition.

\subsection{Mass conservation verification for Eulerian-Lagrangian transitions}
\label{subsec:mass_conservation_transition}

\begin{figure}[H]
\centering
\includegraphics[width=0.4\linewidth]{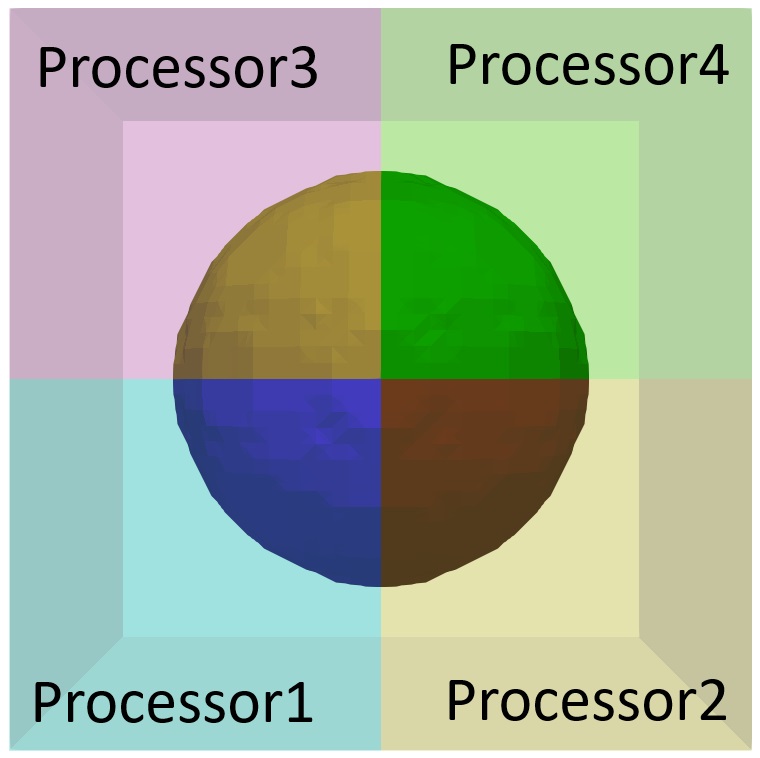}
\caption{Parallel domain-decomposition setup for the switching-schedule conservation case of Section~\ref{subsec:mass_conservation_transition}. The computational domain is decomposed into four processor subdomains, and the spherical vapor region at the domain center initially spans all four processors (Processors~1-4).}
\label{fig:kernel4_parallel_setup}
\end{figure}

A three-dimensional study is conducted to assess volume (mass) conservation under repeated bidirectional transitions between a resolved Eulerian cavity representation and an unresolved Lagrangian bubble representation. The test is designed to isolate the mapping and kernel-deposition operations used during conversion, rather than to reproduce a specific physical collapse or growth process. The computational domain, mesh, and fluid properties are identical to those described in Section~\ref{subsec:collapse}. A spherical Eulerian vapor region of radius $R_0=400~\mu\mathrm{m}$ is prescribed at the domain center. Time integration is performed with a constant time step $\Delta t=5\times10^{-8}~\mathrm{s}$, and the simulation is advanced to $t=5\times10^{-5}~\mathrm{s}$. This case is run in parallel with the domain decomposed into four processor subdomains, with the vapor region centered on the shared decomposition boundary so that it initially spans all four subdomains. Figure~\ref{fig:kernel4_parallel_setup} shows this domain-decomposition setup.

\begin{figure}[t]
\centering

\includegraphics[width=\linewidth]{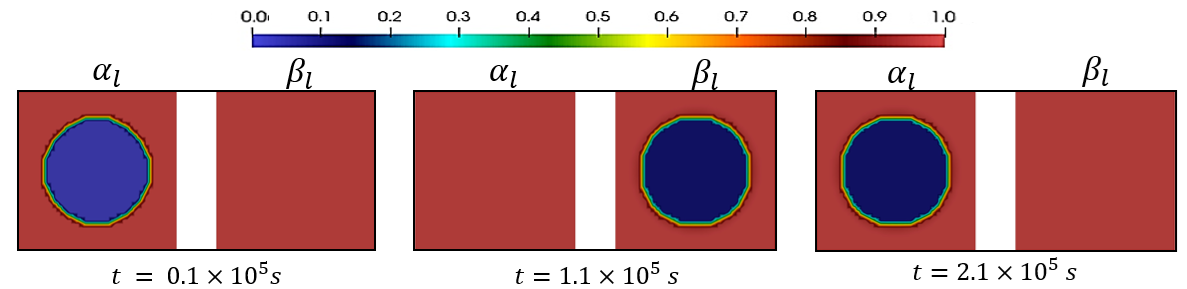}

\par\vspace{4mm}

\includegraphics[width=0.8\linewidth]{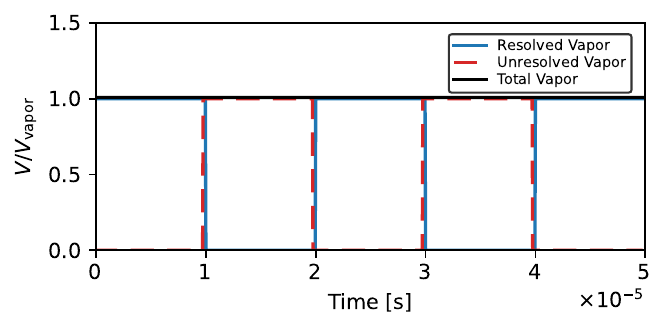}

\caption{Time evolution of the vapor volume components (normalized by the total vapor volume) during scheduled hybrid transitions, and a 2D profile of the resolved and unresolved liquid volume fractions at different times on the plane $z=0$.}
\label{fig:kernel4}
\end{figure}

To directly interrogate conservation across conversions, the hybrid algorithm is driven by a prescribed switching schedule that repeatedly enforces transitions between the Eulerian and Lagrangian representations. Specifically, Eulerian-to-Lagrangian conversions are imposed at $t=1\times10^{-5}~\mathrm{s}$ and $t=3\times10^{-5}~\mathrm{s}$, while Lagrangian-to-Eulerian conversions are imposed at $t=2\times10^{-5}~\mathrm{s}$ and $t=4\times10^{-5}~\mathrm{s}$.. The outcome of this scheduled switching test is summarized in Figure~\ref{fig:kernel4} through the time histories of the resolved vapor volume, the unresolved (Lagrangian) vapor volume, and their combined total, each normalized by the total vapor volume. The resolved and unresolved vapor volumes exchange in a stepwise manner consistent with the imposed schedule, while their sum, the total vapor volume, remains constant throughout the repeated conversions. In addition, Figure~\ref{fig:kernel4} visualizes the planar distribution of the Eulerian volume fraction on the mid-plane ($z=0$) at selected times, highlighting the kernel-based deposition footprint during the mapping operation. Collectively, these results indicate that the kernel-based mapping and the associated occupancy/coverage controls provide a conservative transfer of vapor content between the Lagrangian description and the Eulerian volume-fraction field.

\subsection{Mass and momentum conservation during moving-bubble transition}
\label{subsec:mass_momentum_translation_transition}

A second verification case is introduced to confirm that the hybrid transition preserves not only mass but also momentum when the bubble is in motion. In this test, a resolved Eulerian vapor bubble is initialized in a three-dimensional channel and convected by a uniform background liquid flow. The channel extends from $x=-1.0~\mathrm{mm}$ to $x=1.0~\mathrm{mm}$ with cross-sectional bounds $y,z\in[-0.3,0.3]~\mathrm{mm}$, and is discretized using $60\times20\times20$ cells. The flow field is initialized with a uniform streamwise velocity $\mathbf{U}=(2,0,0)~\mathrm{m/s}$. A spherical Eulerian bubble of radius $75~\mu\mathrm{m}$ is placed initially at $(-0.5~\mathrm{mm},0,0)$ and is transported in the positive $x$ direction.

\begin{figure}[H]
    \centering
    \includegraphics[width=0.6\linewidth]{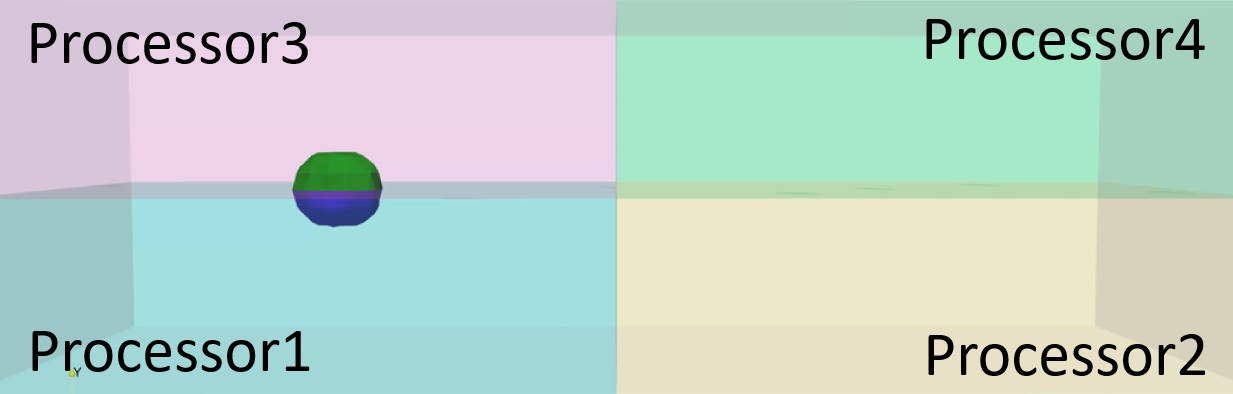}
    \caption{Parallel setup for the moving-bubble transition case. The computational domain is decomposed into four processor subdomains, and the bubble/transition support overlaps multiple processors during the conversion. The conservative redistribution is therefore assembled from local contributions on Processors~1-4.}
    \label{fig:moving_parallel_setup}
\end{figure}

This moving-bubble test is also executed in parallel. The domain is decomposed into four processor subdomains, and the bubble/transition support overlaps multiple ranks during the conversion process. Accordingly, the conservative redistribution and the associated state handoff are carried out from subdomain-local contributions while remaining globally synchronized. Figure~\ref{fig:moving_parallel_setup} shows the processor decomposition used for this parallel moving-bubble case.

\begin{figure}[H]
    \centering
    \includegraphics[width=0.8\linewidth]{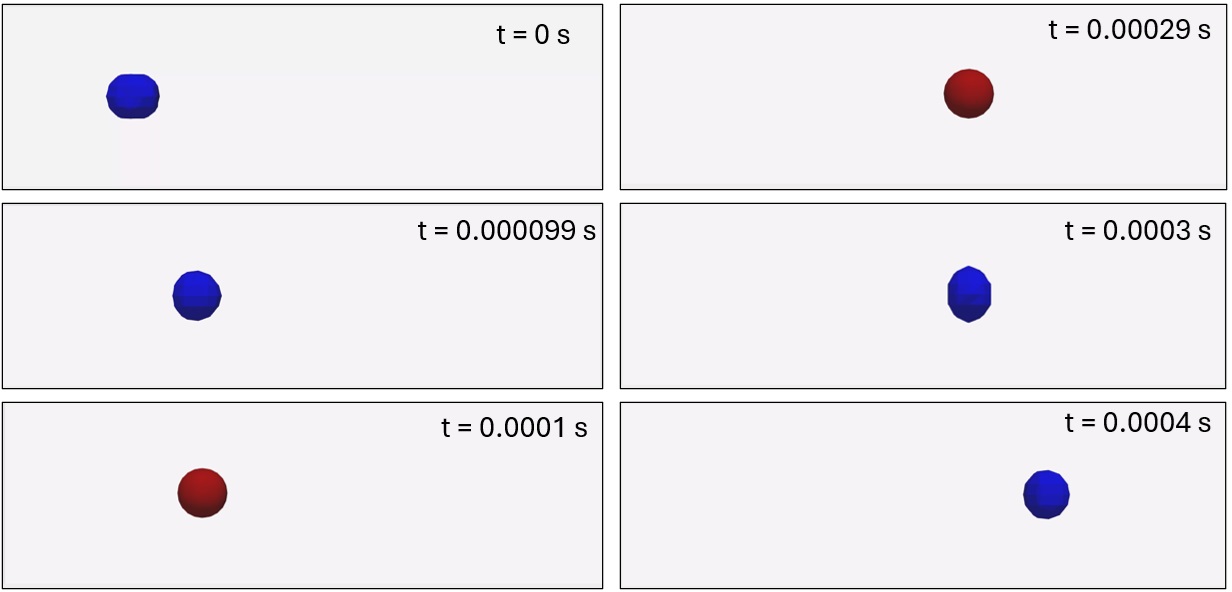}
    \caption{Representative snapshots for the moving-bubble transition verification. The bubble is initially resolved in the Eulerian field (blue), transitions to the Lagrangian representation at $t=1.0\times10^{-4}~\mathrm{s}$ (red), remains Lagrangian while convecting downstream, and transitions back to the Eulerian representation at $t=3.0\times10^{-4}~\mathrm{s}$ (blue).}
    \label{fig:moving_transition_snapshots}
\end{figure}

To isolate the representation change, the transition algorithm is imposed by a prescribed schedule. As shown in Figure~\ref{fig:moving_transition_snapshots}, the bubble begins as a resolved Eulerian structure (blue), is converted to a Lagrangian bubble at $t=1.0\times10^{-4}~\mathrm{s}$ (red), and is then converted back to an Eulerian structure at $t=3.0\times10^{-4}~\mathrm{s}$ (blue). The visual sequence confirms that the bubble remains continuous as a physical entity while only its numerical representation changes.

\begin{figure}[t!]
    \centering
    \includegraphics[width=0.6\linewidth]{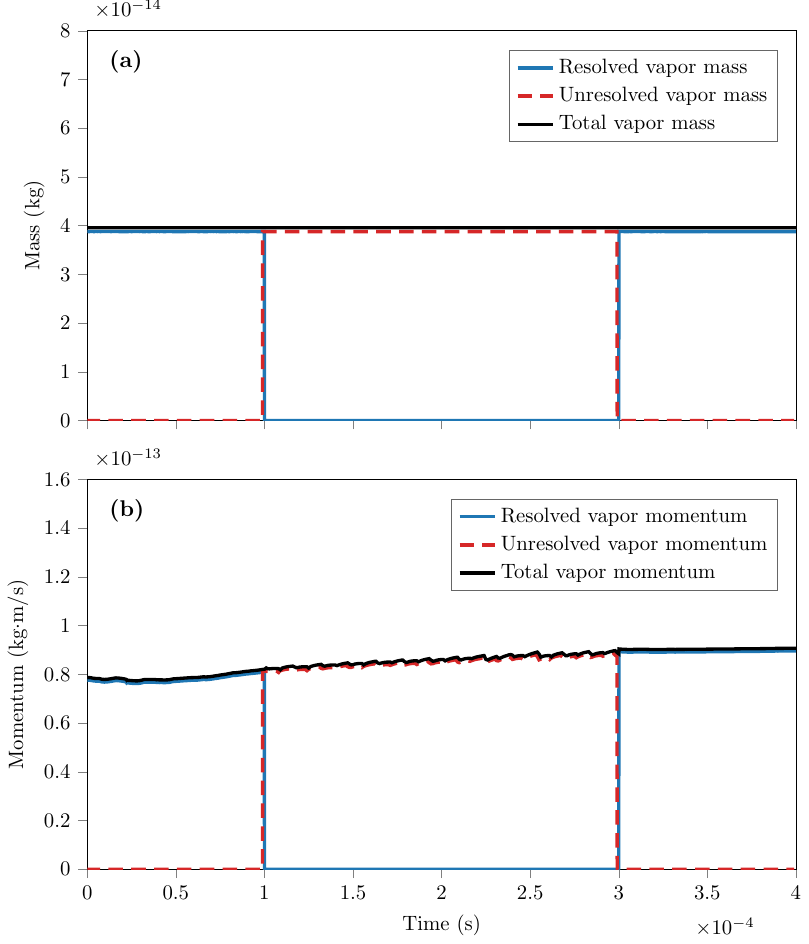}
    \caption{Mass and momentum conservation during the moving-bubble transition case. (a) Time history of the resolved (Eulerian) vapor mass, the unresolved (Lagrangian) vapor mass, and their total. (b) Time history of the corresponding summed momentum measure $\sum m\lvert \mathbf{U} \rvert$. The absence of jumps at the transition times shows that mass and momentum are transferred conservatively between the Eulerian and Lagrangian representations.}
    \label{fig:moving_transition_conservation}
\end{figure}

The quantitative conservation result is shown in Figure~\ref{fig:moving_transition_conservation}. Figure~\ref{fig:moving_transition_conservation}(a) reports the resolved (Eulerian) vapor mass and the unresolved (Lagrangian) vapor mass, together with their total. At the first transition, the resolved vapor mass decreases exactly as the unresolved vapor mass increases; at the reverse transition, the unresolved vapor mass is transferred back to the resolved representation in the same conservative manner. The mass remains constant throughout the simulation. Figure~\ref{fig:moving_transition_conservation}(b) shows the corresponding time history of the summed momentum measure $\sum m\lvert \mathbf{U} \rvert$ for the resolved vapor, the unresolved vapor, and their total. Again, no jump is observed during either transition. Instead, the momentum content is transferred cleanly between the two representations while the total remains continuous. This verification demonstrates that the hybrid transition preserves both mass and momentum during bubble motion, not only for a stationary conversion but also for a convecting bubble whose transported state must be handed consistently from Eulerian to Lagrangian form and back.

%% file: sections/06_conclusions.tex
\section{Conclusions}
\label{sec:conclusions}

This work develops a conservative hybrid Eulerian--Lagrangian framework in which
persistent identity tracking of resolved vapor structures is embedded directly within the
flow solver. The central contribution is not connected-component labeling or
representation transfer in isolation, but the integration of persistent structure history
into the numerical algorithm itself. In conventional hybrid approaches, connected
Eulerian structures may be identified independently at each time step and representation
changes are consequently based largely on instantaneous geometric information. In the
present framework, successive Eulerian observations are instead linked into persistent
physical tracks using prediction-based association, so that the history of a structure is
available at the instant a representation-transition decision is made.

This persistent history has several consequences for hybrid multiscale modeling. It
allows transition eligibility to depend on the recent evolution of a resolved structure
rather than on a single threshold crossing, thereby reducing spurious representation
chattering. It also provides the radius and radius-rate history required to initialize the
Lagrangian bubble-dynamics model when a resolved Eulerian cavity becomes unresolved.
At the same time, birth, death, breakup, coalescence, and representation-transition
events are recorded within a common lineage framework. Structure identity therefore
becomes an active part of the solver rather than a quantity reconstructed retrospectively
for post-processing.

Complementary to the persistent-tracking methodology, conservative
transfer operators are formulated to
preserve vapor mass and linear momentum while preventing simultaneous representation
of the same vapor volume in both descriptions. These two aspects of the framework
address distinct numerical requirements: persistent tracking preserves the temporal
identity and history needed for history-dependent solver decisions, whereas the
representation-transfer operators preserve the conserved quantities during a change of
mathematical description. Their integration provides a consistent computational pathway
for repeated transitions between resolved and unresolved vapor representations.

The tracking methodology is implemented in parallel, beginning with connected-component
labeling whose locally identified structures are reconciled across processor boundaries,
followed by deterministic prediction, gating, association, and event logging. The
verification cases considered here exercise the framework under fragmentation,
coalescence, close crossing, and repeated Eulerian--Lagrangian transitions. 
Together, these results show that persistent structure histories and conservative
representation transfer can be maintained within a parallel hybrid solver rather than
reconstructed or checked only after the simulation.

This distinction is important for large-scale cavitating-flow calculations. Persistent-ID
histories, transition records, and lineage graphs are useful as physical diagnostics only
if they reflect the evolution of the underlying structures rather than numerical labeling,
processor ownership, or traversal order. Embedding the tracker directly in the solver
also allows these histories to influence the evolution of the calculation itself, which
cannot be achieved by an offline tracking procedure. In this sense, persistent identity is
not merely a post-processing convenience but an additional state of the hybrid numerical
framework.


The scope of the present study is deliberately limited to isolating and verifying the
tracking and representation-transfer machinery through canonical configurations chosen
to expose specific numerical failure modes. The next step is to apply the framework to
sheet-to-cloud shedding, tip-vortex cavitation, and other realistic multiscale flows in
which large numbers of structures repeatedly fragment, coalesce, and transition between
representations. Such problems will provide a substantially more demanding test of the
persistent-tracking and conservative-transfer algorithms and will determine how the
additional structure-history information can be exploited in predictive cavitation
simulations.

%% file: sections/appendix_original_equations.tex
\section{Hybrid Model's Consistency with other Models}\label{appA}

Setting $\beta_{\mathrm{l}}=1$ implies that the entire domain is described by the homogeneous mixture model and that the unresolved vapor phase is inactive. To prove that, by definition (see Eqs.~\eqref{eqn:mbubble2} and \eqref{eqn:force2}), all Lagrangian source terms vanish:
\begin{equation}
  \bigl[\dot m_{\mathrm{b}}\bigr]=0, 
  \qquad 
  \Bigl[\tfrac{\mathrm{D}\beta_{\mathrm{l}}}{\mathrm{D}t}\Bigr]=0,
  \qquad
  \Bigl[\tfrac{F}{V}\Bigr]=0.
\end{equation}

\noindent Substituting these into Eqs.~\eqref{eqn:hybrid-non-divergence-free2} and \eqref{eqn:hybrid-mom2} with $\beta_{\mathrm{l}}=1$ yields
\begin{equation}
  \frac{\partial u_i}{\partial x_i} 
  \;=\;
  \left( \frac{1}{\rho_{\mathrm{l}}} - \frac{1}{\rho_{\mathrm{v}}} \right) \dot{m},
  \label{eqn:hybrid-non-divergence-free4}
\end{equation}

\begin{equation}
  \left[ 
    \frac{\partial \bigl( \alpha_{\mathrm{l}} \rho_{\mathrm{l}} + (1-\alpha_{\mathrm{l}})\rho_{\mathrm{v}} \bigr) u_i}{\partial t} 
    + \frac{\partial \bigl( \alpha_{\mathrm{l}} \rho_{\mathrm{l}} + (1-\alpha_{\mathrm{l}})\rho_{\mathrm{v}} \bigr) u_i u_j}{\partial x_j} 
  \right] =  
  - \frac{\partial p}{\partial x_i} 
  + \frac{\partial \tau_{ij}}{\partial x_j} 
  + \rho_{\mathrm{m}} g_i,
  \label{eqn:hybrid-mom3a}
\end{equation}

\noindent where $\rho_{\mathrm{m}} = \alpha_{\mathrm{l}} \rho_{\mathrm{l}} + (1-\alpha_{\mathrm{l}})\rho_{\mathrm{v}}$ is the mixture density. Hence, these equations reduced to the standard homogeneous mixture model form \cite{GNANASKANDAN20191743, Vallier2013}.
Alternatively, following Maeda and Colonius \cite{MAEDA2018994}, one may consider only the resolved liquid and unresolved vapor. For $\alpha_{\mathrm{l}}=1$, the continuity equation \eqref{eqn:hybrid-non-divergence-free} and momentum equation \eqref{eqn:hybrid-mom} reduce, respectively, to

\begin{align}
  &\cancel{\left[\frac{\partial \rho_{\mathrm{l}}}{\partial t} + \frac{\partial \left( \rho_{\mathrm{l}} u_i \right)}{\partial x_i} \right]}
  + \left[\frac{\partial \left(\beta_{\mathrm{l}} \rho_{\mathrm{l}} + (1-\beta_{\mathrm{l}}) \rho_{\mathrm{v}} \right)}{\partial t} 
  + \frac{\partial \left(\beta_{\mathrm{l}} \rho_{\mathrm{l}} u_i + (1-\beta_{\mathrm{l}}) \rho_{\mathrm{v}} u_{Li} \right)}{\partial x_i} \right] \notag \\
  &\qquad - \cancel{\left[
  \frac{\partial \rho_{\mathrm{l}}}{\partial t}
  + \frac{\partial (\rho_{\mathrm{l}} u_i)}{\partial x_i}
  \right]} = 0,
  \label{eqn:cont2}
\end{align}
\begin{align}
  &\cancel{\left[ 
    \frac{\partial  \rho_{\mathrm{l}}  u_i}{\partial t} 
    + \frac{\partial  \rho_{\mathrm{l}}  u_i u_j}{\partial x_j} 
  \right]}
  - \left[ 
    \frac{\partial \left( (\cancel{1}-\beta_{\mathrm{l}}) \rho_{\mathrm{l}} u_i \right)}{\partial t} 
    + \frac{\partial \left( (\cancel{1}-\beta_{\mathrm{l}}) \rho_{\mathrm{l}} u_i u_j \right)}{\partial x_j} 
  \right] \notag \\
  &\qquad= - \left[ 
    \frac{\partial \left( (1-\beta_{\mathrm{l}}) \rho_{\mathrm{v}} u_{Li} \right)}{\partial t} 
    + \frac{\partial \left( (1-\beta_{\mathrm{l}}) \rho_{\mathrm{v}} u_{Li} u_j \right)}{\partial x_j} 
  \right] - \frac{\partial p}{\partial x_i} 
  + \frac{\partial \tau_{ij}}{\partial x_j} 
  + \rho_{\mathrm{m}} g_i.
  \label{eqn:hybrid-mom2b}
\end{align}

\noindent To further evaluate, based on the \cite{MAEDA2018994} assumptions, let $(1-\beta_{\mathrm{l}})=\beta_{\mathrm{v}}$, neglect gravity ($g_i=0$), and assume $\rho_{\mathrm{l}}\gg \rho_{\mathrm{v}}$. Then, Eqn.~\eqref{eqn:cont2} and Eqn.~\eqref{eqn:hybrid-mom2b} further simplify to

\begin{align}
  &\left[
    \frac{\partial \Bigl((1-\beta_{\mathrm{v}})\rho_{\mathrm{l}} + \cancel{\beta_{\mathrm{v}} \rho_{\mathrm{v}}}\Bigr)}{\partial t}
    + \frac{\partial \Bigl((1-\beta_{\mathrm{v}})\rho_{\mathrm{l}} u_i + \cancel{\beta_{\mathrm{v}} \rho_{\mathrm{v}}\, u_{Li}}\Bigr)}{\partial x_i}
  \right] \notag \\
  &\qquad = 
  (1-\beta_{\mathrm{v}})\left[\frac{\partial \rho_{\mathrm{l}}}{\partial t} + \frac{\partial (\rho_{\mathrm{l}} u_i)}{\partial x_i} \right]
  - \rho_{\mathrm{l}} \left[\frac{\partial \beta_{\mathrm{v}}}{\partial t} + u_i\frac{\partial \beta_{\mathrm{v}}}{\partial x_i}\right]
  \;=\; 0,
  \label{eqn:cont3}
\end{align}
\begin{align}
  &\left[
    \frac{\partial \bigl( (1-\beta_{\mathrm{v}})\rho_{\mathrm{l}} u_i \bigr)}{\partial t}
    + \frac{\partial \bigl( (1-\beta_{\mathrm{v}})\rho_{\mathrm{l}} u_i u_j \bigr)}{\partial x_j}
  \right] \notag \\
  &\qquad = (1-\beta_{\mathrm{v}})\left[
    \frac{\partial \left( \rho_{\mathrm{l}} u_i \right)}{\partial t}
    + \frac{\partial \left( \rho_{\mathrm{l}} u_i u_j \right)}{\partial x_j}
  \right] 
  + \rho_{\mathrm{l}} u_i \left[
    \frac{\partial \left( 1-\beta_{\mathrm{v}} \right)}{\partial t}
    + u_j \frac{\partial  \left( 1-\beta_{\mathrm{v}}  \right)}{\partial x_j}
  \right] \notag \\
  &\qquad = - \left[
    \frac{\partial \left( \cancel{\beta_{\mathrm{v}}\rho_{\mathrm{v}} u_{Li}} \right)}{\partial t}
    + \frac{\partial \left( \cancel{\beta_{\mathrm{v}}\rho_{\mathrm{v}} u_{Li} u_j} \right)}{\partial x_j}
  \right]  
  - \left[\frac{\partial p}{\partial x_i}
  -  \frac{\partial \tau_{ij}}{\partial x_j}\right].
  \label{eqn:hybrid-mom3b}
\end{align}

\noindent Collecting Eulerian terms on the left side and unresolved vapor contributions on the right gives
\begin{equation}
  \left[\frac{\partial \rho_{\mathrm{l}}}{\partial t} + \frac{\partial (\rho_{\mathrm{l}} u_i)}{\partial x_i} \right]
  \;=\; \frac{\rho_{\mathrm{l}}}{\,1-\beta_{\mathrm{v}}\,}
  \left[\frac{\partial \beta_{\mathrm{v}}}{\partial t} + u_i\frac{\partial \beta_{\mathrm{v}}}{\partial x_i}\right],
  \label{eqn:cont4}
\end{equation}
\begin{equation}
  \left[
    \frac{\partial \left( \rho_{\mathrm{l}} u_i \right)}{\partial t}
    + \frac{\partial \left( \rho_{\mathrm{l}} u_i u_j \right)}{\partial x_j}
  \right]
  \;=\;
  \frac{ \rho_{\mathrm{l}} u_i}{\,1-\beta_{\mathrm{v}}\,}
  \left[
    \frac{\partial \beta_{\mathrm{v}} }{\partial t}
    + u_j \frac{\partial  \beta_{\mathrm{v}} }{\partial x_j}
  \right]
  - \frac{1}{\,1-\beta_{\mathrm{v}}\,}
  \left[\frac{\partial p}{\partial x_i}
  -  \frac{\partial \tau_{ij}}{\partial x_j}\right].
  \label{eqn:hybrid-mom4a}
\end{equation}
or, equivalently,
\begin{equation}
\begin{aligned}
  \left[
    \frac{\partial \left( \rho_{\mathrm{l}} u_i \right)}{\partial t}
    + \frac{\partial \left( \rho_{\mathrm{l}} u_i u_j \right)}{\partial x_j}
    + \frac{\partial p}{\partial x_i}
    -  \frac{\partial \tau_{ij}}{\partial x_j}
  \right]
  \;=\;&
  \frac{ \rho_{\mathrm{l}} u_i}{\,1-\beta_{\mathrm{v}}\,}
  \left[
    \frac{\partial \beta_{\mathrm{v}} }{\partial t}
    + u_j \frac{\partial  \beta_{\mathrm{v}} }{\partial x_j}
  \right] \\
  &- \frac{\beta_{\mathrm{v}}}{\,1-\beta_{\mathrm{v}}\,}
  \left[\frac{\partial p}{\partial x_i}
  -  \frac{\partial \tau_{ij}}{\partial x_j}\right],
\end{aligned}
  \label{eqn:hybrid-mom4b}
\end{equation}
which match Eqs.~(9)-(10) of \cite{MAEDA2018994} under the stated assumptions.

%% file: sections/appendix_vortex_verification.tex
\section{Verification of the Lagrangian Bubble Kinematics}\label{appVortex}

To assess the robustness of the bubble-tracking algorithm and the fidelity of the numerical framework in predicting bubble advection in vortical flows, we perform a validation study based on the classical reference of Oweis et al. \cite{Oweis2005} and the numerical investigations of Qin et al. \cite{QIN2025105142} and Wang et al. \cite{WANG2025105412}. In these studies, bubble dynamics are examined in a two-dimensional Gaussian vortex, where the carrier-phase velocity and pressure fields are prescribed analytically according to the Gaussian vortex formulation:

\begin{equation}
u_\theta(r) = \frac{\Gamma}{2\pi r} \left( 1 - e^{-\eta_1 (r/r_c)^2} \right),
\label{eq:vortex_velocity}
\end{equation}
\begin{equation}
p(r) - p_\infty = - \int_r^{\infty} \rho_l \frac{u_\theta^2(r')}{r'}  dr',
\label{eq:vortex_pressure}
\end{equation}

\begin{figure}[h]
\centering
\includegraphics[width=0.5\linewidth]{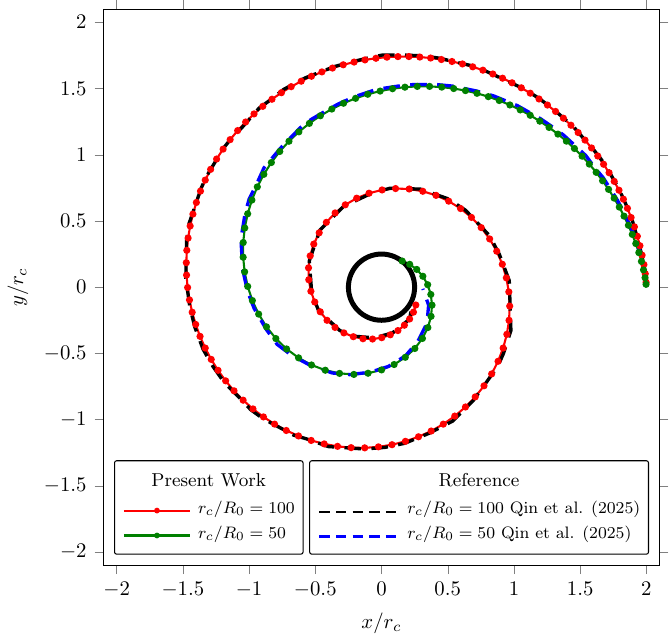}
\caption{Comparison between simulated bubble trajectories (present work) and experimental reference trajectories (Qin et al. \cite{QIN2025105142}) in a Gaussian vortex.}
\label{fig:vortexmoving}
\end{figure}

\noindent Here, $u_\theta(r)$ denotes the tangential velocity, $\Gamma = 0.29~\mathrm{m^2/s}$ is the circulation strength, $r_c = 5.6~\mathrm{mm}$ is the vortex core radius, $\eta_1 = 1.255$ is a shape parameter, and $p_\infty = 10^{5}~\mathrm{Pa}$ is the ambient pressure. The velocity and pressure fields are initialized at $t=0$ using these expressions, thereby ensuring a divergence-free velocity field and a force-balanced initial state. The computational domain spans $[-10\,r_c,\,10\,r_c]$ in both the $x$- and $y$-directions (two-dimensional configuration) and is discretized using a uniform $600 \times 600$ Cartesian grid. The bubble is initialized at $(x,y)=(-2\,r_c,0)$. Two initial bubble radii, $R_0=56~\mu\mathrm{m}$ $(r_c/R_0=100)$ and $R_0=112~\mu\mathrm{m}$ $(r_c/R_0=50)$, are considered to quantify the influence of bubble size on vortex entrainment. All hydrodynamic force contributions described above are included. The carrier-phase properties correspond to a liquid-gas system with surface tension $\sigma=0.072~\mathrm{N/m}$, liquid kinematic viscosity $\nu_l=1.0\times 10^{-6}~\mathrm{m^2/s}$ and density $\rho_l=1000~\mathrm{kg/m^3}$, and gas kinematic viscosity $\nu_g=1.0\times 10^{-5}~\mathrm{m^2/s}$ and density $\rho_g=1~\mathrm{kg/m^3}$. As illustrated in Figure~\ref{fig:vortexmoving}, the results demonstrate that the implemented Lagrangian framework accurately predicts the bubble advection dynamics and trajectory shape under vortex-induced forces. This agreement confirms that the coupling between the vortex velocity field and bubble kinematics is modeled with sufficient fidelity to reproduce the observed behavior.

A complementary metric for quantifying the bubble kinematics is the capture time, defined as the physical time required for a bubble released at an initial radial location $r_0$ to be transported by the vortex-induced flow to a prescribed capture radius $r_{\mathrm{cap}}$ in the vicinity of the vortex core. Consistent with \cite{Oweis2005,QIN2025105142}, the capture radius is set to $r_{\mathrm{cap}} = 0.25\,r_c$. In vortex-bubble interaction studies (including \cite{Oweis2005,QIN2025105142}), the corresponding non-dimensional capture time $t^*$ is typically formed using the peak tangential velocity $u_c$ of the Gaussian vortex as the characteristic velocity scale. The definition is given by

\begin{equation}
t^* = \frac{t_{\mathrm{cap}} u_c}{r_c},
\label{eq:capturetine}
\end{equation}

\noindent where $t_{\mathrm{cap}}$ is the physical capture time, $u_c$ is the maximum tangential velocity of the vortex, and $r_c$ is the vortex core radius. For the vortex, the peak azimuthal velocity occurs at $r = r_c$ and is expressed analytically as

\begin{equation}
u_c = \frac{\Gamma}{2\pi r_c} \left( 1 - e^{-\eta_1} \right),
\label{eq:maxvelcapturetine}
\end{equation}

\noindent Substituting this expression into the definition of $t^*$ gives

\begin{equation}
t^* = \frac{t_{\mathrm{cap}} \Gamma}{2\pi r_c^2} \left( 1 - e^{-\eta_1} \right).
\label{eq:finalmaxvelcapturetine}
\end{equation}

\begin{figure}[h]
\centering
\includegraphics[width=0.7\linewidth]{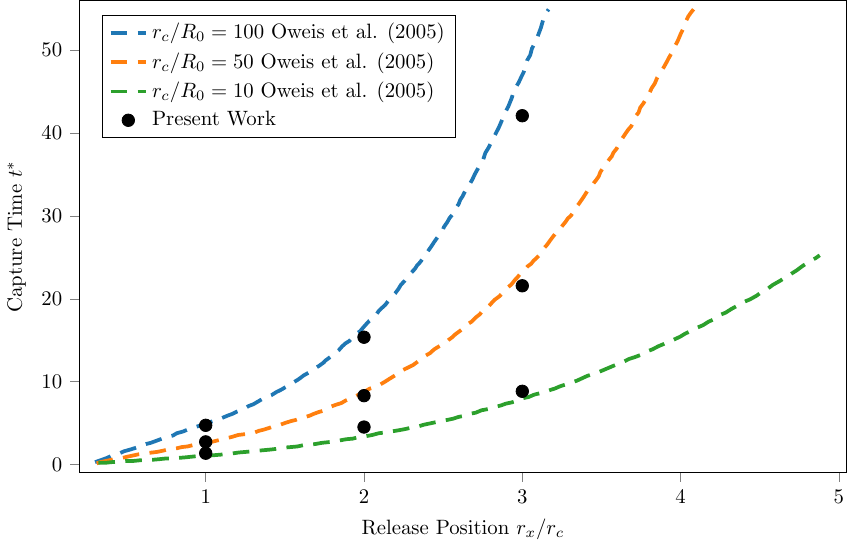}
\caption{Comparison of non-dimensional bubble capture time $t^*$ as a function of the initial release position $r_x/r_c$ for different ratios of $r_c/R_0$. The dashed curves correspond to the reference data of Oweis et al. \cite{Oweis2005} for $r_c/R_0 = 100$, $50$, and $10$. The solid black markers represent the present numerical results obtained from the Lagrangian tracking simulations.}
\label{fig:vortexmoving2}
\end{figure}

\noindent Figure~\ref{fig:vortexmoving2} shows the non-dimensional bubble capture time $t^*$ as a function of the normalized initial release radius $r_x/r_c$ for different size ratios $r_c/R_0$. The dashed curves correspond to the reference results of Oweis et al. \cite{Oweis2005} for three bubble-size ratios ($r_c/R_0 = 100$, $50$, and $10$), demonstrating that larger bubbles (i.e., smaller $r_c/R_0$) are captured more rapidly by the vortex. The black circular markers denote the present results and exhibit close agreement with the reported trends over the examined range of release positions. This agreement supports the correctness of the implemented Lagrangian tracking algorithm and its capability to accurately reproduce vortex-induced bubble entrainment dynamics.

%% file: sections/appendix_rp_derivation.tex
\section{Derivation of the localized Rayleigh-Plesset Equation}\label{appB}

The Rayleigh-Plesset equation couples the bubble-wall motion to the far-field liquid pressure, $p_\infty(t)$, under the assumptions of incompressible, irrotational, spherically symmetric flow. In complex, spatially varying environments (e.g.\ shear and vortical pressure pockets), the use of a single uniform $p_\infty$ may under-represent near-field hydrodynamics. We therefore consider a localized Rayleigh-Plesset equation model in which the forcing is the resolved pressure on a finite spherical shell of radius $r=kR(t)$ with $k>1$. This choice embeds local hydrodynamic effects while preserving the analytical tractability of the RP structure. The derivation assumes an inviscid, irrotational, and spherically symmetric liquid motion, where $ u = \nabla \phi$ and $u(r,t) = \phi_r$. The liquid is considered incompressible with constant density $\rho$ for the core derivation. The bubble has a radius $R(t)$ with wall velocity $\dot R$ and acceleration $\ddot R$. At the bubble wall, the normal-stress balance is imposed as

\begin{equation}
p(R)= p_b(t) - \frac{2\sigma}{R} - \frac{4\mu\dot R}{R},
\label{eq:wall-bc}
\end{equation}

\noindent here, $p_b$ is the internal bubble pressure, $\sigma$ is the surface tension, and $\mu$ is the liquid viscosity. Under incompressibility and spherical symmetry the flux through any sphere is constant

\begin{equation}
  4\pi r^2 u(r,t) \;=\; 4\pi R^2 \dot R
  \quad\Rightarrow\quad
  u(r,t)=\frac{R^2\dot R}{r^2}.
  \label{eq:radial-u}
\end{equation}

\noindent Since $u=\phi_r$, integration in $r$ gives

\begin{align}
  \phi(r,t) &= -\,\frac{R^2\dot R}{r}, \label{eq:phi}\\[0.5em]
  \phi_t(r,t) &= -\frac{1}{r}\,\frac{d}{dt}\!\big(R^2\dot R\big)
  = -\frac{1}{r}\,\Big(2R\dot R^2 + R^2\ddot R\Big).
  \label{eq:phit}
\end{align}

\noindent The unsteady Bernoulli relation for incompressible, irrotational flow reads

\begin{equation}
  \phi_t + \tfrac12|\nabla\phi|^2 + \frac{p}{\rho} = f(t),
  \label{eq:bernoulli}
\end{equation}

\noindent here, $f(t)$ is a gauge (time-dependent reference). Subtracting \eqref{eq:bernoulli} between the bubble wall $r=R(t)$ and a concentric shell $r=kR(t)$ eliminates $f(t)$

\begin{equation}
  \underbrace{\big[\phi_t(kR)-\phi_t(R)\big]}_{\mathcal{A}}
  \;+\;
  \underbrace{\tfrac12\big[u^2(kR)-u^2(R)\big]}_{\mathcal{B}}
  \;+\;
  \frac{p(kR)-p(R)}{\rho}
  \;=\; 0.
  \label{eq:bern-diff}
\end{equation}

\noindent Using \eqref{eq:radial-u}-\eqref{eq:phit}, we arrive at

\begin{align}
  \mathcal{A} &= \Big(1-\frac{1}{k}\Big)\big(2\dot R^2 + R\ddot R\big),\\
  \mathcal{B} &= -\frac{1}{2}\,\dot R^2\Big(1-\frac{1}{k^4}\Big).
\end{align}

\noindent Rearranging yields the localized Rayleigh-Plesset core

\begin{equation}
  \Big(1-\frac{1}{k}\Big)R\ddot R
  \;+\;
  \Big(\frac{3}{2}-\frac{2}{k}+\frac{1}{2k^4}\Big)\dot R^{\,2}
  \;=\; \frac{p(R)-p(kR)}{\rho}.
  \label{eq:finite-core}
\end{equation}

\noindent 
Inserting the wall boundary condition \eqref{eq:wall-bc} into the governing formulation yields the localized Rayleigh-Plesset equation

\begin{equation}
  \Big(1 - \frac{1}{k}\Big) R \ddot{R}
  \;+\;
  \Big(\frac{3}{2} - \frac{2}{k} + \frac{1}{2k^4}\Big) \dot{R}^{\,2}
  \;=\; \frac{1}{\rho}\left(
      p_b - \frac{2\sigma}{R} - \frac{4\mu\,\dot R}{R}
      - p(kR)
  \right),
  \label{eq:finite-RP-app}
\end{equation}

\noindent where \( k \) denotes the non-dimensional distance of the wall (or the location where the pressure is evaluated) relative to the bubble radius. This formulation is consistent with the incompressible limit of the multiscale model presented by Madabhushi and Mahesh \cite{Madabhushi_Mahesh_2023} when the speed of sound tends to infinity.

As a sanity check, Eqn.~\eqref{eq:finite-RP-app} recovers the classical Rayleigh-Plesset equation in the far-field limit. Specifically, as \( k \to \infty \), the coefficients

\begin{equation}
  \Big(1 - \frac{1}{k}\Big) \to 1, 
  \qquad
  \Big(\frac{3}{2} - \frac{2}{k} + \frac{1}{2k^4}\Big) \to \frac{3}{2},
  \label{eq:finite-RP-checked}
\end{equation}

\noindent and the pressure term \( p(kR) \) approaches the ambient far-field pressure \( p_\infty \). Therefore, Eqn.~\eqref{eq:finite-RP-app} reduces to

\begin{equation}
  R \ddot{R} + \frac{3}{2} \dot{R}^2
  = \frac{1}{\rho}\left(
      p_b - \frac{2\sigma}{R} - \frac{4\mu\,\dot R}{R} - p_\infty
    \right),
  \label{eq:RP-classical}
\end{equation}

\noindent which is the standard incompressible Rayleigh-Plesset equation. This confirms that the localized formulation is asymptotically consistent with the classical far-field model as \( k \) becomes large. For a finite wall distance \( k = 2 \), Eqn.~\eqref{eq:finite-RP-app} simplifies to

\begin{equation}
  \frac{1}{2} R \ddot{R} + \frac{17}{32} \dot{R}^2
  = \frac{1}{\rho}\left(
      p_b - \frac{2\sigma}{R} - \frac{4\mu\,\dot R}{R} - p(2R)
    \right),
  \label{eq:RP-wall}
\end{equation}

\noindent which corresponds to the localized Rayleigh-Plesset equation introduced by Ghahramani et al. \cite{GHAHRAMANI2019339}. This form captures the influence of a nearby boundary or localized pressure evaluation region, providing a more accurate representation of near-wall bubble dynamics than the classical far-field approximation. Eqn.~\eqref{eq:finite-RP-app} preserves the simplicity of a single ODE for $R(t)$ while allowing the ambient forcing to be taken from the resolved near field. It reduces artificial sensitivity to domain-size choices for $p_\infty$, better captures interaction with nearby structures or bubbles, and suppresses spurious numerical pulses associated with pointwise sampling.

%% file: sections/appendix_rp_verification.tex
\section[Rayleigh-Plesset verification as k approaches infinity]{Rayleigh-Plesset verification as $k\to\infty$}\label{subsec:rp_kinf_verification}

The Rayleigh-Plesset equation is integrated using a 3/4-order Rosenbrock method \cite{Hairer1996,Shampine1982}. This linearly implicit scheme pairs a third-order solution with an embedded fourth-order error estimator, enabling adaptive time stepping via step-size control based on the estimated local truncation error. Owing to its $L$-stability, the method is well suited for stiff regimes and remains stable when disparate time scales are present, which is common in cavitation bubble dynamics. Two benchmark cases, following Vallier \cite{Vallier2013}, are employed to verify the implementation of the bubble dynamics solver and to assess its numerical accuracy.

\begin{figure}[t]
	\centering
	\includegraphics[width=0.7\textwidth]{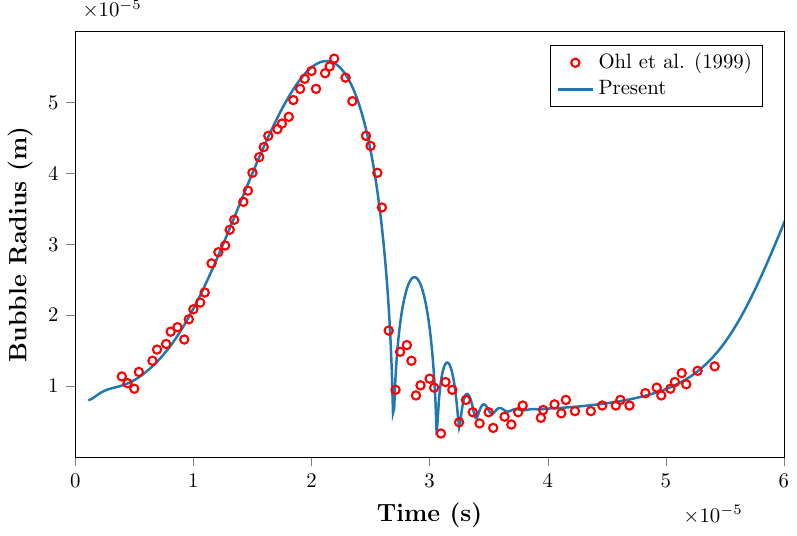} 
	\caption{The temporal evolution of the bubble radius}
 	\label{fig:bubble-2}
\end{figure}

The first validation case is based on the experimental measurements of Ohl et al. \cite{Ohl1999}, which are compared against numerical predictions obtained using the parameter set $R_0 = 8 \times 10^{-6}$, $p_v = 2500$, $\sigma_{st} = 0.07$, $\mu = 0.006~\mathrm{kg/(m\,s)}$, and $\kappa = 1.33$. The imposed liquid pressure is prescribed as $p_L(\boldsymbol{x}_B,t) = p_0 + \sin(2\pi f_0 t),$ with $p_0 = 100~\mathrm{kPa}$ and $f_0 = 21.4~\mathrm{kHz}$. To account for energy dissipation associated with liquid compressibility, which is not represented in the classical Rayleigh-Plesset equation, the parameter set proposed by Ohl et al. \cite{Ohl1999} is modified following the procedure outlined by Vallier \cite{Vallier2013}. In particular, the dynamic viscosity is increased to $\mu = 0.015~\mathrm{kg/(m\,s)}$, thereby effectively augmenting the damping during the collapse phase. Figure~\ref{fig:bubble-2} reports the results obtained with this adjusted parameter set. The predicted initial growth phase agrees closely with the experimental measurements; however, the subsequent expansion phase exhibits an overshoot relative to the experimental data. This discrepancy is likely attributable to the omission of compressibility effects in the Rayleigh-Plesset formulation.

\begin{figure}[H]
	\centering
	\includegraphics[width=0.7\textwidth]{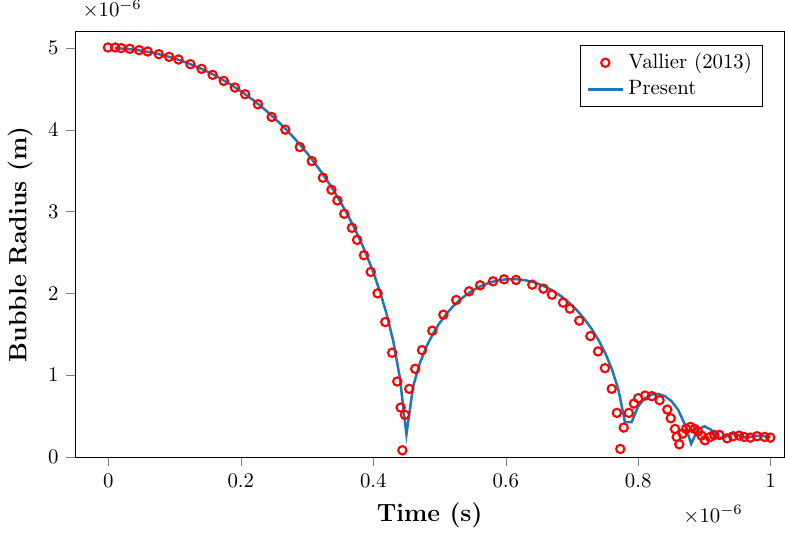}
	\caption{The temporal evolution of the bubble radius}
 	\label{fig:bubble-3}
\end{figure}

The second validation case follows the numerical study of Vallier \cite{Vallier2013}, which considers the relaxation of a bubble radius toward its equilibrium state. The bubble is initialized in a strongly non-equilibrium configuration, while the equilibrium reference state is characterized by the parameters $R_0$ and $p_0$. Specifically, the initial radius is prescribed as $R(0) = 20\,R_0 = 5~\mu\mathrm{m},$
and the ambient liquid pressure is held constant, $p_L(\boldsymbol{x}_B,t) = p_0 = 70~\mathrm{kPa}.$
As shown in Figure \ref{fig:bubble-3}, the bubble radius rapidly relaxes toward the equilibrium radius $R_0$, reaching equilibrium in less than $1~\mu\mathrm{s}$ in the present simulations. The predicted transient evolution exhibits good agreement with the results reported by Vallier \cite{Vallier2013}.

%% file: main.bbl
\begin{thebibliography}{10}
\expandafter\ifx\csname url\endcsname\relax
  \def\url#1{\texttt{#1}}\fi
\expandafter\ifx\csname urlprefix\endcsname\relax\def\urlprefix{URL }\fi
\expandafter\ifx\csname href\endcsname\relax
  \def\href#1#2{#2} \def\path#1{#1}\fi

\bibitem{Karimi1986}
A.~Karimi, J.~L. Martin,
  \href{https://journals.sagepub.com/doi/abs/10.1179/imtr.1986.31.1.1}{Cavitation
  erosion of materials}, International Metals Reviews 31~(1) (1986) 1--26.
\newblock \href
  {http://arxiv.org/abs/https://journals.sagepub.com/doi/pdf/10.1179/imtr.1986.31.1.1}
  {\path{arXiv:https://journals.sagepub.com/doi/pdf/10.1179/imtr.1986.31.1.1}},
  \href {https://doi.org/10.1179/imtr.1986.31.1.1}
  {\path{doi:10.1179/imtr.1986.31.1.1}}.
\newline\urlprefix\url{https://journals.sagepub.com/doi/abs/10.1179/imtr.1986.31.1.1}

\bibitem{kim2014}
K.-H. Kim, G.~Chahine, J.-P. Franc, A.~Karimi, Advanced experimental and
  numerical techniques for cavitation erosion prediction, Vol. 106, Springer,
  2014.

\bibitem{Feng2015}
Y.~Feng, L.~Zhao, G.~ter Haar, M.~Wan,
  \href{https://doi.org/10.1007/978-94-017-7255-6_9}{Cavitation Mechanobiology
  and Applications}, Springer Netherlands, Dordrecht, 2015, pp. 457--503.
\newblock \href {https://doi.org/10.1007/978-94-017-7255-6_9}
  {\path{doi:10.1007/978-94-017-7255-6_9}}.
\newline\urlprefix\url{https://doi.org/10.1007/978-94-017-7255-6_9}

\bibitem{GNANASKANDAN20191743}
A.~Gnanaskandan, C.-T. Hsiao, G.~Chahine,
  \href{https://www.sciencedirect.com/science/article/pii/S0301562919300912}{Modeling
  of microbubble-enhanced high-intensity focused ultrasound}, Ultrasound in
  Medicine \& Biology 45~(7) (2019) 1743--1761.
\newblock \href
  {https://doi.org/https://doi.org/10.1016/j.ultrasmedbio.2019.02.022}
  {\path{doi:https://doi.org/10.1016/j.ultrasmedbio.2019.02.022}}.
\newline\urlprefix\url{https://www.sciencedirect.com/science/article/pii/S0301562919300912}

\bibitem{KOKSAL2021}
Çağatay Sabri~Köksal, O.~Usta, B.~Aktas, M.~Atlar, E.~Korkut,
  \href{https://www.sciencedirect.com/science/article/pii/S0029801821011793}{Numerical
  prediction of cavitation erosion to investigate the effect of wake on marine
  propellers}, Ocean Engineering 239 (2021) 109820.
\newblock \href
  {https://doi.org/https://doi.org/10.1016/j.oceaneng.2021.109820}
  {\path{doi:https://doi.org/10.1016/j.oceaneng.2021.109820}}.
\newline\urlprefix\url{https://www.sciencedirect.com/science/article/pii/S0029801821011793}

\bibitem{Kawanami1997}
Y.~Kawanami, H.~Kato, H.~Yamaguchi, M.~Tanimura, Y.~Tagaya,
  \href{https://doi.org/10.1115/1.2819499}{Mechanism and control of cloud
  cavitation}, Journal of Fluids Engineering 119~(4) (1997) 788--794.
\newblock \href
  {http://arxiv.org/abs/https://asmedigitalcollection.asme.org/fluidsengineering/article-pdf/119/4/788/5757046/788\_1.pdf}
  {\path{arXiv:https://asmedigitalcollection.asme.org/fluidsengineering/article-pdf/119/4/788/5757046/788\_1.pdf}},
  \href {https://doi.org/10.1115/1.2819499} {\path{doi:10.1115/1.2819499}}.
\newline\urlprefix\url{https://doi.org/10.1115/1.2819499}

\bibitem{ganesh2016bubbly}
H.~Ganesh, S.~A. M{\"a}kiharju, S.~L. Ceccio, Bubbly shock propagation as a
  mechanism for sheet-to-cloud transition of partial cavities, Journal of Fluid
  Mechanics 802 (2016) 37--78.

\bibitem{LavariPOF2025}
M.~Lavari, A.~Gnanaskandan,
  \href{https://doi.org/10.1063/5.0304009}{Investigation of sheet to cloud
  cavitation dynamics using modal decomposition}, Physics of Fluids 37~(12)
  (2025) 123326.
\newblock \href {https://doi.org/10.1063/5.0304009}
  {\path{doi:10.1063/5.0304009}}.
\newline\urlprefix\url{https://doi.org/10.1063/5.0304009}

\bibitem{Dhruv2026}
D.~Apte, M.~Lavari, D.~V.~Revelo, A.~Gnanaskandan, O.~Coutier-Delgosha,
\href{https://doi.org/10.1063/5.0301584}{Investigation of cavitation over a wedge using different cavitation and turbulence models in a cross-platform study},
Physics of Fluids 38~(1) (2026) 013345.
\newblock \href{https://doi.org/10.1063/5.0301584}
{\path{doi:10.1063/5.0301584}}.
\newline\urlprefix\url{https://doi.org/10.1063/5.0301584}

\bibitem{ABDELMAKSOUD20101065}
M.~AbdelMaksoud, D.~Hänel, U.~Lantermann,
  \href{https://www.sciencedirect.com/science/article/pii/S0142727X10000998}{Modeling
  and computation of cavitation in vortical flow}, International Journal of
  Heat and Fluid Flow 31~(6) (2010) 1065--1074, 7th World Conference on
  Experimental Heat Transfer, Fluid Mechanics and Thermodynamics (ExHFT-7),
  Krakow and The Conference on Modelling Fluid Flow (CMFF ’09), Budapest.
\newblock \href
  {https://doi.org/https://doi.org/10.1016/j.ijheatfluidflow.2010.05.010}
  {\path{doi:https://doi.org/10.1016/j.ijheatfluidflow.2010.05.010}}.
\newline\urlprefix\url{https://www.sciencedirect.com/science/article/pii/S0142727X10000998}

\bibitem{Shams2011}
E.~Shams, J.~Finn, S.~V. Apte,
  \href{https://onlinelibrary.wiley.com/doi/abs/10.1002/fld.2452}{A numerical
  scheme for euler–lagrange simulation of bubbly flows in complex systems},
  International Journal for Numerical Methods in Fluids 67~(12) (2011)
  1865--1898.
\newblock \href
  {http://arxiv.org/abs/https://onlinelibrary.wiley.com/doi/pdf/10.1002/fld.2452}
  {\path{arXiv:https://onlinelibrary.wiley.com/doi/pdf/10.1002/fld.2452}},
  \href {https://doi.org/https://doi.org/10.1002/fld.2452}
  {\path{doi:https://doi.org/10.1002/fld.2452}}.
\newline\urlprefix\url{https://onlinelibrary.wiley.com/doi/abs/10.1002/fld.2452}

\bibitem{Yakubov2011}
S.~Yakubov, B.~Cankurt, T.~Maquil, P.~Schiller, M.~Abdel-Maksoud, T.~Rung,
  Euler-euler and euler-lagrange approaches to cavitation modelling in marine
  applications, MARINE 2011 - Computational Methods in Marine Engineering IV
  (2011) 544--555.

\bibitem{YAKUBOV2013365}
S.~Yakubov, B.~Cankurt, M.~Abdel-Maksoud, T.~Rung,
  \href{https://www.sciencedirect.com/science/article/pii/S0045793012000291}{Hybrid
  mpi/openmp parallelization of an euler–lagrange approach to cavitation
  modelling}, Computers \& Fluids 80 (2013) 365--371, selected contributions of
  the 23rd International Conference on Parallel Fluid Dynamics ParCFD2011.
\newblock \href
  {https://doi.org/https://doi.org/10.1016/j.compfluid.2012.01.020}
  {\path{doi:https://doi.org/10.1016/j.compfluid.2012.01.020}}.
\newline\urlprefix\url{https://www.sciencedirect.com/science/article/pii/S0045793012000291}

\bibitem{Ma2015a}
J.~Ma, C.-T. Hsiao, G.~L. Chahine,
  \href{https://doi.org/10.1115/1.4028853}{Euler–lagrange simulations of
  bubble cloud dynamics near a wall}, Journal of Fluids Engineering 137~(4)
  (2015) 041301.
\newblock \href {https://doi.org/10.1115/1.4028853}
  {\path{doi:10.1115/1.4028853}}.
\newline\urlprefix\url{https://doi.org/10.1115/1.4028853}

\bibitem{Ma2015b}
J.~Ma, C.-T. Hsiao, G.~L. Chahine,
  \href{https://doi.org/10.1115/1.4030919}{Shared-memory parallelization for
  two-way coupled euler–lagrange modeling of cavitating bubbly flows},
  Journal of Fluids Engineering 137~(12) (2015) 121106.
\newblock \href {https://doi.org/10.1115/1.4030919}
  {\path{doi:10.1115/1.4030919}}.
\newline\urlprefix\url{https://doi.org/10.1115/1.4030919}

\bibitem{MAEDA2018994}
K.~Maeda, T.~Colonius,
  \href{https://www.sciencedirect.com/science/article/pii/S0021999118303358}{Eulerian–lagrangian
  method for simulation of cloud cavitation}, Journal of Computational Physics
  371 (2018) 994--1017.
\newblock \href {https://doi.org/https://doi.org/10.1016/j.jcp.2018.05.029}
  {\path{doi:https://doi.org/10.1016/j.jcp.2018.05.029}}.
\newline\urlprefix\url{https://www.sciencedirect.com/science/article/pii/S0021999118303358}

\bibitem{Apte2009}
S.~V. Apte, E.~Shams, J.~Finn, A hybrid lagrangian--eulerian approach for
  simulation of bubble dynamics, in: Proceedings of the 7th International
  Symposium on Cavitation (CAV2009), no.~74, 2009.

\bibitem{Vallier2013}
A.~Vallier, Simulations of cavitation: From the large vapour structures to the
  small bubble dynamics, Doctoral thesis (compilation), Fluid Mechanics (2013).

\bibitem{HSIAO2017102}
C.-T. Hsiao, J.~Ma, G.~L. Chahine,
  \href{https://www.sciencedirect.com/science/article/pii/S0301932216304220}{Multiscale
  tow-phase flow modeling of sheet and cloud cavitation}, International Journal
  of Multiphase Flow 90 (2017) 102--117.
\newblock \href
  {https://doi.org/https://doi.org/10.1016/j.ijmultiphaseflow.2016.12.007}
  {\path{doi:https://doi.org/10.1016/j.ijmultiphaseflow.2016.12.007}}.
\newline\urlprefix\url{https://www.sciencedirect.com/science/article/pii/S0301932216304220}

\bibitem{Hsiao2015a}
C.-T. Hsiao, J.~ma, G.~Chahine, Simulation of sheet and tip vortex cavitation
  on a rotating propeller using a multiscale two-phase flow model, in: 4th
  International Symposium on Marine Propulsors (SMP'15), Austin, TX, USA, 2015.

\bibitem{MA201768}
J.~Ma, C.-T. Hsiao, G.~L. Chahine,
  \href{https://www.sciencedirect.com/science/article/pii/S0045793016303930}{A
  physics based multiscale modeling of cavitating flows}, Computers \& Fluids
  145 (2017) 68--84.
\newblock \href
  {https://doi.org/https://doi.org/10.1016/j.compfluid.2016.12.010}
  {\path{doi:https://doi.org/10.1016/j.compfluid.2016.12.010}}.
\newline\urlprefix\url{https://www.sciencedirect.com/science/article/pii/S0045793016303930}

\bibitem{Lidtke2017}
A.~Lidtke, Predicting radiated noise of marine propellers using acoustic
  analogies and hybrid eulerian-lagrangian cavitation models, Ph.D. thesis,
  University of Southampton (06 2017).

\bibitem{GHAHRAMANI2019339}
E.~Ghahramani, M.~H. Arabnejad, R.~E. Bensow,
  \href{https://www.sciencedirect.com/science/article/pii/S0301932218303197}{A
  comparative study between numerical methods in simulation of cavitating
  bubbles}, International Journal of Multiphase Flow 111 (2019) 339--359.
\newblock \href
  {https://doi.org/https://doi.org/10.1016/j.ijmultiphaseflow.2018.10.010}
  {\path{doi:https://doi.org/10.1016/j.ijmultiphaseflow.2018.10.010}}.
\newline\urlprefix\url{https://www.sciencedirect.com/science/article/pii/S0301932218303197}

\bibitem{Ghahramani2021}
E.~Ghahramani, H.~Str{\"o}m, R.~Bensow, Numerical simulation and analysis of
  multi-scale cavitating flows, Journal of Fluid Mechanics 922 (2021) A22.
\newblock \href {https://doi.org/10.1017/jfm.2021.424}
  {\path{doi:10.1017/jfm.2021.424}}.

\bibitem{Madabhushi_Mahesh_2023}
A.~Madabhushi, K.~Mahesh, A compressible multi-scale model to simulate
  cavitating flows, Journal of Fluid Mechanics 961 (2023) A6.
\newblock \href {https://doi.org/10.1017/jfm.2023.192}
  {\path{doi:10.1017/jfm.2023.192}}.

\bibitem{Zhao2024}
X.~Zhao, H.~Cheng, B.~Ji, L.~Li, R.~E. Bensow,
  \href{https://link.aps.org/doi/10.1103/PhysRevFluids.9.104304}{Insights into
  the characteristics of sheet/cloud cavitation and tip-leakage cavitation
  based on a compressible euler-lagrange model}, Phys. Rev. Fluids 9 (2024)
  104304.
\newblock \href {https://doi.org/10.1103/PhysRevFluids.9.104304}
  {\path{doi:10.1103/PhysRevFluids.9.104304}}.
\newline\urlprefix\url{https://link.aps.org/doi/10.1103/PhysRevFluids.9.104304}

\bibitem{Wang2024}
X.~Wang, M.~Song, H.~Cheng, B.~Ji, L.~Li, A multiscale euler–lagrange model
  for high-frequency cavitation noise prediction, Journal of Fluids Engineering
  146~(6) (2024) 061502.
\newblock \href {https://doi.org/10.1115/1.4064296}
  {\path{doi:10.1115/1.4064296}}.

\bibitem{WANG2025120398}
Z.~Wang, H.~Cheng, X.~Luo, B.~Ji,
  \href{https://www.sciencedirect.com/science/article/pii/S0029801825001131}{Multiscale
  numerical investigation of bubble distribution characteristics and cavitation
  erosion risk evolution with special emphasis on the influence of cavitation
  number}, Ocean Engineering 321 (2025) 120398.
\newblock \href
  {https://doi.org/https://doi.org/10.1016/j.oceaneng.2025.120398}
  {\path{doi:https://doi.org/10.1016/j.oceaneng.2025.120398}}.
\newline\urlprefix\url{https://www.sciencedirect.com/science/article/pii/S0029801825001131}

\bibitem{Wang2025ijfe}
X.~Wang, Y.~Wang, H.~Cheng, B.~Ji,
  \href{https://doi.org/10.1063/5.0238990}{Multiscale modeling of wake-induced
  propeller cavity bursting}, International Journal of Fluid Engineering 2~(1)
  (2025) 013502.
\newblock \href {https://doi.org/10.1063/5.0238990}
  {\path{doi:10.1063/5.0238990}}.
\newline\urlprefix\url{https://doi.org/10.1063/5.0238990}

\bibitem{Yang2024}
Z.~Yang, X.~Wang, M.~Song, H.~Cheng, B.~Ji,
  \href{https://doi.org/10.1063/5.0239033}{Numerical investigation of tip
  vortex cavitation noise with an emphasis on environmental nucleation effect},
  Physics of Fluids 36~(12) (2024) 122134.
\newblock \href {https://doi.org/10.1063/5.0239033}
  {\path{doi:10.1063/5.0239033}}.
\newline\urlprefix\url{https://doi.org/10.1063/5.0239033}

\bibitem{QIN2025105142}
X.~Qin, Y.~Chen, X.~Feng, X.~Shao, J.~Deng,
  \href{https://www.sciencedirect.com/science/article/pii/S0301932225000205}{A
  numerical investigation concerning wall-nucleation effects of the inception
  of sheet cavitation}, International Journal of Multiphase Flow 185 (2025)
  105142.
\newblock \href
  {https://doi.org/https://doi.org/10.1016/j.ijmultiphaseflow.2025.105142}
  {\path{doi:https://doi.org/10.1016/j.ijmultiphaseflow.2025.105142}}.
\newline\urlprefix\url{https://www.sciencedirect.com/science/article/pii/S0301932225000205}

\bibitem{Zhang2025}
B.~Zhang, X.~Qin, Y.~Chen, X.~Shao, J.~Deng, Numerical analysis of nucleation
  effects on cloud cavitation using multiscale euler–lagrange model, Physics
  of Fluids 37~(2) (2025) 022103.
\newblock \href {https://doi.org/10.1063/5.0249952}
  {\path{doi:10.1063/5.0249952}}.

\bibitem{LavariFEDSM2025}
M.~Lavari, A.~Gnanaskandan, \href{https://doi.org/10.1115/FEDSM2025-155755}{A
  hybrid euler-lagrange multiscale model for nuclei-initiated cavitation}, in:
  Proceedings of the ASME Fluids Engineering Division Summer Meeting
  (FEDSM2025), ASME, 2025, p. V002T07A001.
\newblock \href {https://doi.org/10.1115/FEDSM2025-155755}
  {\path{doi:10.1115/FEDSM2025-155755}}.
\newline\urlprefix\url{https://doi.org/10.1115/FEDSM2025-155755}

\bibitem{Herrmann2010}
M.~Herrmann, A parallel {E}ulerian interface tracking/{L}agrangian point
  particle multi-scale coupling procedure, Journal of Computational Physics
  229~(3) (2010) 745--759.
\newblock \href {https://doi.org/10.1016/j.jcp.2009.10.009}
  {\path{doi:10.1016/j.jcp.2009.10.009}}.

\bibitem{Gao2021ON}
Q.~Gao, G.~B. Deane, H.~Liu, L.~Shen, A robust and accurate technique for
  {L}agrangian tracking of bubbles and drops and detecting fragmentation and
  coalescence, International Journal of Multiphase Flow 135 (2021) 103523.
\newblock \href {https://doi.org/10.1016/j.ijmultiphaseflow.2020.103523}
  {\path{doi:10.1016/j.ijmultiphaseflow.2020.103523}}.

\bibitem{Chan2021Lineage}
W.~H.~R. Chan, M.~S. Dodd, P.~L. Johnson, P.~Moin, Identifying and tracking
  bubbles and drops in simulations: A toolbox for obtaining sizes, lineages,
  and breakup and coalescence statistics, Journal of Computational Physics 432
  (2021) 110156.
\newblock \href {https://doi.org/10.1016/j.jcp.2021.110156}
  {\path{doi:10.1016/j.jcp.2021.110156}}.

\bibitem{Mangani2022}
F.~Mangani, G.~Soligo, A.~Roccon, A.~Soldati, Influence of density and
  viscosity on deformation, breakage, and coalescence of bubbles in turbulence,
  Physics Review Fluids 7~(5) (2022) 053601.
\newblock \href {https://doi.org/10.1103/PhysRevFluids.7.053601}
  {\path{doi:10.1103/PhysRevFluids.7.053601}}.

\bibitem{Bussmann2022Tracking}
A.~Bu{\ss}mann, J.~Buchmeier, M.~S. Dodd, S.~Adami, I.~Bermejo-Moreno, Tracking
  and analysis of interfaces and flow structures in multiphase flows, Computers
  \& Fluids 248 (2022) 105665.
\newblock \href {https://doi.org/10.1016/j.compfluid.2022.105665}
  {\path{doi:10.1016/j.compfluid.2022.105665}}.

\bibitem{GonzalezDiaz2018Topological}
R.~Gonz{\'a}lez-D{\'i}az, M.-J. Jim{\'e}nez, B.~Medrano, Topological tracking
  of connected components in image sequences, Journal of Computer and System
  Sciences 95 (2018) 134--142.
\newblock \href {https://doi.org/10.1016/j.jcss.2018.01.004}
  {\path{doi:10.1016/j.jcss.2018.01.004}}.

\bibitem{Li2022Multiscale}
L.~Li, B.~Jiang, G.~Wei, X.~Li, Z.~Zhu, Multiscale multiphase flow simulations
  using interface capturing and lagrangian particle tracking, Physics of Fluids
  34 (2022) 121801.
\newblock \href {https://doi.org/10.1063/5.0134102}
  {\path{doi:10.1063/5.0134102}}.

\bibitem{HEINRICH2020}
M.~Heinrich, R.~Schwarze,
  \href{https://www.sciencedirect.com/science/article/pii/S2352711020300303}{3d-coupling
  of volume-of-fluid and lagrangian particle tracking for spray atomization
  simulation in openfoam}, SoftwareX 11 (2020) 100483.
\newblock \href {https://doi.org/https://doi.org/10.1016/j.softx.2020.100483}
  {\path{doi:https://doi.org/10.1016/j.softx.2020.100483}}.
\newline\urlprefix\url{https://www.sciencedirect.com/science/article/pii/S2352711020300303}

\bibitem{Liu2024Kuhn}
Q.~Liu, Y.~Wu, N.~Gui, X.~Yang, J.~Tu, S.~Jiang, Bubble tracking method based
  on {K}uhn--{M}unkres algorithm for boiling two-phase flow study,
  International Journal of Heat and Mass Transfer 226 (2024) 125436.
\newblock \href {https://doi.org/10.1016/j.ijheatmasstransfer.2024.125436}
  {\path{doi:10.1016/j.ijheatmasstransfer.2024.125436}}.

\bibitem{Rubel2019Owkes}
C.~Rubel, M.~Owkes, Extraction of droplet genealogies from high-fidelity
  atomization simulations, Atomization and Sprays 29~(8) (2019) 709--739.
\newblock \href {https://doi.org/10.1615/AtomizSpr.2019032023}
  {\path{doi:10.1615/AtomizSpr.2019032023}}.

\bibitem{Gaylo2022}
D.~B. Gaylo, K.~Hendrickson, D.~K.-P. Yue, An {E}ulerian label advection method
  for conservative volume-based tracking of bubbles/droplets, Journal of
  Computational Physics 470 (2022) 111560.
\newblock \href {https://doi.org/10.1016/j.jcp.2022.111560}
  {\path{doi:10.1016/j.jcp.2022.111560}}.

\bibitem{Basak2026ON}
S.~Basak, U.~C. Bitencourt, Q.~Gao, G.~B. Deane, M.~D. Stokes, L.~Shen, A
  parallel algorithm for detection and tracking of non-binary fragmentation and
  coalescence of droplets and bubbles in numerical simulations of multiphase
  flows, Journal of Computational Physics 545 (2026) 114441.
\newblock \href {https://doi.org/10.1016/j.jcp.2025.114441}
  {\path{doi:10.1016/j.jcp.2025.114441}}.

\bibitem{KUNZ2000}
R.~F. Kunz, D.~A. Boger, D.~R. Stinebring, T.~S. Chyczewski, J.~W. Lindau,
  H.~J. Gibeling, S.~Venkateswaran, T.~R. Govindan,
  \href{https://www.sciencedirect.com/science/article/pii/S0045793099000390}{A
  preconditioned navier–stokes method for two-phase flows with application to
  cavitation prediction}, Computers \& Fluids 29~(8) (2000) 849--875.
\newblock \href {https://doi.org/https://doi.org/10.1016/S0045-7930(99)00039-0}
  {\path{doi:https://doi.org/10.1016/S0045-7930(99)00039-0}}.
\newline\urlprefix\url{https://www.sciencedirect.com/science/article/pii/S0045793099000390}

\bibitem{Merkle1998}
C.~L. Merkle, J.~Feng, P.~E.~O. Buelow,
  \href{https://api.semanticscholar.org/CorpusID:116987965}{Computational
  modeling of the dynamics of sheet cavitation}, in: Proceedings of the 3rd
  International Symposium on Cavitation, Grenoble, France, 1998, pp. 307--311.
\newline\urlprefix\url{https://api.semanticscholar.org/CorpusID:116987965}

\bibitem{Singhal2002}
A.~K. Singhal, M.~M. Athavale, H.~Li, Y.~Jiang,
  \href{https://doi.org/10.1115/1.1486223}{Mathematical basis and validation of
  the full cavitation model}, Journal of Fluids Engineering 124~(3) (2002)
  617--624.
\newblock \href
  {http://arxiv.org/abs/https://asmedigitalcollection.asme.org/fluidsengineering/article-pdf/124/3/617/5901243/617\_1.pdf}
  {\path{arXiv:https://asmedigitalcollection.asme.org/fluidsengineering/article-pdf/124/3/617/5901243/617\_1.pdf}},
  \href {https://doi.org/10.1115/1.1486223} {\path{doi:10.1115/1.1486223}}.
\newline\urlprefix\url{https://doi.org/10.1115/1.1486223}

\bibitem{saito2007numerical}
Y.~Saito, R.~Takami, I.~Nakamori, T.~Ikohagi, Numerical analysis of unsteady
  behavior of cloud cavitation around a naca0015 foil, Computational Mechanics
  40 (2007) 85--96.

\bibitem{Schnerr2001}
G.~H. Schnerr, J.~Sauer, Physical and numerical modeling of unsteady cavitation
  dynamics, in: 4th International Conference on Multiphase Flow, New Orleans,
  LA, USA, 2001.

\bibitem{AmsdenORourkeButler1989KIVAII}
A.~A. Amsden, P.~J. O'Rourke, T.~D. Butler,
  \href{https://www.osti.gov/biblio/6228444}{Kiva-ii: A computer program for
  chemically reactive flows with sprays}, Tech. Rep. LA-11560-MS, Los Alamos
  National Laboratory, Los Alamos, NM (May 1989).
\newline\urlprefix\url{https://www.osti.gov/biblio/6228444}

\bibitem{Plesset1977}
M.~S. Plesset, A.~Prosperetti,
  \href{https://www.annualreviews.org/content/journals/10.1146/annurev.fl.09.010177.001045}{Bubble
  dynamics and cavitation}, Annual Review of Fluid Mechanics 9~(Volume 9, 1977)
  (1977) 145--185.
\newblock \href
  {https://doi.org/https://doi.org/10.1146/annurev.fl.09.010177.001045}
  {\path{doi:https://doi.org/10.1146/annurev.fl.09.010177.001045}}.
\newline\urlprefix\url{https://www.annualreviews.org/content/journals/10.1146/annurev.fl.09.010177.001045}

\bibitem{weller1998tensorial}
H.~G. Weller, G.~Tabor, H.~Jasak, C.~Fureby, A tensorial approach to
  computational continuum mechanics using object-oriented techniques, Computers
  in physics 12~(6) (1998) 620--631.

\bibitem{Lavari2024}
M.~Lavari, interphasechangebubblefoam: A hybrid eulerian-lagrangian solver to
  capture nuclei effects on hydrodynamic cavitation, in: H.~Nilsson (Ed.),
  Proceedings of CFD with OpenSource Software, 2024.
\newblock \href {https://doi.org/http://dx.doi.org/10.17196/OS_CFD#YEAR_2024}
  {\path{doi:http://dx.doi.org/10.17196/OS_CFD#YEAR_2024}}.

\bibitem{greenshieldsweller2022}
C.~Greenshields, H.~Weller, Notes on Computational Fluid Dynamics: General
  Principles, CFD Direct Ltd, Reading, UK, 2022.

\bibitem{Nhan2021}
T.~A. Nhan, N.~Madden, \href{https://doi.org/10.1007/s12190-020-01390-z}{An
  analysis of diagonal and incomplete cholesky preconditioners for singularly
  perturbed problems on layer-adapted meshes}, Journal of Applied Mathematics
  and Computing 65~(1) (2021) 245--272.
\newblock \href {https://doi.org/10.1007/s12190-020-01390-z}
  {\path{doi:10.1007/s12190-020-01390-z}}.
\newline\urlprefix\url{https://doi.org/10.1007/s12190-020-01390-z}

\bibitem{Lavari2024SheetCloud}
M.~Lavari, D.~Vaca-Revelo, A.~Gnanaskandan, Multiscale modelling of sheet to
  cloud cavitation transition, in: Proceedings of the 12th International
  Cavitation Symposium – CAV2024, MAICh Conference Centre in Chania, Greece,
  2024.

\bibitem{Oweis2005}
G.~F. Oweis, I.~E. van~der Hout, C.~Iyer, G.~Tryggvason, S.~L. Ceccio,
  \href{https://doi.org/10.1063/1.1834916}{Capture and inception of bubbles
  near line vortices}, Physics of Fluids 17~(2) (2005) 022105.
\newblock \href {https://doi.org/10.1063/1.1834916}
  {\path{doi:10.1063/1.1834916}}.
\newline\urlprefix\url{https://doi.org/10.1063/1.1834916}

\bibitem{WANG2025105412}
Z.~Wang, D.~Liu, X.~Luo,
  \href{https://www.sciencedirect.com/science/article/pii/S0301932225002885}{Analysis
  of compressibility effects on multiscale cavitating flow in the nozzle using
  eulerian-lagrangian method}, International Journal of Multiphase Flow 193
  (2025) 105412.
\newblock \href
  {https://doi.org/https://doi.org/10.1016/j.ijmultiphaseflow.2025.105412}
  {\path{doi:https://doi.org/10.1016/j.ijmultiphaseflow.2025.105412}}.
\newline\urlprefix\url{https://www.sciencedirect.com/science/article/pii/S0301932225002885}

\bibitem{Hairer1996}
E.~Hairer, S.~P. Nørsett, G.~Wanner, Solving Ordinary Differential Equations
  II: Stiff and Differential-Algebraic Problems, 2nd Edition, Springer-Verlag,
  Berlin, 1996.

\bibitem{Shampine1982}
L.~F. Shampine, Implementation of rosenbrock methods, ACM Transactions on
  Mathematical Software 8 (1982) 93--113.

\bibitem{Ohl1999}
C.~D. Ohl, T.~Kurz, R.~Geisler, O.~Lindau, W.~Lauterborn, Bubble dynamics,
  shock waves and sonoluminescence, Philosophical Transactions of the Royal
  Society of London A. Mathematical, Physical and Engineering Sciences
  357~(357) (1999) 269--294.

\bibitem{Lavari2026SNH}
M.~Lavari, A.~Gnanaskandan, J.~Lindau, D.~Leonard, X.~Yang, R.~Kunz, F.~Thomas,
  A.~Pelster, Lagrangian-Eulerian modeling of cavitation with robust Eulerian
  numerics, in: Proceedings of the 36th Symposium on Naval Hydrodynamics,
  Busan, Korea, 2026.

\bibitem{Lavari2026USNCTAM}
M.~Lavari, R.~Kunz, A.~Gnanaskandan, Bridging microscale inception and
  largescale cavitation via a hybrid Eulerian-Lagrangian approach, in:
  Proceedings of the 20th U.S. National Congress on Theoretical and Applied
  Mechanics (USNC/TAM 2026), Pasadena, California, USA, 2026.

\bibitem{MENON2026134117}
S.~S. Menon, M.~Lavari, A.~Kokernak, J.~Mathew, C.~A. Jagtap, J.~Jayachandran,
  A.~Gnanaskandan, A.~D. Jagtap,
  \href{https://www.sciencedirect.com/science/article/pii/S0925231226015158}{Intelligent
  fluid flows: A survey of deep learning methods for turbulent flows,
  multiphase flows, and combustion}, Neurocomputing 697 (2026) 134117.
\newblock \href{https://doi.org/10.1016/j.neucom.2026.134117}
  {\path{doi:https://doi.org/10.1016/j.neucom.2026.134117}}.
\newline\urlprefix\url{https://www.sciencedirect.com/science/article/pii/S0925231226015158}


\bibitem{Brackbill1992}
J.~U. Brackbill, D.~B. Kothe, C.~Zemach, A continuum method for modeling
  surface tension, Journal of Computational Physics 100~(2) (1992) 335--354.
\newblock \href {https://doi.org/10.1016/0021-9991(92)90240-Y}
  {\path{doi:10.1016/0021-9991(92)90240-Y}}.

\end{thebibliography}
